\documentclass[twocolumn]{aastex63}

\accepted{May 13, 2021}

\shorttitle{AGN-Driven Outflows in NGC 4051}
\shortauthors{Meena et al.}

\usepackage{float}
\usepackage{color}
\usepackage{comment}
\usepackage{graphics}
\usepackage{epstopdf}
\usepackage{microtype}
\usepackage{subfigure}
\usepackage[normalem]{ulem}
\graphicspath{{./}{figures/}}
\usepackage{capt-of}
\usepackage{wasysym}
\usepackage{scalerel}

\newcommand{\othree}{[O~III]~}
\newcommand{\halpha}{H$\alpha$~}
\newcommand{\hbeta}{H$\beta$~}
\newcommand{\hst}{{\it HST}~}

\newcommand{\kms}{km~s$^{-1}$~}
\newcommand{\app}{$\approx$~}
\definecolor{malachite}{rgb}{0.01, 0.8, 0.24}

\usepackage{soul,xcolor}
\setstcolor{red}

\begin{document}

\title{Radiative Driving of the AGN Outflows in the Narrow-Line Seyfert 1 Galaxy NGC~4051\footnote{Based on observations made with the NASA/ESA Hubble Space Telescope, obtained from the Data Archive at the Space Telescope Science Institute, which is operated by the Association of Universities for Research in Astronomy, Inc., under NASA contract NAS 5-26555. These observations are associated with program No. 8253, 12212.}\footnote{Based in part on observations obtained with the Apache Point Observatory 3.5-meter telescope, which is owned and operated by the Astrophysical Research Consortium.}}

\correspondingauthor{Beena Meena}
\email{bmeena@astro.gsu.edu}

\author[0000-0001-8658-2723]{Beena Meena}
\affil{Department of Physics and Astronomy,
Georgia State University,
25 Park Place, Suite 605,
Atlanta, GA 30303, USA}

\author[0000-0002-6465-3639]{D. Michael Crenshaw}
\affil{Department of Physics and Astronomy,
Georgia State University,
25 Park Place, Suite 605,
Atlanta, GA 30303, USA}

\author[0000-0003-2450-3246]{Henrique R. Schmitt}
\affil{Naval Research Laboratory,
Washington, DC 20375, USA}

\author[0000-0002-4917-7873]{Mitchell Revalski}
\affil{Space Telescope Science Institute, 3700 San Martin Drive, Baltimore, MD 21218, USA}

\author[0000-0002-3365-8875]{Travis C. Fischer}
\affiliation{AURA for ESA, Space Telescope Science Institute, 3700 San Martin Drive, Baltimore, MD 21218, USA}

\author[0000-0001-5862-2150]{Garrett E. Polack}
\affil{Department of Physics and Astronomy,
Georgia State University,
25 Park Place, Suite 605,
Atlanta, GA 30303, USA}

\author[0000-0002-6928-9848]{Steven B. Kraemer}
\affil{Institute for Astrophysics and Computational Sciences,
Department of Physics,
The Catholic University of America,
Washington, DC 20064, USA}

\author[0000-0001-8122-0037]{Dzhuliya Dashtamirova}
\affil{Space Telescope Science Institute, 3700 San Martin Drive, Baltimore, MD 21218, USA}

\begin{abstract}
We explore the properties of ionized gas in the nuclear and circumnuclear environment of the narrow-line Seyfert 1 galaxy NGC~4051 using spectroscopic and imaging observations from the {\it Hubble Space Telescope} (\textit{HST}) and Apache Point Observatory (APO)'s ARC 3.5m Telescope. We identify an unresolved moderate-density intermediate width component and a high-density broad component in the optical emission lines from the active nucleus, as well as spatially-resolved emission extending up to $\sim$1 kpc in the AGN ionized narrow-line region (NLR) and $\sim$8 kpc in the stellar ionized host galaxy. The \hst narrow-band image reveals a distinct conical structure in \othree emission towards the NE, and the ionized gas kinematics shows up to two blueshifted velocity components, indicating outflows along the edges of a cone. We introduce an improved model of biconical outflow, with our line of sight passing through the wall of the cone, which suggests that the large number of outflowing UV absorbers seen in NGC~4051 are NLR clouds in absorption.
Using the de-projection factors from the biconical geometry, we measure true outflow velocities up to 680 \kms at a distance of $\sim$350 pc, however, we do not find any rotational signature inside a projected distance $\leq$ 10\arcsec\ ($\sim$800 pc) from the nucleus.
We compare the gas kinematics with analytical models based on a radiation-gravity formalism, which show that most of the observed NLR outflows are launched within $\sim$0.5 pc of the nucleus and can travel up to $\sim$1 kpc from this low-luminosity AGN.
\end{abstract}


\vspace{7ex}
\section{INTRODUCTION}\label{sec:intro}

\subsection{AGN Driven Feedback} \label{subsec:feedback}

\begin{figure*}[ht!]
\centering
\includegraphics[width=\textwidth]{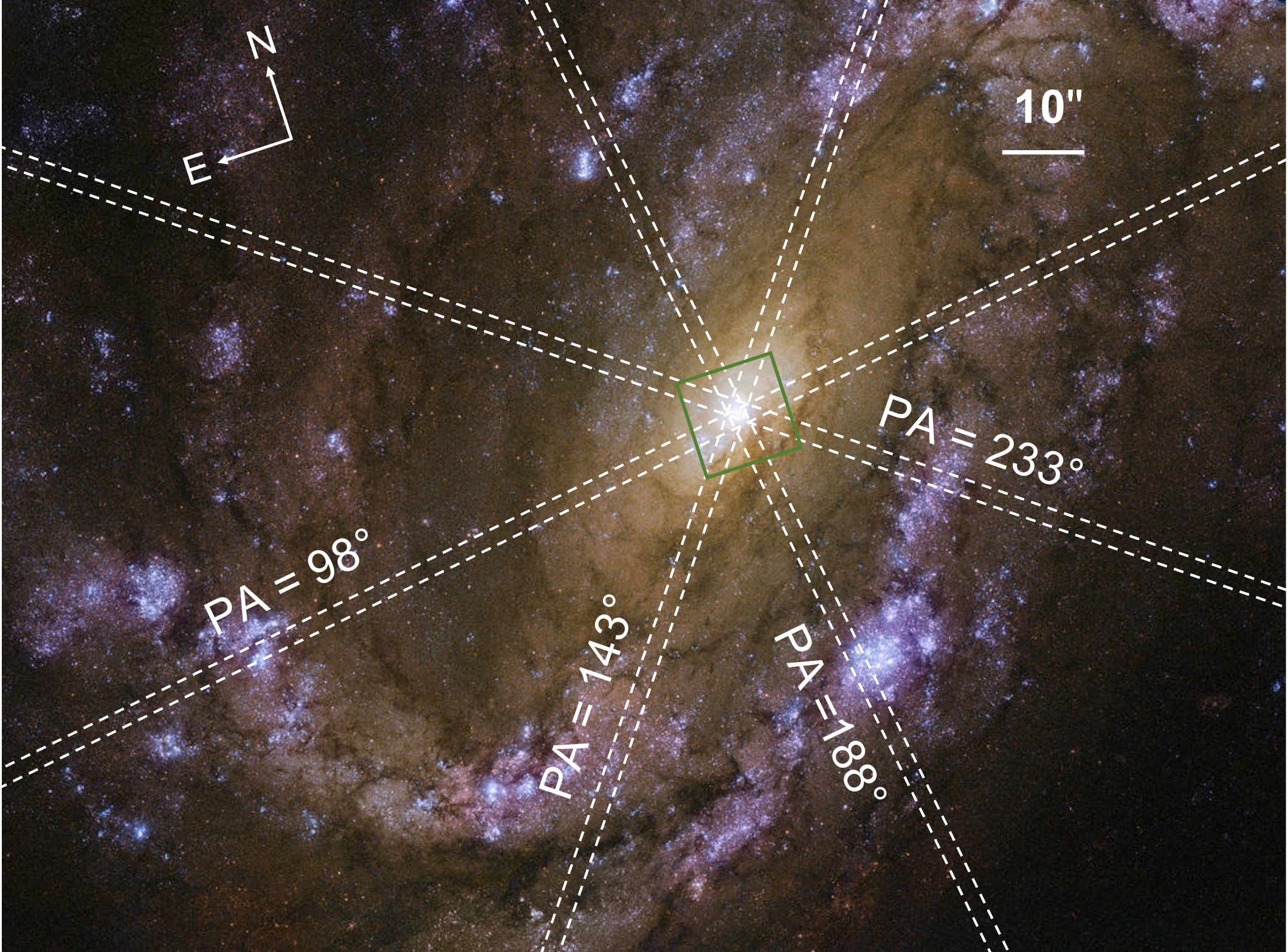}
\caption{A 2.60\arcmin\ $\times$ 1.93\arcmin\ color-composite image of NGC~4051 using multiple {\it HST} WFC3 filters. The image shows the orientation of the host galaxy, spiral arms, and the star forming regions. The [O~III] ionized gas inside the 12\arcsec\ $\times$ 12\arcsec\ green box at the center is shown in Figure~\ref{fig:F502N}. North is 18\arcdeg\ counter-clockwise, and the 2\arcsec-wide APO slits are overlaid and labeled with their position angles. Image credit: NASA/ESA/Hubble, D.~M. Crenshaw and O. Fox. \label{fig:ngc4051_galaxy}}
\end{figure*}

Active Galactic Nuclei (AGN) are powered by supermassive black holes (SMBHs) with masses $\geq$ 10$^{6} M_{\odot}$ that are feeding on the surrounding gas at the centers of their host galaxies. This feeding process creates massive amounts of electromagnetic radiation from the accretion disk of the SMBH that can lead to AGN feedback. This feedback is usually detected in the form of jets and/or winds \citep{Crenshaw2003,Fabian2012,King2015,Blandford2019}. Jets are collimated beams of extremely high energy particles that travel at relativistic speeds and can impact large-scale extragalactic environments. AGN driven winds are outflows of ionized \citep{Storchi-Bergmann2010,Laha2021}), neutral \citep{Rupke2005,Rupke2017}, and/or molecular  \citep{Veilleux2017,Herrera-Camus2019} gas that can be produced by the interaction of radiation from the accretion disk around the SMBH with the surrounding matter. They predominantly influence the nuclear and circumnuclear regions of their host galaxies. These outflows can push the fueling gas away from the nucleus, and propagate kinetic energy into the interstellar medium, thereby disrupting  potentially star-forming gas. Consequently, this feeding-feedback mechanism may regulate the growth of SMBHs and the properties of the  galactic bulge, possibly explaining empirical relationships such as the BH mass - bulge luminosity (M$_{\bullet}$ - $L_{bulge}$:  \citealp{Kormendy1995,Magorrian1998, Marconi2003}) and BH mass - bulge stellar velocity dispersion ($M_{\bullet}$ - $\sigma_{\star}$:\citealp{Ferrarese2000,Gebhardt2000,Kormendy2013}) of the present day universe.

The ionized gas outflows are often observed in X-ray $\&$ UV absorption \citep{crenshaw2005,Costantini2010,Crenshaw2012_n} and optical emission lines in the narrow-line regions (NLRs) of nearby Seyfert galaxies \citep{Veilleux1987,Veilleux1991,Colbert1996,Nelson2000,Fischer2013}. These outflows have been seen in a roughly biconical or hourglass structure \citep{Antonucci1985,Pogge1988,Schmitt1996,Arribas1996,Hutchings1998,Bergmann2018,May2020}, shaped by the thick torus of gas and dust around the central engine. Spectroscopic observations have shown that the central AGN can radiatively excite and drive the outflowing clouds to hundreds of parsecs or more with velocities up to $\sim$2000 \kms \citep{Crenshaw2000a,Crenshaw2000b,Das2005,Muller2011}. At larger distances, the AGN may ionize gas that is part of the galaxy without pushing it outwards or creating outflows, in a region that has been conventionally defined as the extended narrow-line region (ENLR, \citealp{Unger1987}).

It is important to quantify the outflow properties in order to investigate the feedback efficiency of the AGN. The physical conditions of the gas in the NLR and ENLR, such as density and temperature, can be estimated using photoionization models \citep{Ferland1998,Kraemer2008}, while calculations of mass outflow rates \citep{Crenshaw2015,Revalski2018a,Revalski2018b,Trindade2020,Revalski2021} give us direct measurements of gas removal and kinetic energy deposition into the ambient medium of the host galaxy bulges. These studies require accurate measurements of travel distances and space velocities to calculate the gas evacuation time scales.

The anatomy of AGN driven outflows has been widely studied using optical imaging and spatially resolved kinematics of emission lines such as \othree $\lambda$5007 and \halpha $\lambda$6563. Using these emission lines, \cite{Fischer2013} identified non-circular velocities in the NLR and derived geometric models \citep{Crenshaw2000a,Crenshaw2000b,Das2005} of ionized gas outflows for a sample of nearby Seyfert galaxies. Recent studies have found that the radial extent of the photoionized gas (NLR + ENLR) roughly scales with the NLR luminosity \citep{Bennert2002,Schmitt2003,Ganguly2008,Mullaney2013,Karouzos2016,Bergmann2018};
however the sizes of AGN ionized outflows are shown to be smaller than the entire emission line regions and bulge radii \citep{Karouzos2016,Kang2018,Fisher2018}. By measuring the velocity dispersion of \othree ionized gas, \cite{Fisher2018} suggested that even though nearby AGN driven outflows may not be powerful enough to clear the full extents of the bulges \citep{Revalski2018a,Revalski2018b,Trindade2020}, they could still disrupt and redistribute a significant fraction of potential star forming gas in the system. 

Previously, we have explored the origins of AGN outflows by developing dynamical models \citep{Das2007} based on radiation pressure mechanisms and the gravitational drag of the host galaxy. Using these models, \cite{Fischer2017}, showed that outflows in the high luminosity Seyfert 2 galaxy Mrk~573 (log($L_{bol}$) = $45.5\pm0.6$ erg s$^{-1}$: \citealp{Melendez2008a,Melendez2008b,Kraemer2009}) can be launched from as far out as 500 pc due to in-situ ionization and acceleration of gas in the plane of the host galaxy. They later found that these launch distances are even larger \citep{Fischer2019} for the lower mass Seyfert 2 galaxy 2MASX J04234080+0408017, abbreviated as 2MASX J0423  (log($L_{bol}$) = $45.55\pm0.3$ erg s$^{-1}$), where the outflows can be driven from tens of thousands of parsecs encompassing the entire emitting field. Similar methods have been recently followed by \citep{Garcia2021} for the intermediate luminosity AGN of NGC~5643 (log($L_{bol}$) = 43.9 erg s$^{-1}$), showing that almost all of the observed gas originated from the inner 16 pc.

To expand our understating of outflow properties and determine whether an AGN mass/luminosity scaling factor is involved, we systematically study the morphology of ionized gas outflows and radiative efficiency of a low mass, low luminosity ($L_{bol} \approx 10^{43}$ erg s$^{-1}$) AGN in the nearby narrow-line Seyfert 1 (NLS1) galaxy NGC~4051. In the future, we will include more targets in our sample and carry out a similar analysis to further explore the relationships between AGN, host galaxy, and outflow parameters.

\subsection{NGC~4051}\label{subsec:literature}

NGC~4051 is an SAB(rs)bc galaxy at a redshift of $z =$ 0.00234 \citep{deVaucouleurs1991, Verheijen2001}. A color image of the galaxy is shown in Figure~\ref{fig:ngc4051_galaxy}. The distance measurements for this galaxy range from 8.8 Mpc to 17.0 Mpc (NASA/IPAC Extragalactic Database). For our work, we adopted a distance of 16.6$\pm$0.3 Mpc from Cepheid-based measurements \citep{Yuan2020} so that its transverse scale corresponds to $\sim$80 pc/\arcsec\ on the plane of the sky. 
It is classified as a NLS1 based on the low [O~III]/\hbeta (broad plus narrow) ratio in the nuclear regions \citep{Osterbrock1985,Leighly1999} and the observed width of broad line region (BLR) \hbeta emission line with full-width at half-maximum (FWHM) $<$ 3000 km s$^{-1}$, which is smaller than typical type 1 Seyfert galaxies. The AGN of this galaxy hosts a SMBH with a mass of log$(M_{\mathrm{BH}})$ $=$ 6.13$^{+0.12}_{-0.15}$ $M_{\odot}$ \citep{Bentz2018} and is highly variable in optical and X-ray observations  \citep{Peterson2000,Breedt2010}. An estimated bolometric luminosity of log($L_{bol}$) $=$ 42.95 erg s$^{-1}$ \citep{Denney2009,Bentz2009,Bentz2013} has been reported based on $\lambda L_{\lambda}$(5100\AA) measurements. With this mass and luminosity, the AGN is radiating at approximately 5\% of the Eddington luminosity ($L/L_{edd}$ \app 0.053) as noted by \cite{Peterson2004}.

The AGN driven outflows in NGC~4051 has been extensively studied in X-ray and UV absorption lines \citep{Collinge2001,Kraemer2012,Crenshaw2012_n,Laha2014}. In addition to the warm absorbers, X-ray ultra fast outflows (UFOs) and shocked outflows have been reported \citep{Pounds2011,Pounds2013,Laha2016, King2015} in momentum-driven winds. In the optical, the ionized gas outflows have been detected in both narrow Balmer and forbidden emission lines. Ground-based \othree  images show an unresolved nucleus with faint emission \citep{Haniff1988,Pogge1989} that co-aligns with the strong nuclear radio structure at 100\arcdeg\ in the SE. \cite{Schmitt1996} verify similar emission in an \textit{HST} Wide Field Planetary Camera (WFPC) - F502N image. Using a Manchester Echelle Spectrograph (MES) image, \cite{Christopoulou1997} discovered a 9\arcsec\ long wedge shaped ionized emission in \othree that bisects at 33\arcdeg. Spectroscopic observations using the same instrument provided a model for conical outflow that is inclined at 50\arcdeg\ from line of sight with a half opening angle of 25\arcdeg. GEMINI GMOS-IFU observations show outflows with strong flux in the west to east direction close to the nucleus, but more extended towards the north with velocities reaching up to 500 \kms \citep{Barbosa2009}. \cite{Fischer2013} developed a kinematic model for a biconical outflow in NGC~4051 using \textit{HST} STIS spectra that is directed towards position angle (PA) 80\arcdeg\ NE with a maximum half opening angle of 25\arcdeg.

In this work we present our improved understanding of the NLR in NGC~4051, including the extent and morphology of the outflows, and investigate the significance of AGN radiation pressure driving. We describe the imaging and spectroscopic observations from \hst and APO (\S \ref{sec:obs}), spectroscopic analysis (\S \ref{sec:analysis}), results including the ionized gas kinematics (\S \ref{subsec:gas_kinematics} and \ref{subsec:roatation}), emission line diagnosis (\S \ref{subsec:BPT}), and the revised outflow model (\S \ref{subsec:outflow}). Finally we compare the kinematics with analytical models of radiative acceleration-gravitation deceleration (\S \ref{subsec:radiative_driving}), discuss (\S \ref{sec:discuss}) and conclude (\S \ref{sec:conclusion}) the observed results.

\section{OBSERVATIONS}\label{sec:obs}

\subsection{Hubble Space Telescope (HST)}\label{subsec:stisobs}

\setlength{\tabcolsep}{0.025in}
\renewcommand{\arraystretch}{1.1}
\tabletypesize{\small}
\begin{deluxetable*}{hhcccccccccccc}[ht!]
\tablenum{1}
\tablecaption{{\it HST} Observations of NGC~4051 \label{tab:stisobs}}
\tablehead{
\nocolhead{Target} & \nocolhead{Observing} & \colhead{Instrument} & \colhead{Proposal} & \colhead{Observation} & \colhead{Date} & \colhead{Filter /} & \colhead{Slit} & \colhead{Exposure} & \colhead{Spectral} & \colhead{Wavelength} & \colhead{Spatial} & \colhead{Position} & \colhead {Spatial} \vspace{-2ex}\\
\nocolhead{Name} & \nocolhead{Facility} & \colhead{Name} & \colhead{ID} & \colhead{ID} & \colhead{(UT)} & \colhead{Grating} & \colhead{Name} & \colhead{Time} & \colhead{Dispersion} & \colhead{Range} & \colhead{Scale} & \colhead{Angle} & \colhead {Offset} \vspace{-2ex}\\
\nocolhead{} & \nocolhead{} & \colhead{}  & \colhead{} & \colhead{} & \colhead{} & \colhead{} & \colhead{}  & \colhead{(s)}& \colhead{(\AA~pix$^{-1}$)} & \colhead{(\AA)} & \colhead{($\arcsec$~pix$^{-1}$)} & \colhead{(deg)} & \colhead {($\arcsec$)}
}
\startdata
NGC~4051 &HST & WFC3 & 12212 & IBGU11BSQ & 2011 Jul 07 & F502N & ... & 1554 & ... & 4963-5059 & 0.039 & ... & ...\\
NGC~4051 &HST & WFC3 & 12212 & IBGU11BXQ  & 2011 Jul 07 & F547M & ... & 779 & ... & 5039-5909 & 0.039 & ... & ...\\
NGC~4051 & HST & STIS & 8253 & O5G402010 & 2000 Apr 15 & G430M & A & 1796 & 0.28 & 4818-5104 & 0.051 & 89.85 & 0.05 S\\
NGC~4051 &HST & STIS & 8253 & O5G402010 & 2000 Apr 15 & G430M & B & 600 & 0.28 & 4818-5104 & 0.051 & 88.85 & 0.20 N\\
\enddata
\tablecomments{A summary of the \hst observations used in this study. The columns list the (1) \hst Instrument, (2) Program ID, (3) Observation ID, (4) observation date, (5) filter (imaging) or grating (spectra), (6) assigned labels (names) for the two long slits, (7) total exposure time for each data set, (8) spectral dispersion of the long slit grating, (9) wavelength range (for spectra) or bandpass (for imaging), (10) spatial scale of the image/spectra, (11) PA of the STIS slits, (12) their spatial offsets from nucleus. The values in column (8)-(10) were obtained from their respective instrument handbooks \citep{wfc3ihb,stisihb} where the exact STIS spatial scale is given as 0.05078$\arcsec$~pix$^{-1}$.}
\end{deluxetable*}

We used archival observations of NGC~4051 from the {\it Hubble Space Telescope (HST)}'s Space Telescope Imaging Spectrograph (STIS) and Wide Field Camera 3 (WFC3). We retrieved the calibrated data from the Mikulski Archive at the Space Telescope Science Institute (MAST) and processed and combined them into the final data files using the Interactive Data Language (IDL). Additional details of post-data reduction and calibration are given in \cite{Fischer2013}. The WFC3 combined and drizzled images from the Hubble legacy Archive (HLA) were analysed using SAOImage DS9 and the Astropy library in Python \citep{ds92000, astropy:2013}. The STIS observations were retrieved under Hubble program ID 8253 (PI: M. Whittle) and the WFC3 images from program ID 12212 (PI: D. M. Crenshaw). A summary of the \hst observations is provided in Table~\ref{tab:stisobs}.

\begin{figure*}[ht!]
\centering
\includegraphics[width=\textwidth]{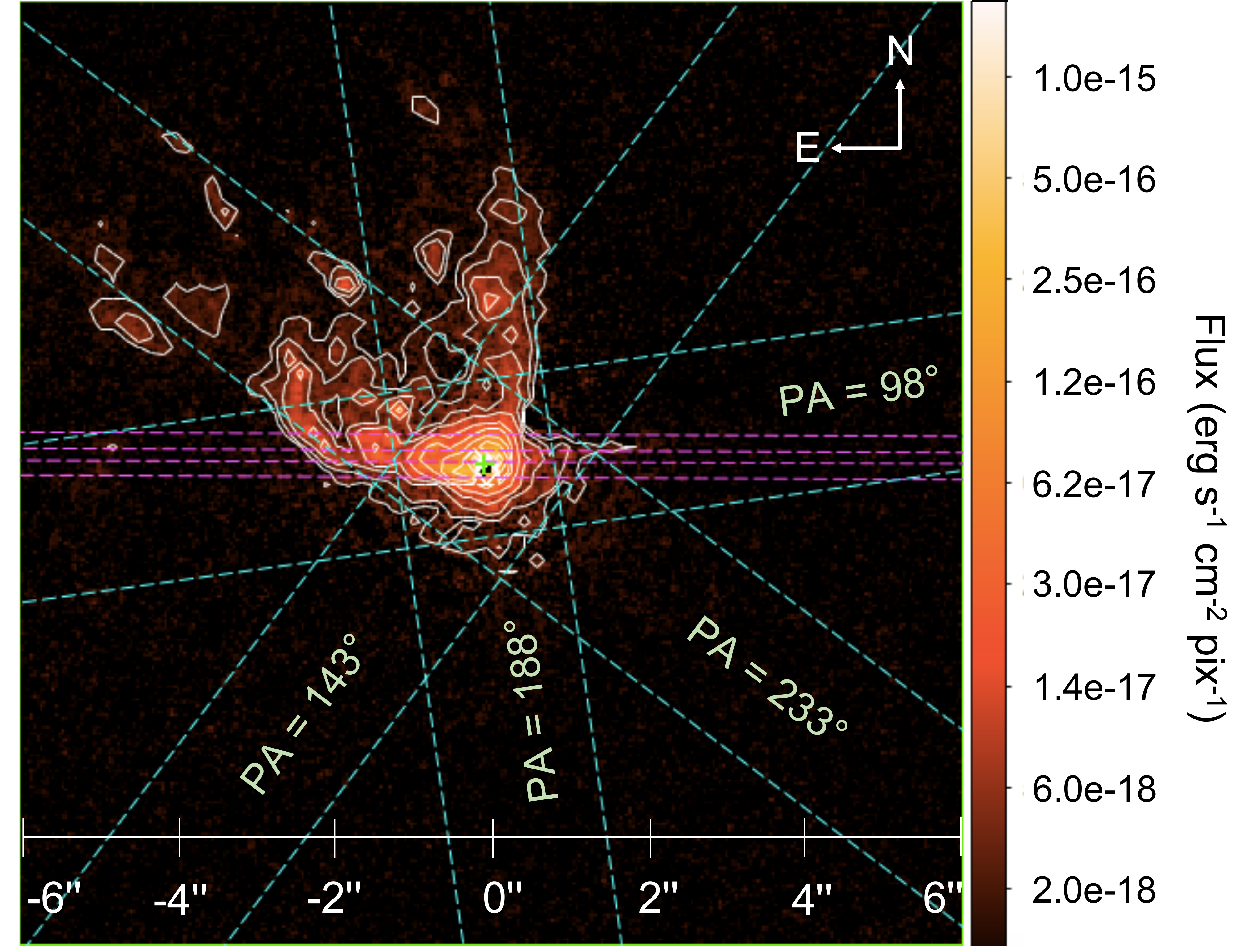}
\label{f502n}
\caption{A $12\arcsec \times 12\arcsec$ continuum-subtracted  \othree image of NGC~4051 using WFPC3 camera on {\it HST}. The APO (cyan) and STIS (magenta) slits are overlaid, with the APO PA labeled. The PA for both STIS slits is 89.8$\degr$. The flux contours are generated at 3$\sigma$ above the background and increase in powers of 2 $\times$ 3$\sigma$ for each inner contour. The color bar on the right shows the flux variation in the units of erg s$^{-1}$ cm$^{-2}$ pix$^{-1}$. The standard deviation ($\sigma$) was calculated from a fraction of the background image outside the galaxy. The green cross marks the continuum centroid, associated with the highest flux position in the F547M continuum image. The dark spots are saturated pixels in F502N image due to the bright central point source. The STIS slits ``A'' and ``B'' are 0\farcs05 south and 0\farcs2 north of the nucleus, respectively.  \label{fig:F502N}}
\end{figure*}

To investigate the kinematics of emission line clouds in the NLR of NGC~4051, we used the long-slit spectra from the medium dispersion G430M grating (resolving power R $\approx$ 9000), which provides a spectral resolution of 0.56~\AA\ in the dispersion direction and a spatial resolution of 0\farcs10 in the cross-dispersion direction. Two parallel $52\arcsec \times 0\farcs2$ long slits observations were available along a PA of 89.8\arcdeg. For an extended source filling the slit in the dispersion direction, the effective spectral resolving power is reduced to R $\approx$ 4500 (FWHM $\approx$ 67 km s$^{-1}$). The two slits are at offsets of $0\farcs05$ south (labeled ``A'') and $0\farcs20$ north (labeled ``B'') from the nucleus. The spectra extracted along each slit contain the strong [O~III] $\lambda \lambda$4959,5007 lines, which trace the AGN ionized gas in a part of the NLR with high angular contrast. The STIS slit positions and offsets are shown in Figure~\ref{fig:F502N}.

We supplemented our long-slit spectroscopic observations with a narrow-band image available for the F502N filter on WFC3, displayed in Figure~\ref{fig:F502N}, to study the morphology of the emission line regions. The high resolution (pixel scale of $0\farcs039$ pixel$^{-1}$)~image was taken using the Ultraviolet-Visible (UVIS) detector with a field of view (FOV) of $162\arcsec \times 162\arcsec$. The F502N image contains sufficiently strong [O~III] emission, to compare with our spectroscopic analysis and to map the circumnuclear ionized gas in NGC~4051. We used a F547M image (from the same program ID) to subtract the continuum from the line emission. The final image was cropped to $12\arcsec \times 12\arcsec$ as shown in Figure~\ref{fig:F502N} to highlight the structure of the [O~III] emission in the NLR.

\subsection{Apache Point Observatory (APO)}\label{subsec:apo-obs}

We have previously observed extended ionized emission using relatively wide long slits in ground-based observations of Mrk~573 \citep{Fischer2017, Revalski2018a}, Mrk~34 \citep{Revalski2018b}, Mrk 3 \citep{Gnilka2020}, and Mrk~78 \citep{Revalski2021}. These observations allow us to detect fainter emission in the ENLR and host galaxy from ionized gas at larger distances that were undetectable with narrow \textit{HST} STIS slits.

To accurately measure the extent of the ionized gas and AGN driven outflows, we use $6\arcmin \times 2\arcsec$ wide slits on the Apache Point Observatory (APO)'s 3.5 meter telescope with its Dual Imaging Spectrograph (DIS). The long slit covers the ionized gas from the host galaxy on large scales, allowing us to identify the rotation curve of NGC~4051. The slit PAs on the host galaxy are shown in Figures \ref{fig:ngc4051_galaxy} and \ref{fig:F502N}.

\setlength{\tabcolsep}{0.063in}
\renewcommand{\arraystretch}{1.1}
\tabletypesize{\small}
\begin{deluxetable*}{hcccccccccc}[ht!]
\tablenum{2}
\tablecaption{ARC 3.5m Telescope Observations of NGC~4051 at APO \label{tab:apo-obs}}
\tablehead{
\nocolhead{Target} & \colhead{Instrument} & \colhead{Date} & \colhead{Filter /} & \colhead{Exposure} & \colhead{Spectral} & \colhead{Wavelength} & \colhead{Spatial} & \colhead{Position} & \colhead{Mean} & \colhead {Mean} \vspace{-2ex}\\
\nocolhead{Name} & \colhead{Name} & \colhead{(UT)} & \colhead{Grating} & \colhead{Time} & \colhead{Dispersion} & \colhead{Range} & \colhead{Scale} & \colhead{Angle} & \colhead{Air Mass} &\colhead {Seeing} \vspace{-2ex}\\
\nocolhead{} & \colhead{} & \colhead{} & \colhead{} & \colhead{(s)} & \colhead{(\AA~pix$^{-1}$)} & \colhead{(\AA)} & \colhead{($\arcsec$~pix$^{-1}$)} & \colhead{(deg)} & \colhead{} &\colhead {($\arcsec$)}
}
\startdata
NGC~4051 & ARCTIC & 2020 Mar 23 & J-C B & 190 & ... & 3400-6000 & 0.228 & ... & 1.12 & 1.80\\
NGC~4051 & ARCTIC & 2020 Mar 23 & J-C V & 80 & ... & 4500-7000 & 0.228 & ... & 1.16 & 1.80\\
NGC~4051 & ARCTIC & 2020 Mar 23 & J-C R & 30 & ... & 5400-10000 & 0.228 & ... & 1.18 & 1.80\\
NGC~4051 & DIS & 2017 Feb 27 & B1200 & 2700 & 0.62 & 4373-5633 & 0.42 & 98 & 1.02 & 2.27\\
NGC~4051 & DIS & 2017 Feb 27 & R1200\tablenotemark{$\star$} & 2700 & 0.58 & 6098-7283 & 0.40 & 98 & 1.02 & 2.24\\
NGC~4051 & DIS & 2017 Feb 27 & B1200 & 2700 & 0.62 & 4373-5633 & 0.42 & 143 & 1.07 & 2.27\\
NGC~4051 & DIS & 2017 Feb 27 & R1200\tablenotemark{$\star$} & 2700 & 0.58 & 6098-7283& 0.40 & 143 & 1.07 & 2.24\\
NGC~4051 & DIS & 2017 Feb 27 & B1200 & 2700 & 0.62 & 4373-5633 & 0.42 & 188 & 1.15 & 1.97\\
NGC~4051 & DIS & 2017 Feb 27 & R1200\tablenotemark{$\star$} & 2700 & 0.58 & 6098-7283 & 0.40 & 188 & 1.15 & 1.68\\
NGC~4051 & DIS & 2017 Feb 27 & B1200 & 1800 & 0.62 & 4373-5633 & 0.42 & 233 &1.26 & 1.97\\
NGC~4051 & DIS & 2017 Feb 27 & R1200\tablenotemark{$\star$} & 1800 & 0.58 & 6099-7185 & 0.40 & 233 & 1.26 & 1.68\\
\enddata
\tablecomments{A summary of the ground-based imaging and spectroscopy used in this study. The columns list (1) the APO instruments, (2) the observations dates, (3) filters (for imaging)/ gratings (for spectra) used, (4) the total exposure times for each data set, (5) spectral dispersion (6) wavelength range (for spectra) or bandpass (for imaging), (7) spatial scales of the images/spectra, (8) PAs for the long slits, (9) mean air mass and (10) mean seeing for the nights of observations. \tablenotemark{$\star$}Affected by the instrument scattered light.}
\end{deluxetable*}

The DIS uses a dichroic element to split light into blue and red channels, allowing simultaneous measurements in the H$\beta$ and H$\alpha$ regions of the spectrum. For this study, we used high dispersion gratings B1200 (blue) and R1200 (red) with a resolving power R $\approx$ 4000 -- 5500. Similar to \hst, we used the DIS blue spectra that contain strong \othree lines and DIS red spectra that contain the H$\alpha$ and [N~II] lines to measure kinematics of the circumnuclear and extended ionized gas. The spatial scales of the blue and red channels are $0\farcs42$ pixel$^{-1}$ and $0\farcs40$ pixel$^{-1}$, respectively. APO DIS provides a lower angular contrast than \hst but offers larger spatial coverage that enables us to probe the rotation and possible extended outflow signatures in the host galaxy.

\begin{figure*}[ht!]
\centering
\includegraphics[width=0.7\textwidth]{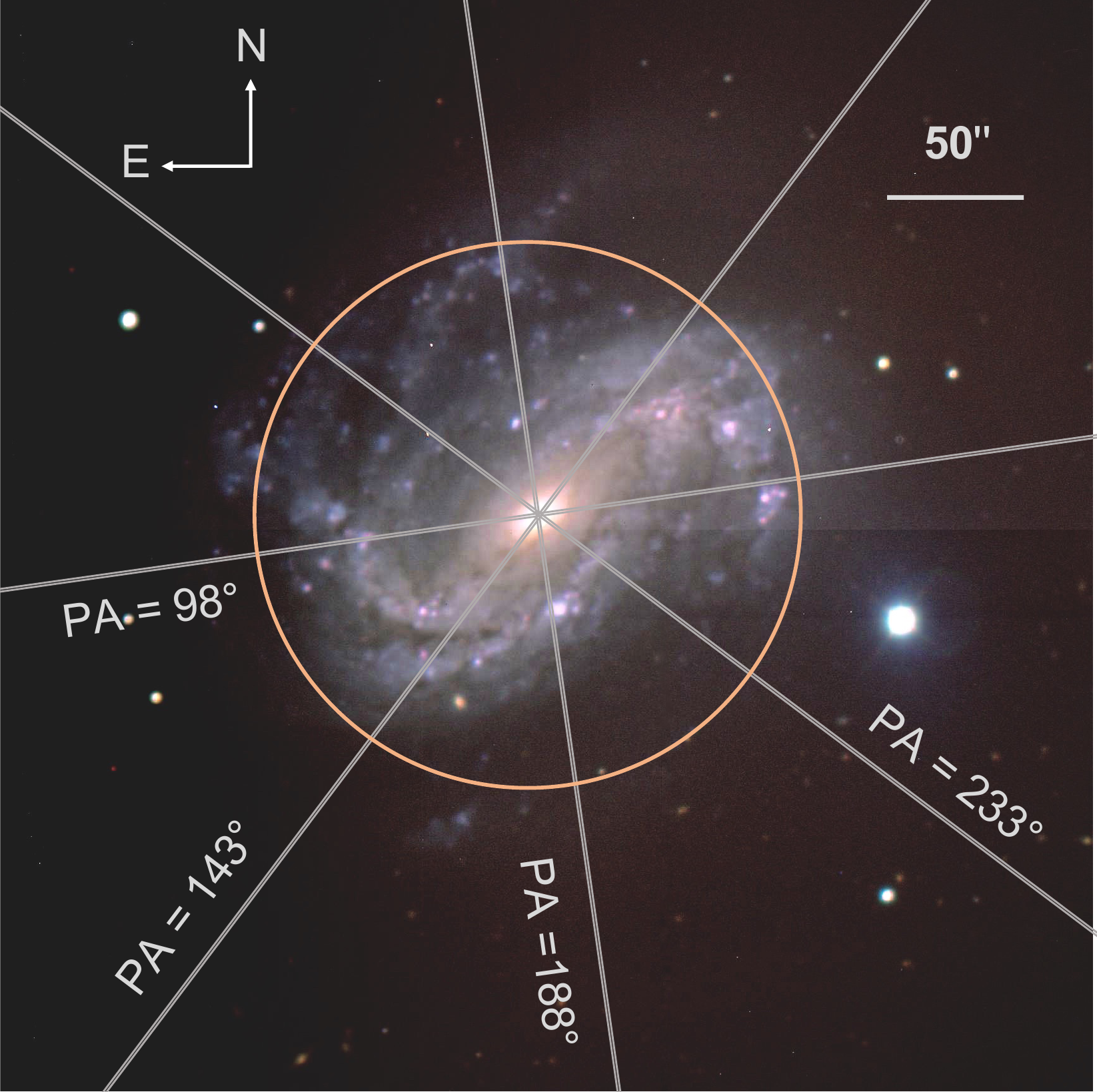}
\caption{A $400\arcsec \times 400\arcsec$ RGB-composite image of NGC~4051 obtained with ARCTIC on APO's 3.5m Telescope. The four DIS slits are shown as grey lines and the circle with a 100\arcsec\ radius corresponds to the extent of H$\alpha$ velocities in Figure~\ref{fig:rotation} and \ref{fig:velmaps_diskfit}.
\label{fig:arctic}}
\end{figure*}

We obtained four long-slit observations corresponding to PAs of 98\arcdeg, 188\arcdeg, 143\arcdeg, and 233\arcdeg\ on the same night. We reduced the observed raw data to the two dimensional spectral images using a standard IRAF routine \citep{Tody1986, Tody1993}, which involves primary calibrations such as bias subtraction, flat-field correction, cosmic ray removal with the IRAF task, LA Cosmic \citep{Vandokkum2001} and combining multiple exposures. We used the arc lamp images that were taken before every science exposure for the wavelength calibrations. Standard star exposures were taken on the same night for flux calibration \citep{Oke1990} and the data were corrected for atmospheric extinction using an APO extinction curve. Additional calibrations were performed using IDL to correct for tilt in the location of the spectrum in the cross-dispersion direction \citep{Gnilka2020} and for sky line subtraction. Seeing for the observing night was measured by calculating the point-spread function (PSF) of the standard star. More information about the instrument and observations are given in Table~\ref{tab:apo-obs}.

As discussed in \cite{Gnilka2020}, the DIS experienced scattered light due to a contamination from condensation in the latter years of its operation. The instrument was serviced periodically for optical correction. We characterized the impact of the scattered light in our data by measuring the emission lines from the standard stars observed on the same nights. Strong broad wings or ``halos" around the emission lines were seen in the observations after 2017 September 19. Fortunately, NGC~4051 was observed in the beginning of 2017 after an upgrade to the instrument. The blue channel shows no sign of scatted light in the spectral image, which provides us the best measurements of spectral lines such as \othree and \hbeta for our kinematics analysis. However, we found faint scattered light close to the nucleus in the red channel (\halpha region) from the bright central NLR. Therefore, the spectra from 3\arcsec\ to 15\arcsec\ on both sides of the nucleus were slightly corrupted, which made it more difficult to separate the kinematic components of the \halpha $+$ [N II] lines. We focused on the \othree lines from the blue channel to measure the ionized gas kinematics and outflow structure in those regions. The spectra extracted for the extended part of the galaxy were unaffected and show strong narrow \halpha emission lines, which we use to measure the rotational kinematics of the host galaxy.

We also observed NGC~4051 with the Astrophysical Research Consortium Telescope Imaging Camera (ARCTIC) on the 3.5m telescope using its broadband Johnson Cousins B, V, and R filters (wavelength range is given in Table \ref{tab:apo-obs}) to increase the spatial coverage and compare with the spectroscopic observations.
ARCTIC has a field of view of $7.85\arcmin \times $7.85\arcmin. The image was taken with $2 \times 2$ binning, which gives a plate scale of 0.228\arcsec pixel$^{-1}$. The raw images were corrected for darks, bias and flat-fields using an open source automatic reduction pipeline named Acronym, which was developed at APO to reduce ARCTIC images \citep{Weisenburger2017}. We then combined the different exposures using the task IMCOMBINE in IRAF for each bandpass. A $400 \arcsec \times 400\arcsec$ RGB-composite image is shown in Figure~\ref{fig:arctic}, with the APO DIS longs slit locations overlayed.

\subsection {Slit Placements}\label{subsec:slits}
Figure~\ref{fig:F502N} shows a strong bright core of \othree emission around the Seyfert 1 nucleus and additional emission along the walls of a cone-like structure extending to a projected distance of $\sim$6\arcsec\ in the NE. Assuming this is part of a bicone, the SW cone is hidden by dust in the host galaxy as described by \cite{Fischer2013}. The STIS slit positions capture portions of the walls of the cone. The APO DIS slits overlap at the nucleus. The APO slits at PA $=$ 233\arcdeg\ $\&$ 188\arcdeg\ capture the east and north walls of the cone, albeit at low spatial resolution, while the slit at PA $=$ 98\arcdeg\ lies closest to and encompasses the STIS G430M slits in the regions of interest. The slit at PA $=$ 143\arcdeg\ is close to the projected major axis of the host galaxy, as shown in Figures \ref{fig:ngc4051_galaxy} and \ref{fig:arctic}. All of the APO slits cross areas of strong star formation in the spiral arms.

\section{ANALYSIS} \label{sec:analysis}

\subsection{Spectral Fitting in the \othree region} \label{subsec:gaussfit}
\begin{figure*}[ht!]
\centering
\subfigure[]{
\includegraphics[width=0.49\textwidth ]{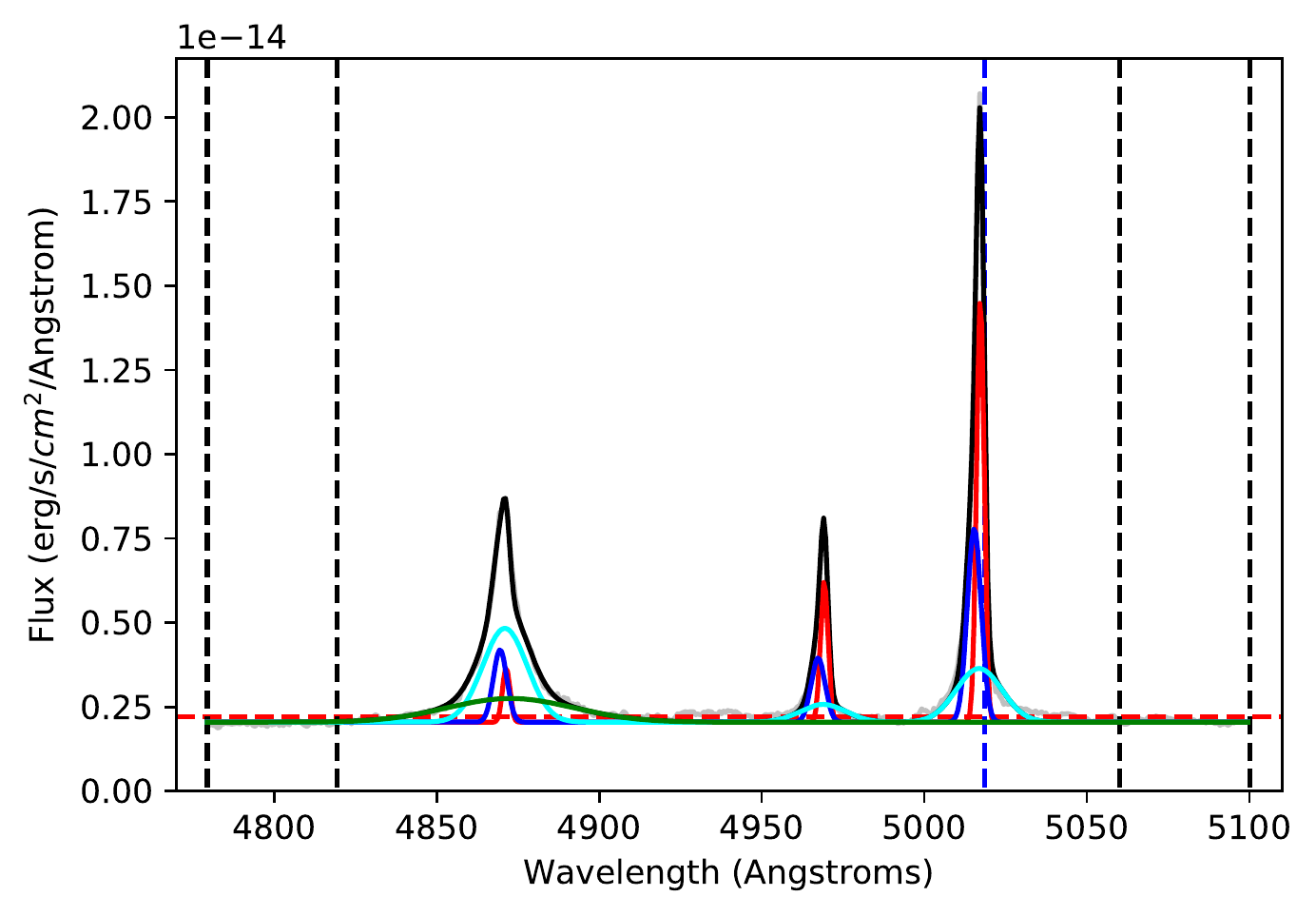}}
\subfigure[]{
\includegraphics[width=0.49\textwidth ]{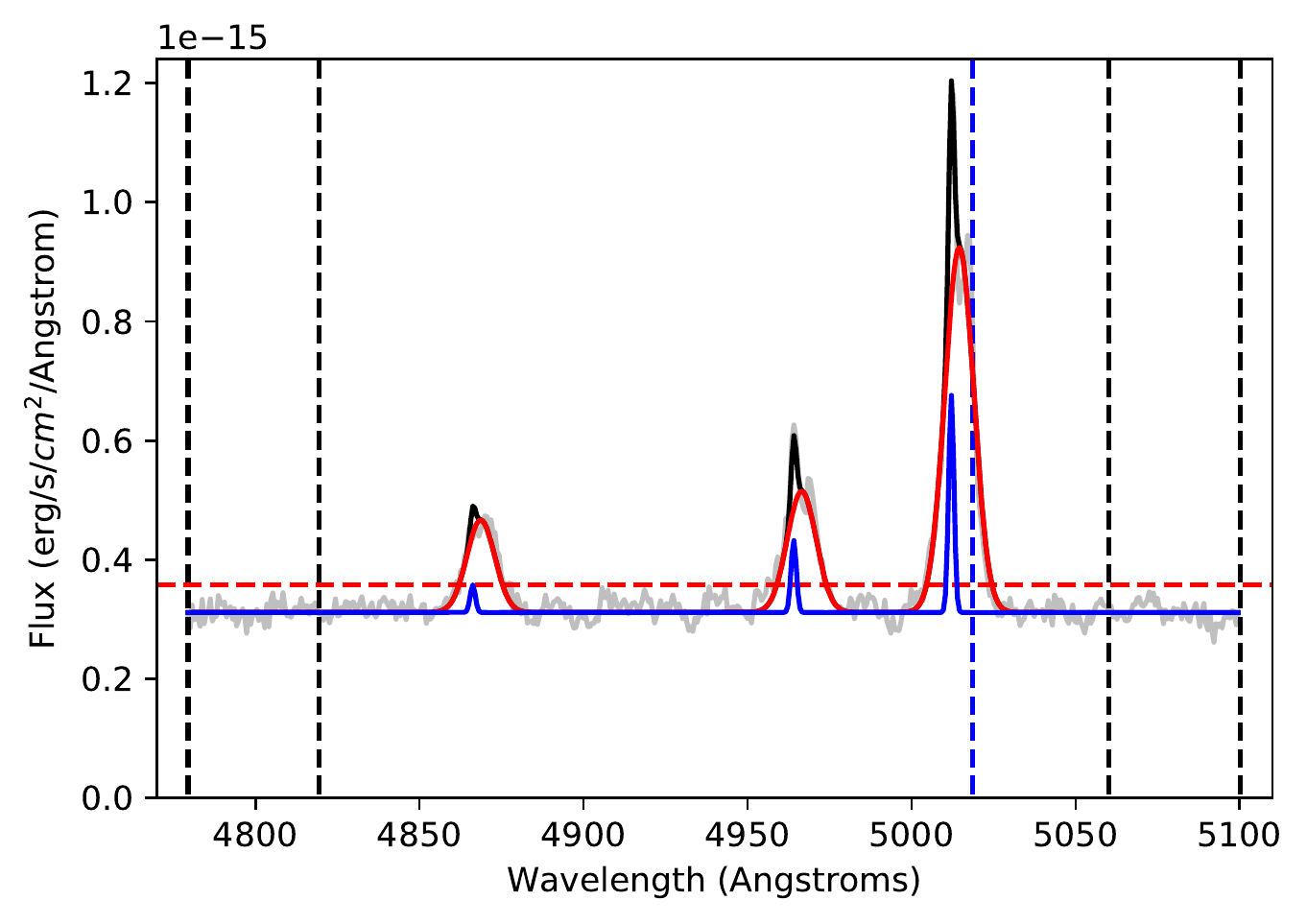}}
\caption{Examples of \hbeta$\lambda4861$ and \othree$\lambda\lambda5007/4959$ emission lines fits with multi-Gaussian components for APO slit PA 188\arcdeg. The observed data is shown in gray and the total flux with the multiple Gaussian fits is shown with a black solid line. (a) Spectral lines extracted at the nucleus and fit with two narrow components (red and blue) as well as a broad \hbeta component (dark green) and an intermediate broad component (cyan), which is present in both the \hbeta and \othree lines. Panel (b) shows a spectrum extracted at 2.5\arcsec\ to the north, which only contains two narrow components and a double-peaked profile. The vertical dashed black lines are the defined continuum regions, while the horizontal dashed red line represents the flux level at $3\sigma$ over the noise. The vertical blue dashed line close to \othree corresponds to the redshift of the galaxy.}
\label{fig:gaussfit}
\end{figure*}

We fit the \othree $\lambda\lambda$4959,5007 lines in the \textit{HST} STIS G430M and APO DIS blue channel spectra with multi-component Gaussian profiles. To fit Gaussians to emission lines along the slits at each position, we used a fitting routine, which employs a multimodal nested sampling algorithm called MultiNest\citep{Feroz2008, Feroz2009, Feroz2019, Buchner2014}. The algorithm uses Bayesian statistics to determine the significant number of components that best fit the emission lines with the least complexity. The Gaussian fits output the wavelength centroid, width and peak flux of each component, which are used to calculate the mean velocity, FWHM and total flux of the ionized gas clouds as shown in Figure~\ref{fig:gaussfit}.

Although the emission lines are often asymmetric and not exactly Gaussian \citep{Heckman1981,Veilleux1991}, employing a Gaussian model provides the simplest and most efficient method to decompose the line profile into different kinematic components, which are often seen as distinct bumps or even multiple peaks in the profiles. A detailed description of the Gaussian fitting routine can be found in the Appendix of \cite{Fischer2017}. Our group has previously employed this technique (\citealp{Revalski2018b,Fisher2018,Gnilka2020,Revalski2021}) to study the gas kinematics in various AGN.

To achieve the best results, we fit the \othree $\lambda\lambda$4959,5007 lines along with the \hbeta $\lambda4861$ line and selected the adjacent regions on either side as continuum points, as shown in Figure~\ref{fig:gaussfit}. We fixed the relative height ratios of $\lambda5007$ and $\lambda4959$ to 3.01 and their wavelength difference to 47.9 \AA\ in the rest frame of the source \citep{Osterbrock_Ferland2006}. We also set a wavelength difference for \othree $\lambda5007$ and \hbeta $\lambda4861$ to 145.515 \AA\ for the narrow emission lines. The mean velocity and velocity widths (FWHM) of each component for both \hbeta and \othree emission lines were fixed under the assumption that both of these lines originated from the same cloud. We restricted the width of the narrow component to vary above a minimum value associated with the spectral resolution (FWHM of the line-spread function) of the instrument.
We also found contributions from an unresolved broad line region (BLR) and an intermediate line region (ILR) component associated with the nucleus. We limited the velocity widths of NLR emission to accommodate these two broad/intermediate components. See section \S {\ref{subsec:broad_fit}} for more details.
 
The peak flux for all the components was allowed to vary from $3\sigma$ above the background noise level up to the maximum height of the emission line, with the standard deviation ($\sigma$) calculated from the average flux measured in the continuum regions free of any strong emission or absorption features.

\subsection{Fitting the Broad and Intermediate Components}\label{subsec:broad_fit}

\begin{figure*}[hbt!]
\centering
\subfigure{
\includegraphics[width=0.68\textwidth]{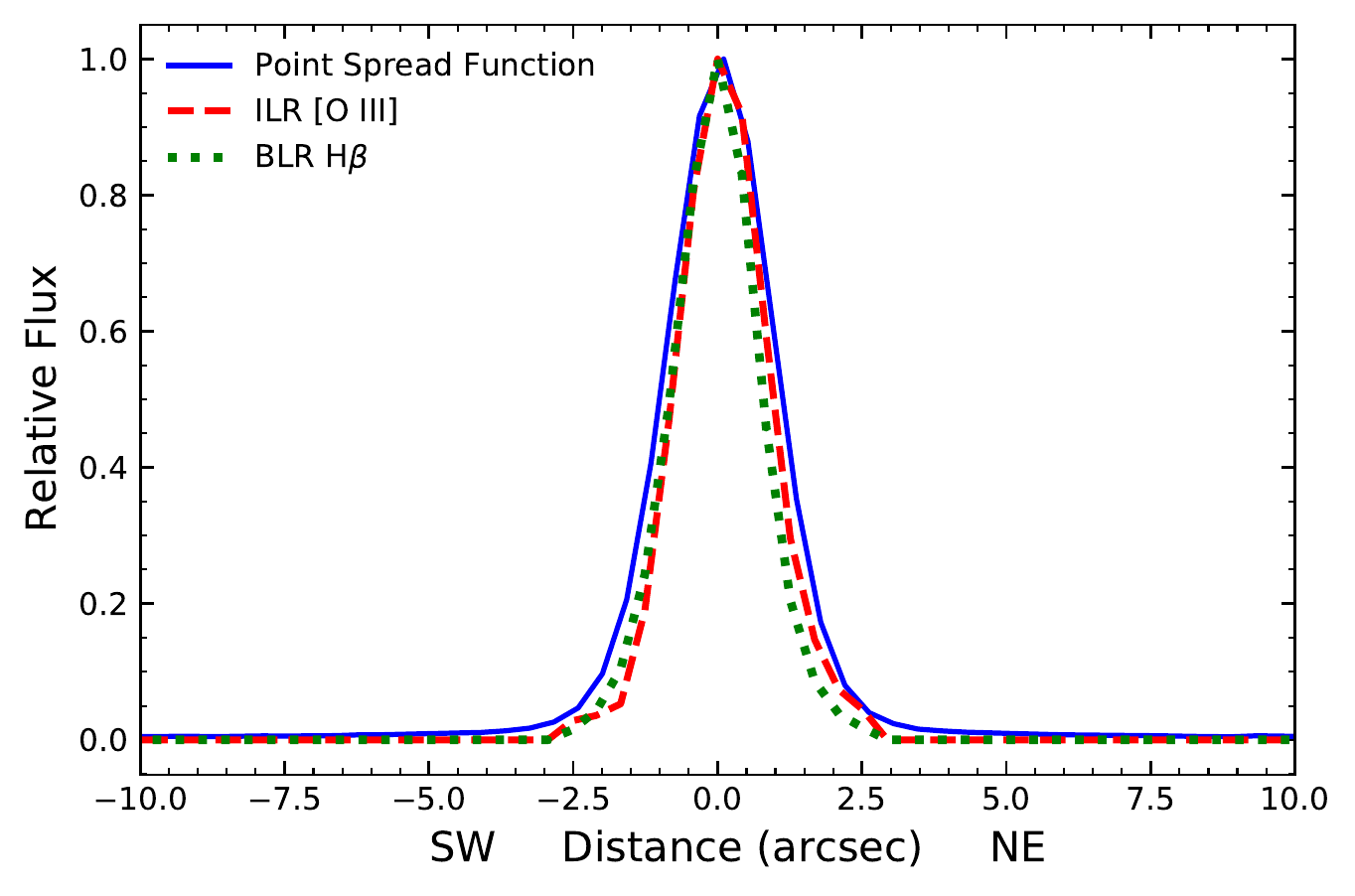}}
\caption{ Normalized brightness profiles along the APO DIS slit giving the PSF from a standard star (solid blue line), the ILR component as seen in [O~III] (dashed red line), and  the BLR \hbeta emission (dotted green line), demonstrating that the ILR and BLR are unresolved in the APO DIS slit. The slightly broader PSF of the standard star likely indicates slightly different seeing at the time and location of its observation.}
\label{fig:psf}
\end{figure*}

NGC~4051 is a narrow-line Seyfert 1 galaxy, which means the BLR emission lines such as \hbeta$\lambda4861$ are relatively narrow compared to most Seyfert 1 galaxies, but still broader than the lines from the NLR. Therefore, we find little contamination of broad \hbeta towards the \othree emission lines. Nevertheless, it is important to identify the contribution from broad \hbeta to separate it from the NLR components while fitting both \othree and \hbeta simultaneously. The \hbeta BLR component was fit well by a single Gaussian unlike more complicated profiles often seen in normal Seyfert 1s \citep{Marziani2009,Park2012,Barth2015}. We measured the centroid and width of the broad component using a Gaussian fit to the \hbeta line at the nucleus.
The wavelength centroid was found close to the systemic velocity of the galaxy and width of the broad emission line was measured as FWHM $=$ 2800 km s$^{-1}$. We fixed the centroid and width of the broad \hbeta line as we moved the spectral fits along the slit. The flux of the broad component decreases smoothly with distance from the nucleus, as expected from the PSF of an unresolved source (the BLR). 

In addition to the narrow emission lines and the broad \hbeta component, we detected a component with an intermediate width and density from an ``intermediate line region'', which has been claimed in a number of other AGN \citep{Crenshaw1986,Crenshaw2007,Mullaney2008,Crenshaw2009,Zhu2009,Adhikari2016}, and specifically for NGC~4051 in He II and \hbeta emission lines \citep{Kraemer2012,Yang2013}.
The intermediate component was found by noticing that the previous fits to the narrow \hbeta and \othree lines near the nucleus in the APO spectra yielded a relatively ``broad'' (FWHM $\approx$ 1000 km s$^{-1}$) component with an unusually small [O~III]/H$\beta$ flux ratio ({$\approx$ 0.55}), indicating significant collisional de-excitation of the level that gives rise to the [O~III] $\lambda\lambda$4959,5007 lines at a critical density of $n_e$ = 6.8 $\times$ 10$^{5}$ cm$^{-3}$ \citep{Osterbrock_Ferland2006}.
The intermediate component is blueshifted from the host galaxy plane with a radial velocity centroid of $=$ $-$110$\pm$15 \kms and has FWHM  $=$ 1010$\pm$70 km s$^{-1}$, which is within the uncertainties of the intermediate H$\beta$ FWHM $=$ 1137$\pm$178 \kms as measured by \cite{Yang2013}.

Similar to the broad \hbeta component, the intermediate component is unresolved in both the STIS and APO spectra, consistent with a location close to the SMBH.
Therefore, in the revised spectral fits, we restricted the FWHM of the narrow components to be $\leq$~900 km s$^{-1}$. At the same time, we fixed the velocity centroid and velocity width of the broad and intermediate components while allowing the fluxes to change along the slit.

A comparison of the radial extents of the broad \hbeta and intermediate \othree components with the PSF, as calculated from standard star brightness profiles, is shown in Figure~\ref{fig:psf}. Due to the simultaneous fit, both \hbeta and \othree ILR components have the same velocity and width as well as radial extent. On the other hand, unlike broad H$\beta$, the \othree does not have a BLR component and therefore does not affect the BLR brightness profile in Figure~\ref{fig:psf}.
Figure \ref{fig:psf} indicates that broad \hbeta and intermediate (\othree and H$\beta$) components are unresolved within the spatial resolution of the APO spectra, and appear in the extended emission due to seeing (see Table \ref{tab:apo-obs}).
The broad \hbeta and intermediate \hbeta $\&$ \othree components were also detected in the STIS spectra as point sources, which limits the projected size of the ILR to $\leq$~0\farcs1 ($\sim$8 pc). We will discuss the ILR size calculations in section \S \ref{subsec:ILR}.

An example of emission line fitting is shown in Figure~\ref{fig:gaussfit} for spectra obtained with one of the APO DIS long slits. 
The double peaked \othree emission line seen in Figure~\ref{fig:gaussfit}b is similar to those in space/ground based spectra of several Seyfert galaxies and can be attributed to the asymmetric distribution of ionized gas in NLR and/or to distinct contributions from the rotating disk and outflowing gas \citep{Rosario2010,Fischer2011,Shen2011,Wylezalek2020}. 

\subsection{Gaussian Fitting to Other Emission Lines}\label{subsec:bpt_linefit}
\begin{figure*}[ht!]
\subfigure[]{
\includegraphics[width=0.46\textwidth]{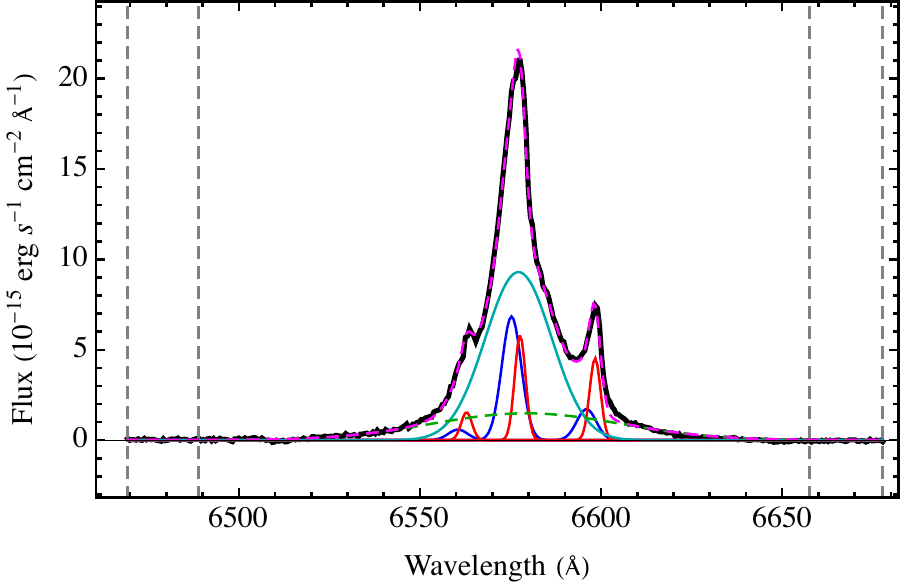}}\hspace{3ex}
\subfigure[]{
\includegraphics[width=0.46\textwidth]{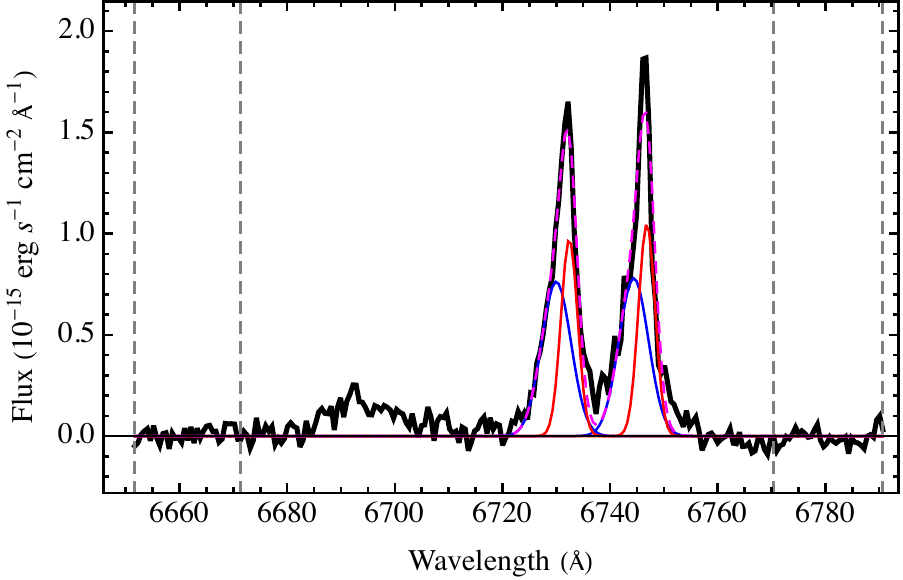}}
\caption{Examples of (a) \halpha $\lambda6563$ + [N~II] $\lambda\lambda$6548,6583  and (b)  [S~II] $\lambda\lambda$6716,6731  emission line fits with multi-Gaussian components for APO slit PA 188\arcdeg\ at the nucleus. The velocity centroid and width of each component are fixed for all emission lines as determined from the [O~III] $\lambda\lambda$4959,5007 fits. The observed spectra are plotted in solid black. The NLR emission Gaussian fits are shown in blue and red for wider and narrower kinematic components. The broad (dashed) green and cyan curves in (a) show the broad line region  and intermediate line region \halpha emission, respectively. The sum of all components is given in dashed magenta.}
\label{fig:gaussfit2}
\end{figure*}

To measure the line ratios for ionization diagnosis (\S \ref{subsec:BPT}) for the APO observations, we fit emission lines \hbeta $\lambda4861$; \halpha $\lambda6563$; [N~II] $\lambda\lambda$6548,6583; and [S~II] $\lambda\lambda$6716,6731 by using the multi-component fits of the \othree lines as a template. The procedure, described in detail in \cite{Revalski2018a,Revalski2018b,Revalski2021}, fixes the positions and widths of the emission-line components based on the [O~III] kinematics, and allows the fluxes to vary, subject to atomic constraints on fixed ratios such as those of the [O~III] and [N~II] doublets. An example of the fits to the emission lines in the red APO spectra is shown in Figure~\ref{fig:gaussfit2}. This analysis was not done for the \textit{HST} STIS spectra, because there are no G750M or G750L spectra available for the same PA as the G430M spectra to compare the line flux of the above lines at the same location in the NLR\footnote{The G750M observations in the \hst archives are short and lie outside of the nominal bicone at PA $=$ 133$\degr$.}.

The DIS blue and red channel spectra permits us to simultaneously observe all of these lines at each location, which allowed us to generate Baldwin-Phillips-Terlevich (BPT) diagrams \citep{Baldwin1981, Veilleux1987,Kewley2001,Kauffmann2003, Kewley2006} and determine the ionization mechanism of the gas in the NLR-ENLR (see section \S \ref{subsec:BPT}).
To fit the \halpha line from the red channel spectra, we used \othree and \hbeta to obtain the wavelength centroid, width, and number of narrow components as well as the broad and intermediate \halpha component along the slit as identified in broad and intermediate H$\beta$. The wavelength centroid and width of each line were scaled to the corresponding emission line under the assumption that emission produced from the same cloud will provide the same mean velocity and dispersion. We also fit \halpha independently for the spectra extracted inside 3\arcsec\ and outside 15\arcsec\ radii of the nucleus for the DIS red channel for the reasons described in the next section. Fits in the inner regions help to compare \othree kinematics with \halpha in the NLR and at larger radii, they provide the rotational kinematics of the host galaxy.

\subsection{Effects of the APO Point-Spread Function}\label{subsec:psf}

\begin{figure*}[hbt!]
\centering
\subfigure[]{
\includegraphics[width=0.48\textwidth]{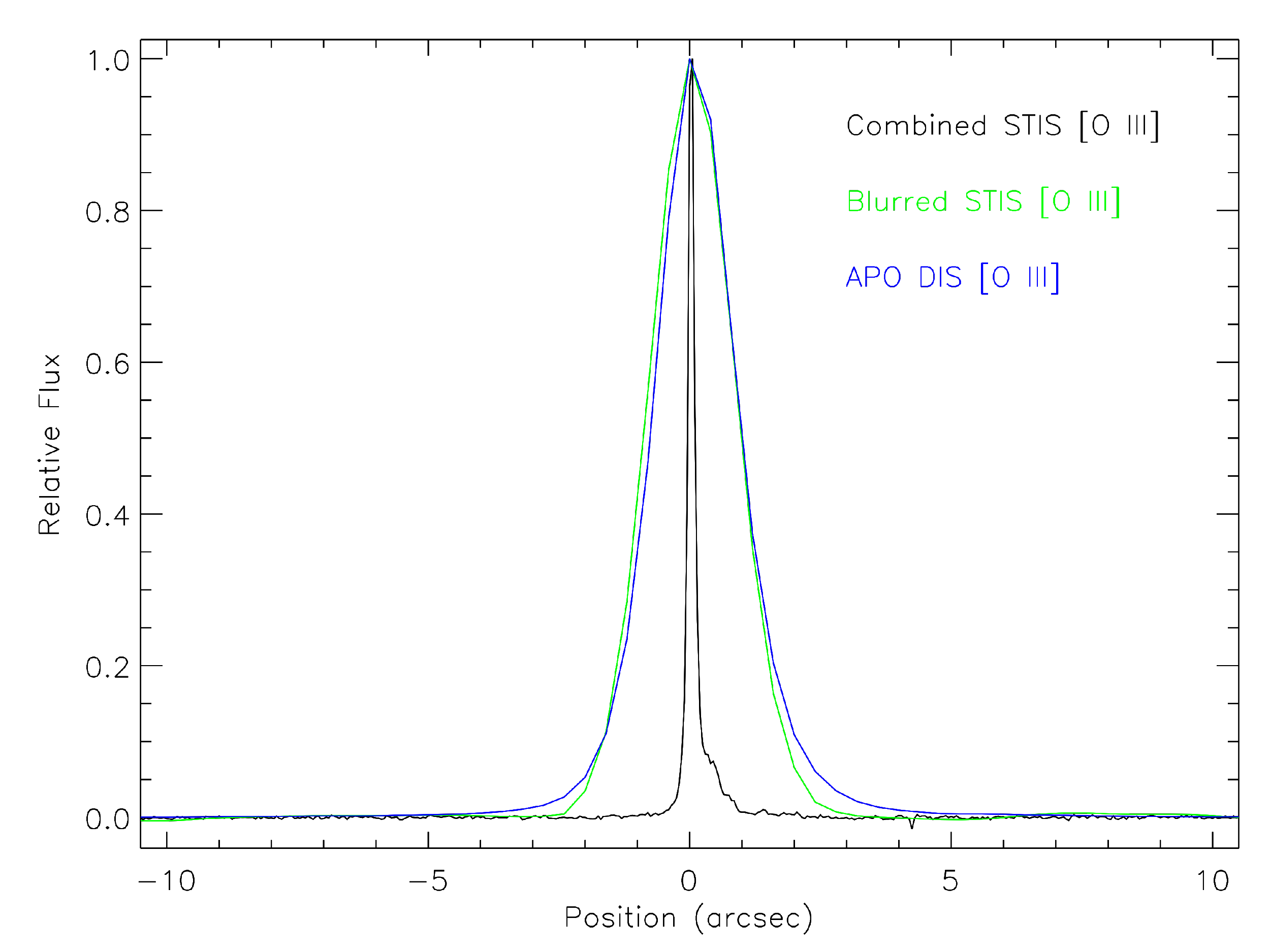}}\hspace{3ex}
\subfigure[]{
\includegraphics[width=0.48\textwidth]{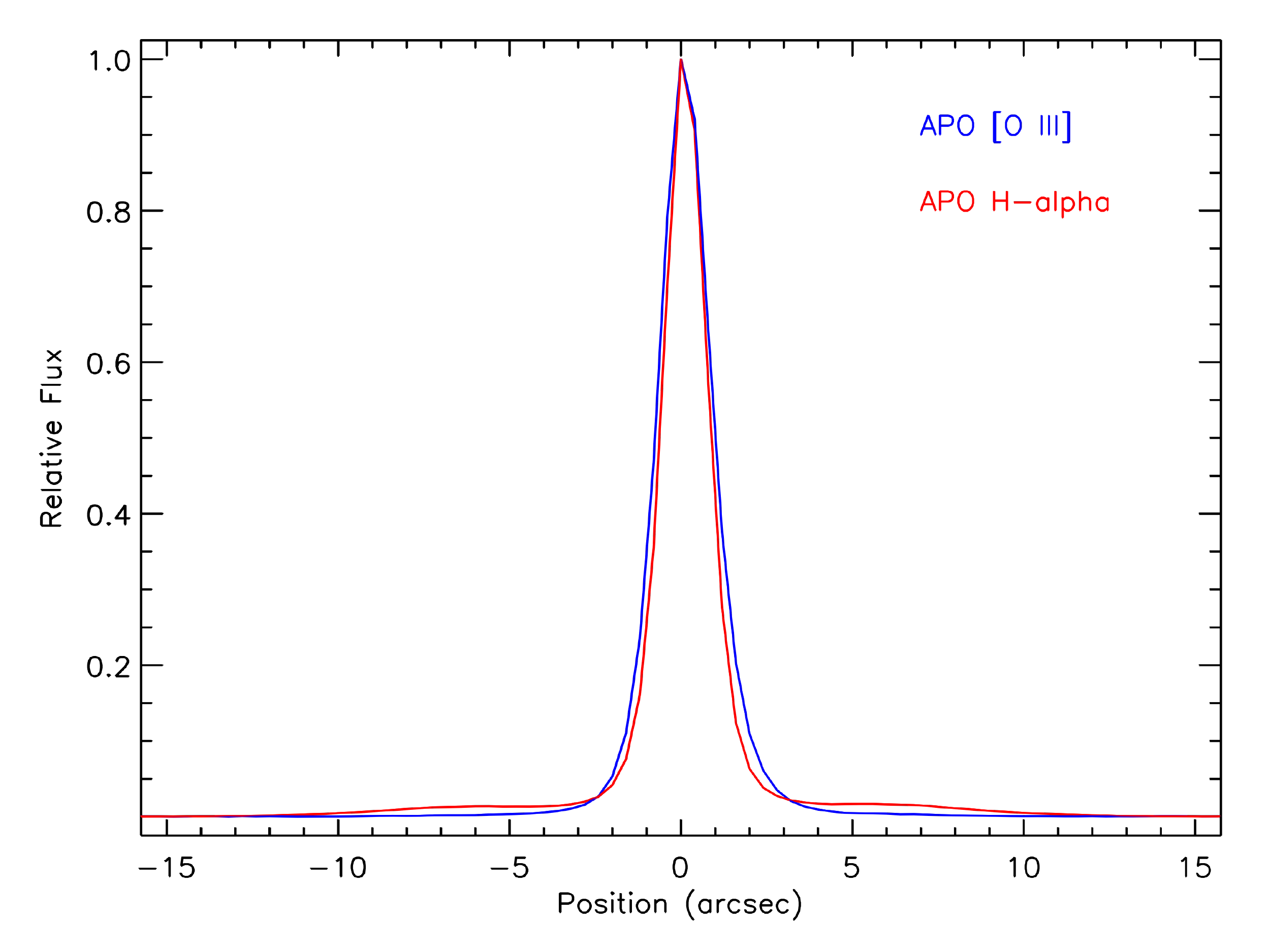}}
\caption{(a) A test to identify the effects of the PSF in the APO observations. We compare the \othree brightness profiles from the combined STIS spectra, APO slit, and SITS spectra convolved with the APO PSF. (b) A comparison of the brightness profiles of the \othree emission (blue channel) and \halpha emission (red channel) along the slit. The enhanced emission in the wings of the \halpha profile at 3\arcsec\ - 15\arcsec\ on both sides of the nucleus.}
\label{fig:scattered}
\end{figure*}

As shown in Figure~\ref{fig:F502N}, the APO slits are larger compared to the size of the NLR, and the bright NLR core convolved with the seeing PSF can affect kinematic measurements of the emission lines beyond $\sim$2\arcsec\ \footnote{The APO PSF also blurs the spatial locations of the kinematic components within $\sim$2\arcsec, but we have these inner regions partially covered with \textit{HST} STIS slits.}. We simulated this effect by combining the STIS spectral images at PA $=$ 89.8\arcdeg\ and degrading them to the spectral and spatial resolutions of the APO DIS blue spectral image at PA $=$ 98\arcdeg. We accomplished the latter by convolving the combined STIS spectra with Gaussians representing the line-spread function of APO in the dispersion direction (FWHM $\approx$ 1.2 \AA) and the PSF seeing at the time of the APO observations (FWHM $\approx$ 2\farcs2) in the cross-dispersion direction, and binning the convolved spectra to the spectral and spatial scales of APO DIS.

Figure~\ref{fig:scattered}a presents a brightness profile of the [O~III] emission along the combined STIS slits after continuum subtraction, which shows the bright central core of emission and the structure due to distinct [O~III] knots to the east of the core (see Figure~\ref{fig:F502N}). The convolution erases this spatial structure and extends the brightness profile outward to match that of the [O~III] emission from the APO spectrum, as expected.
The APO brightness profile has enhanced wings compared to that of the convolved STIS spectrum, likely due to the inclusion of more NLR emission-line knots in the APO DIS slit at PA $=$ 98\degr, as shown in Figure~\ref{fig:F502N}.
To account for this effect in the APO spectra, we identify kinematic components in the extended emission that match those in the NLR core in velocity and width and are consistent with a continuation of the PSF, and exclude those from further consideration.

In the right panel of Figure~\ref{fig:scattered}, we show the brightness profile of the H$\alpha$ emission from the red APO channel compared to that of [O~III] from the blue channel. The H$\alpha$ profile shows enhanced emission from 3\arcsec\ to 15\arcsec\ at the level of 1 - 2\% of the peak flux, due to the scattering problem in the red channel discussed earlier, which is why we exclude the measurements of H$\alpha$ and other emission lines from the red channel in this region.

\section{RESULTS}\label{sec:results}

\subsection{Ionized Gas Kinematics}\label{subsec:gas_kinematics}

\begin{figure*}[ht!]
\centering
\subfigure{
\includegraphics[width=0.305\textwidth]{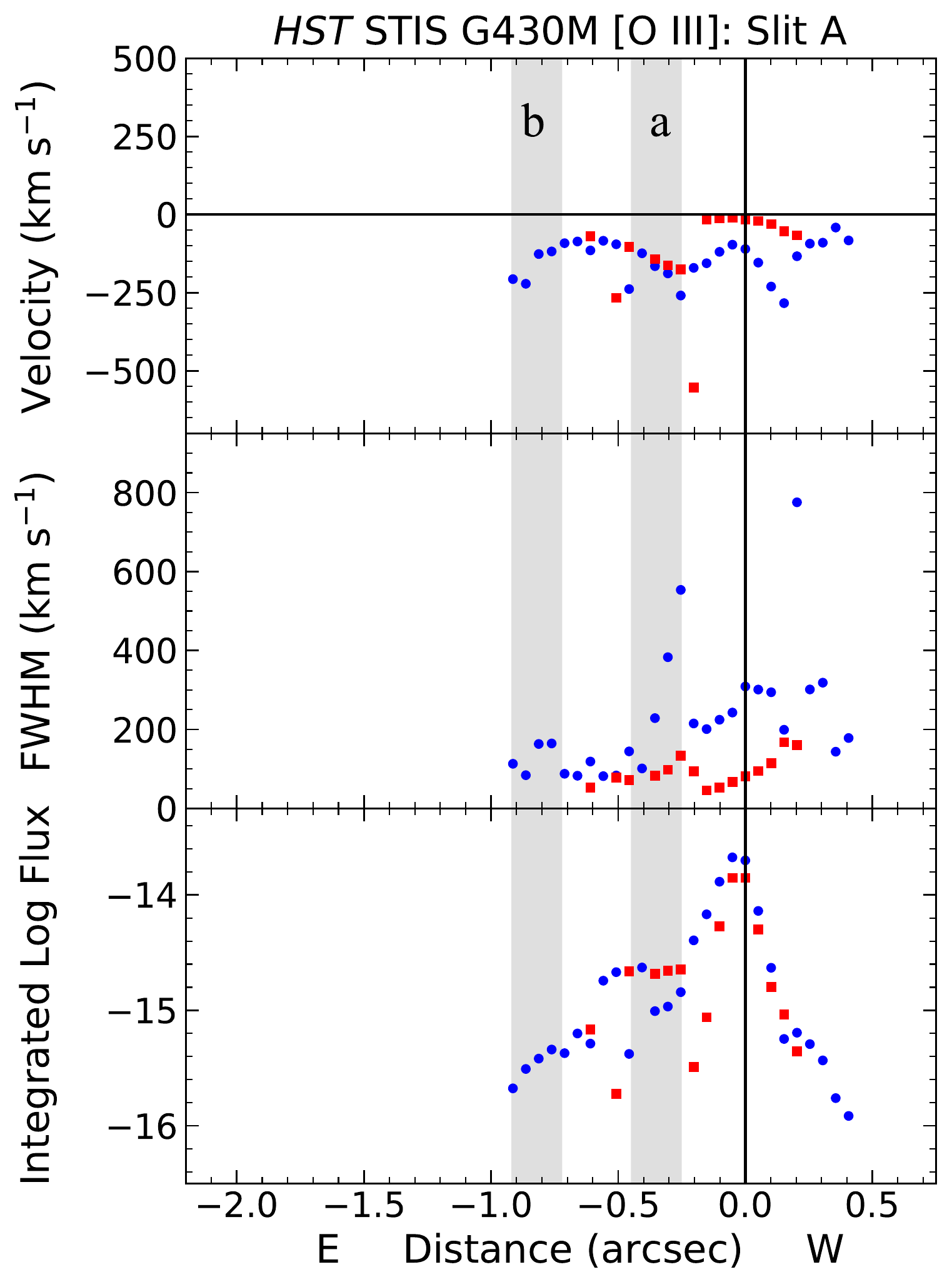}}\hspace{2ex}
\subfigure{
\includegraphics[width=0.305\textwidth]{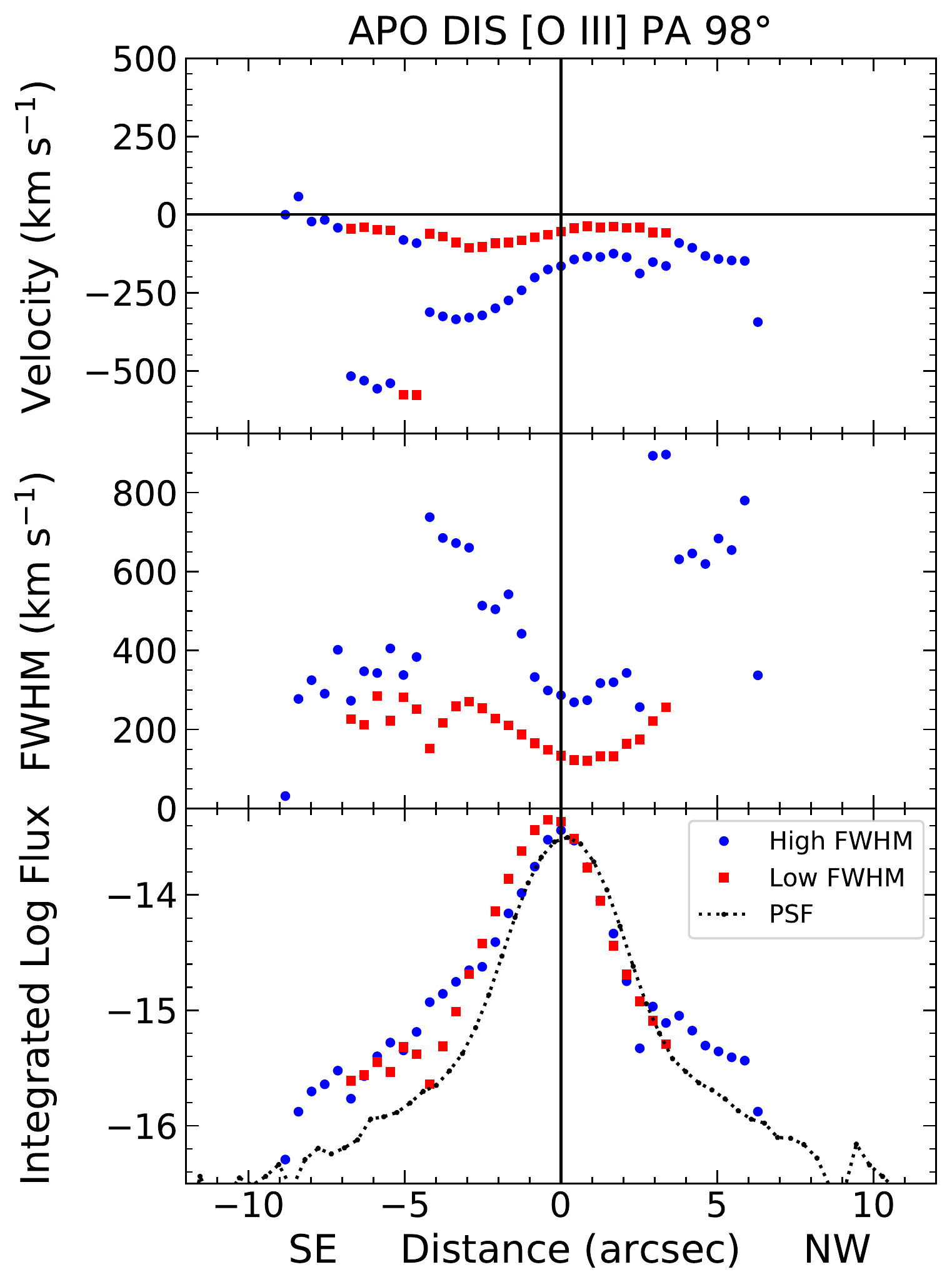}}\hspace{2ex}
\subfigure{
\includegraphics[width=0.305\textwidth]{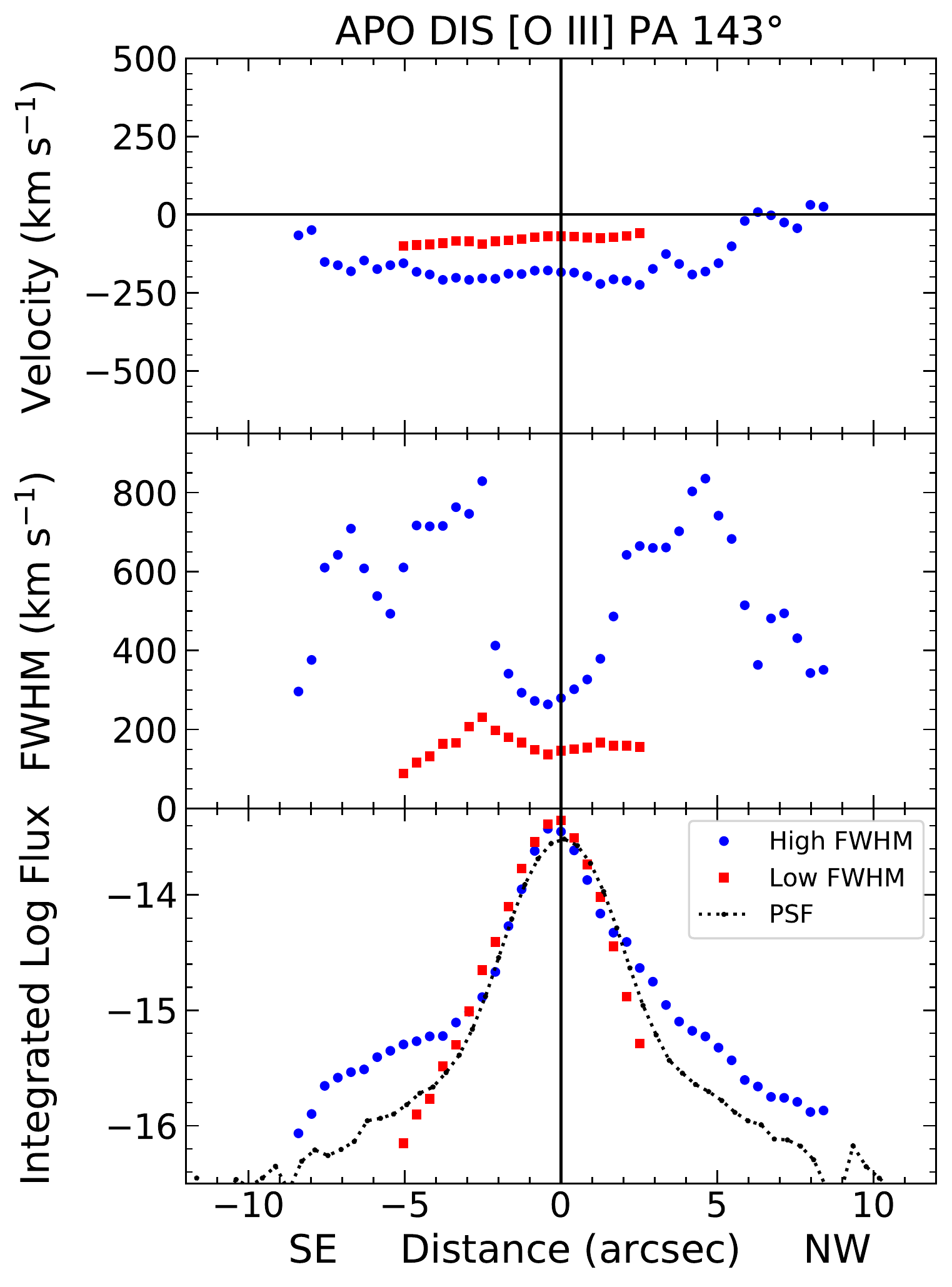}}\hspace{2ex}
\subfigure{
\includegraphics[width=0.305\textwidth]{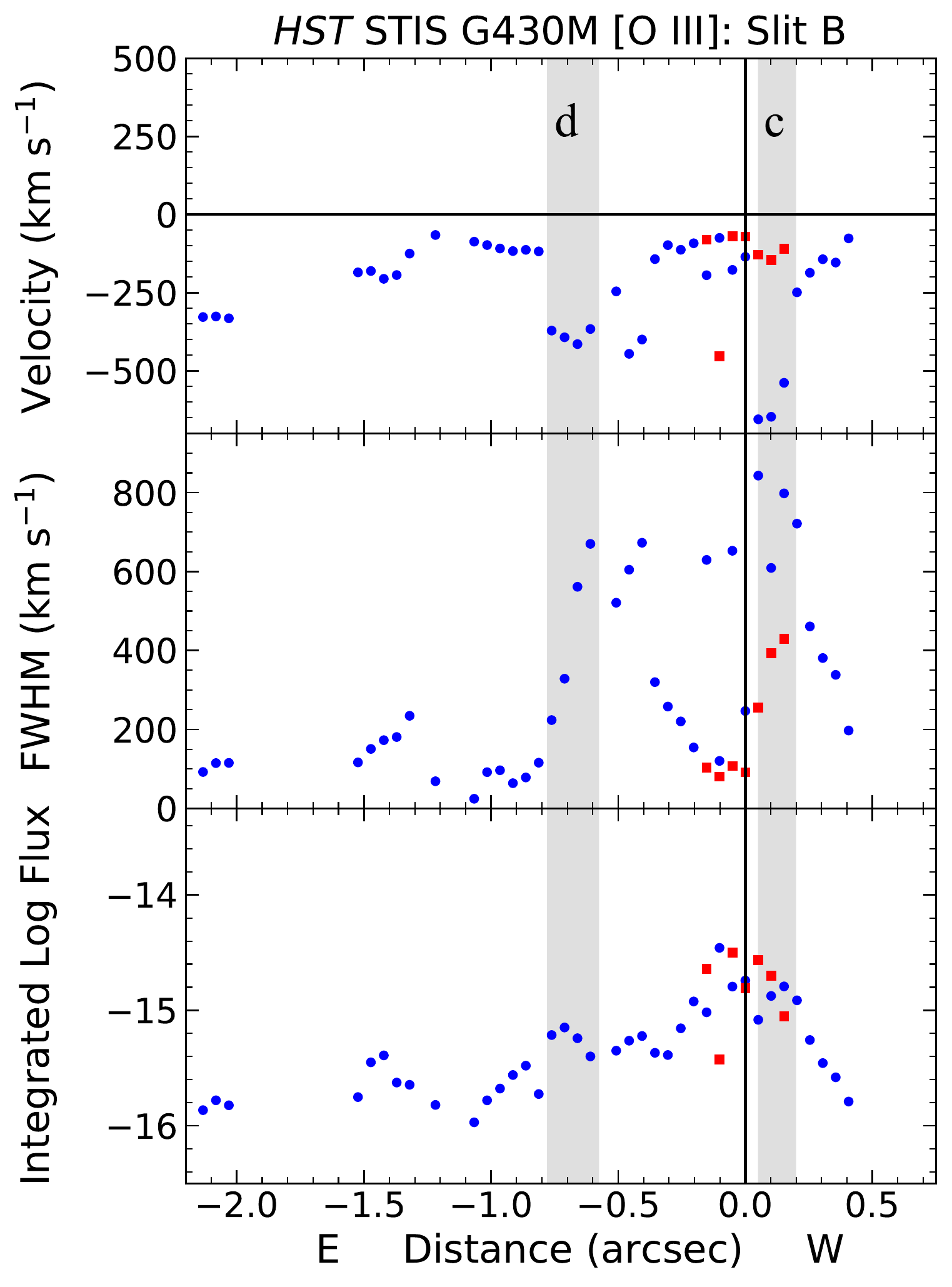}}\hspace{2ex}
\subfigure{
\includegraphics[width=0.305\textwidth]{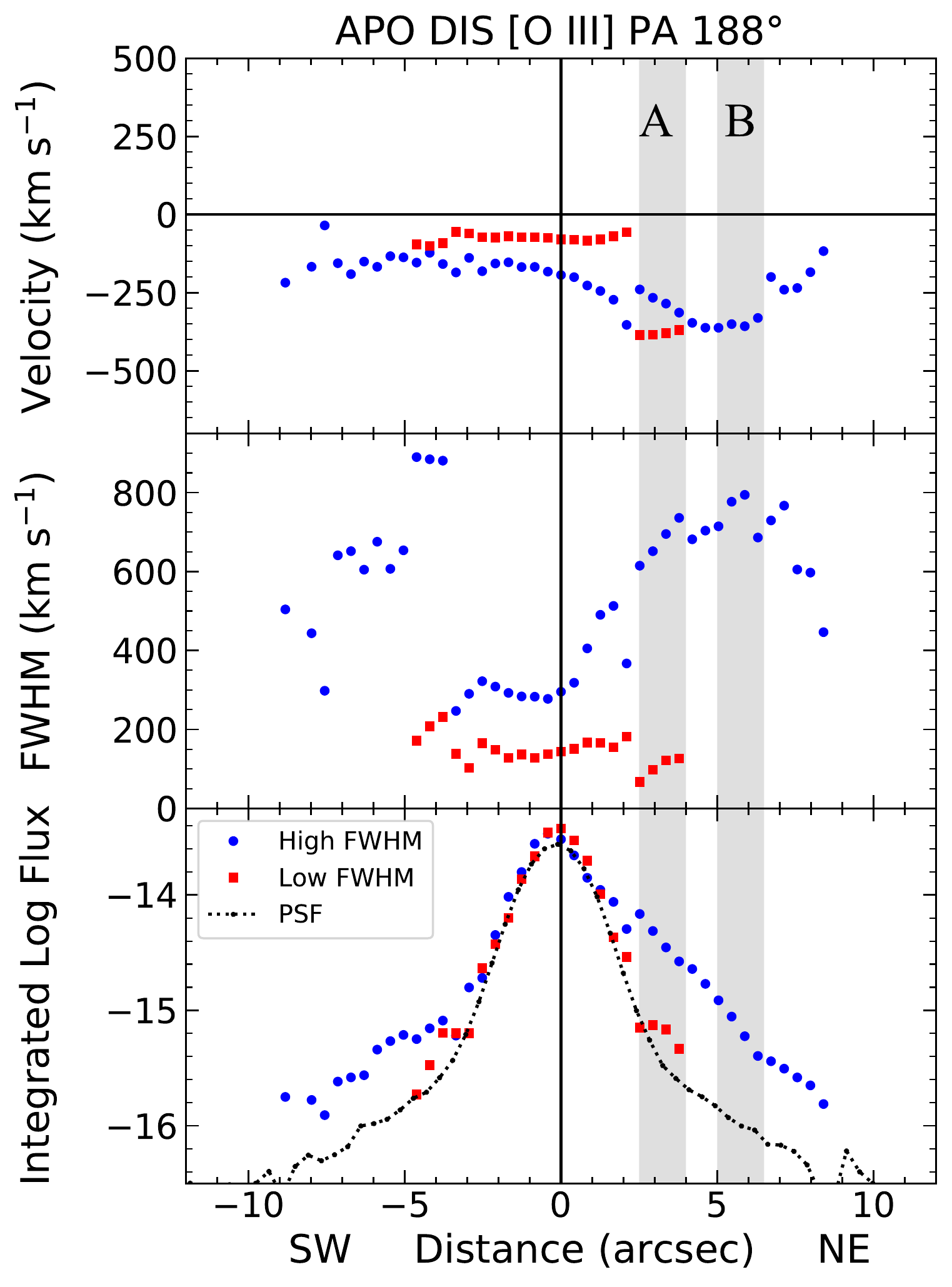}}\hspace{2ex}
\subfigure{
\includegraphics[width=0.305\textwidth]{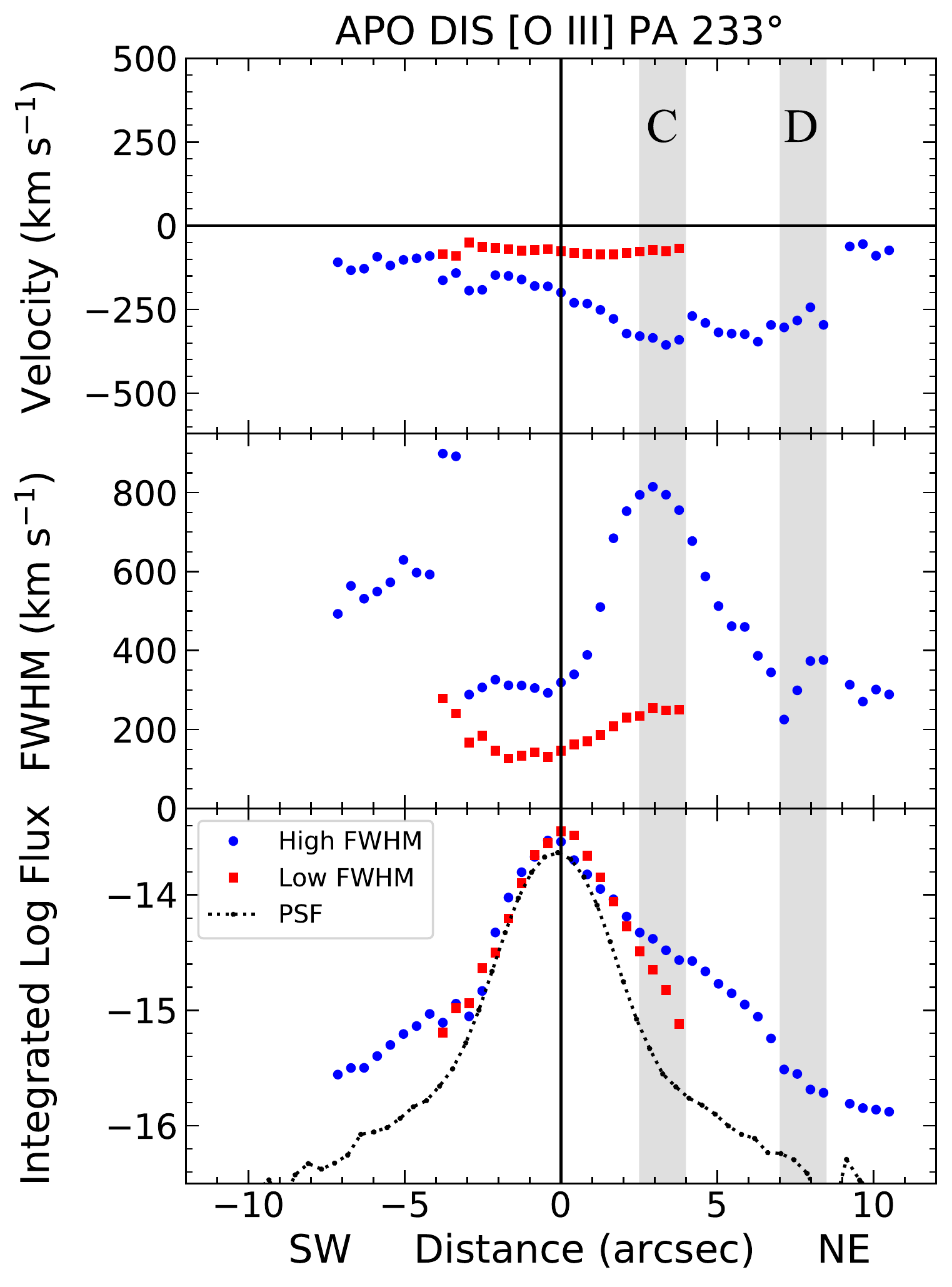}}
\caption{The [O~III] ionized gas kinematics for NGC~4051. The left column presents the observed kinematics inside the two \textit{HST} STIS G430M slits labeled as A (offset -0.05\arcsec) and B (offset +0.20\arcsec) at PA 89\arcdeg. The second and third columns contain the kinematics of the four APO DIS slits (PAs 98\arcdeg, 143\arcdeg, 188\arcdeg, 233\arcdeg). The positions of the \hst and APO slits can be seen in Figure~\ref{fig:F502N}. The top panels of each figure shows the radial velocity distribution of the \othree emitting gas along the slits for the two narrow components. The middle panels show the FWHM distribution of these components and the bottom panels map the radial variation in the integrated flux in logarithm (erg s$^{-1}$ cm$^{-2}$) for each line component. Because the APO slit width (2\arcsec) is 10 times wider than the G430M slit (0.2\arcsec), it captures the integrated emission at larger distances (up to $\sim$10\arcsec) that are too faint (below S/N $<$ 3) to be detected with the narrow G430M slits. The two NLR components are shown in blue circles (high FWHM) and red squares (low FWHM). The black dotted line in the bottom panel outlines the spread the PSF of the instruments as determined by the brightness profile of the standard star observation on the same night. The velocities are shown after subtracting off the systemic velocity of the galaxy. The FWHM values have been corrected to intrinsic values by subtracting the line-spread function in quadrature. The vertical lines at 0\farcs0 correspond to the continuum centroid of each slit. The points inside the shaded regions of the two STIS slits and APO slits PAs 188\arcdeg\ and 233\arcdeg\ are selected as separate knots of outflow based on their clustering in flux and/or kinematics, and will be employed for radiation-gravity model in section \S \ref{subsec:radiative_driving}.}
\label{fig:stis_apo_vel}
\end{figure*}

Figure~\ref{fig:stis_apo_vel} shows the kinematic and flux distributions of \othree ionized gas up to 10$''$ ($\sim$800 pc) from the central SMBH in NGC~4051, from both \hst and APO long slit observations. The first column corresponds to \textit{HST} STIS G430M slits A and B. Slit A (offset 0\farcs05 to the south) is closer to the nucleus and encompasses higher flux clouds than slit B (0\farcs20 to the north), which covers part of the outflow bicone up to a larger distance than slit A, although neither are fully along a bright edge of the bicone.
Our Bayesian algorithm does not fit emission lines below 3$\sigma$ over the noise and therefore shows fewer velocity points than from the manual fitting analysis we performed previously \citep{Fischer2013}. However, the overall kinematic structure is very similar to that presented in \cite{Fischer2013}. The FWHM distribution shows most of the kinematics have FWHM $<$ 400 \kms in slit A. The widths are larger on average for slit B, which is slightly farther from the nucleus.
In general, the clumping together of points in velocity, FWHM, and flux indicates that STIS is resolving the individual knots of [O~III] emission, some of which can be seen in Figure~\ref{fig:F502N}.

Nearly all of the radial velocities in the STIS observations are blueshifted or near zero compared to the systemic velocity of NGC~4051 (at zero in these plots). The emission is more extended in the east due to the relative orientations of the slits and cone, which can also be seen in Figure~\ref{fig:F502N}.
We observe blueshifted velocities up to $\sim$500 \kms inside 1\arcsec\ to the east for higher FWHM ($>$ 200 km s$^{-1}$) clouds. The \othree line splitting as mentioned in \cite{Christopoulou1997} can be seen in the two velocity components separated by $\sim$150 \kms at $\sim$0.5\arcsec. With the exception of a few points, the narrow component is close to the systemic velocity of the host galaxy.

The morphology of the ionized gas in Figure~\ref{fig:F502N} and its location mostly to the NE of the nucleus, together with the presence of mostly blueshifted radial velocities, suggests an outflowing bicone of ionized gas, where the SW, redshifted cone is likely hidden behind the host disk.
In \cite{Fischer2013}, the modelling of the outflow bicone was carried out using the kinematics from these two \hst G430M slits. As seen from the \othree image in Figure~\ref{fig:F502N}, the blueshifted cone is clearly visible and lies to much further extents than the G430M slits coverage, which is also comparable to that seen in the MERLIN Echelle Spectrograph (MES) \othree image \citep{Christopoulou1997} and [S III] observations with GEMINI GMOS/IFU \citep{Barbosa2009}.

To estimate the total extent of AGN ionized gas and the size of AGN driven outflows, we used the APO DIS spectra with wider slits that may detect the low flux emission. As seen from Figure~\ref{fig:F502N}, the four APO slits cover nearly all of the ionized ``cone''. Similar to the G430M observations, we measured the velocities, FWHM and flux of the ionized gas by using Gaussian fits to the emission lines in these fours slits as shown in the second and third column of Figure~\ref{fig:stis_apo_vel}. The narrow \othree emission lines can be fit with two kinematic components, except in more distant, fainter regions where one component is sufficient.
We focus on the emission outside of that covered by STIS ($\gtrsim$ 2$''$ from the nucleus) and kinematic components that are clearly not due to the wings of the PSF from the bright central \othree knots (Figure~\ref{fig:stis_apo_vel}).

The two NLR components are sorted by FWHM and color-coded as blue (high FWHM) and red (low FWHM). In most cases, the points with higher FWHM have higher velocity than the ones with lower FWHM except for a few positions when the narrower component corresponds to a higher velocity centroid.
In a couple of instances, the high and low FWHM components have switched locations in the velocity plots, such as those at $-$5\arcsec\ in the 98\arcdeg\ slit, which is likely due to their relatively similar FWHM derived from the Gaussian fitting routine.
Although a large majority of the APO NLR points have FWHM less than the maximum allowed width (900 km s$^{-1}$), a few points
reach the maximum FWHM. Those at $+$3\arcsec\ in the 98\arcdeg\ slit may be due to contamination from the ILR component, whereas the points at $-$4\arcsec\ are likely real due to their high fluxes above the ILR PSF.
We detect significant jumps in velocity and FWHM curves in both STIS and APO observations (Figure \ref{fig:stis_apo_vel}). This is because we are detecting separate ionized gas knots (as shown in Figure \ref{f502n}) that have their own peculiar velocities and FWHMs, and therefore, the noticeable jumps between two adjacent points are due to transitions between these knots. The clumping of points together in the APO data is due to a convolution of the finite extent of the clouds with the seeing.

The APO observations at PA $=$ 98\arcdeg\ cover the emitting gas in the two \textit{HST} STIS G430M slits but with lower spatial resolution. 
In this and all other slits we see a bright, narrow, slightly blueshifted component at $\sim$ $-$70 \kms extending to $\pm$5$''$ that is likely due to the PSF from the bright core.
There is a high FWHM blueshifted component at $\sim$ $-$300 \kms from 2$''$ -- 4$''$ in the east that is similar to the STIS blueshifted emission closer in and clearly not due to the PSF.
In addition, we observe high blueshifted velocities of $-$550 \kms towards the east between 5$''$ and 7$''$, just outside of the nominal cone, which extends the outflows beyond those seen in the STIS data.
The APO slits at PA $=$ 188\arcdeg\ and 233\arcdeg\ align along the two edges of the cone and encompass the majority of the \othree emission. We see line of sight velocities up to $\sim$ $-$400 km s$^{-1}$, extending the detection of outflows up to a projected distance of $\sim$8\arcsec\ ($\sim$640 pc) in the NE.
At PA $=$ 143\arcdeg, which lies close to the major axis of the host galaxy, we continue to detect two components of blueshifted emission, although no velocities exceed $-$250 km s$^{-1}$. 
We built an outflow bicone model based on kinematics from the slits at PA $=$ 188\arcdeg\ and 233\arcdeg\ in section \S \ref{subsec:outflow}. 

Our observed velocities are in agreement with those of \cite{Christopoulou1997}, \cite{Barbosa2009}, and \cite{Fischer2013} at smaller radii. Similar to the STIS spectra, the narrowest component close to the nucleus is near systemic velocity with a slight blueshift of only $\sim$ $-$70 km s$^{-1}$. It is ambiguous as to whether this is part of the host galaxy rotation or part of outflows close to the plane of the galaxy. The wider component (blue points) clearly traces blueshifted outflows at distances from 2\arcsec\ to 8\arcsec\ with high velocity widths (FWHM $>$ 400 km s$^{-1}$). However the FWHM of emission in the slit along PA 233\arcsec\ starts to fall after 5\arcsec\ while still moving with velocities $\sim$ $-$400 km s$^{-1}$. A few points around 10\arcsec -- 11\arcsec\ in the NE are close to zero radial velocity but with a rather large FWHM of  $\sim$200 km s$^{-1}$, suggesting a limit to the outflows and transition to rotation.

\subsection{Comparison with Rotation}\label{subsec:roatation}

\begin{figure*}[hbt!]
\centering
\subfigure{
\includegraphics[width=0.48\textwidth]{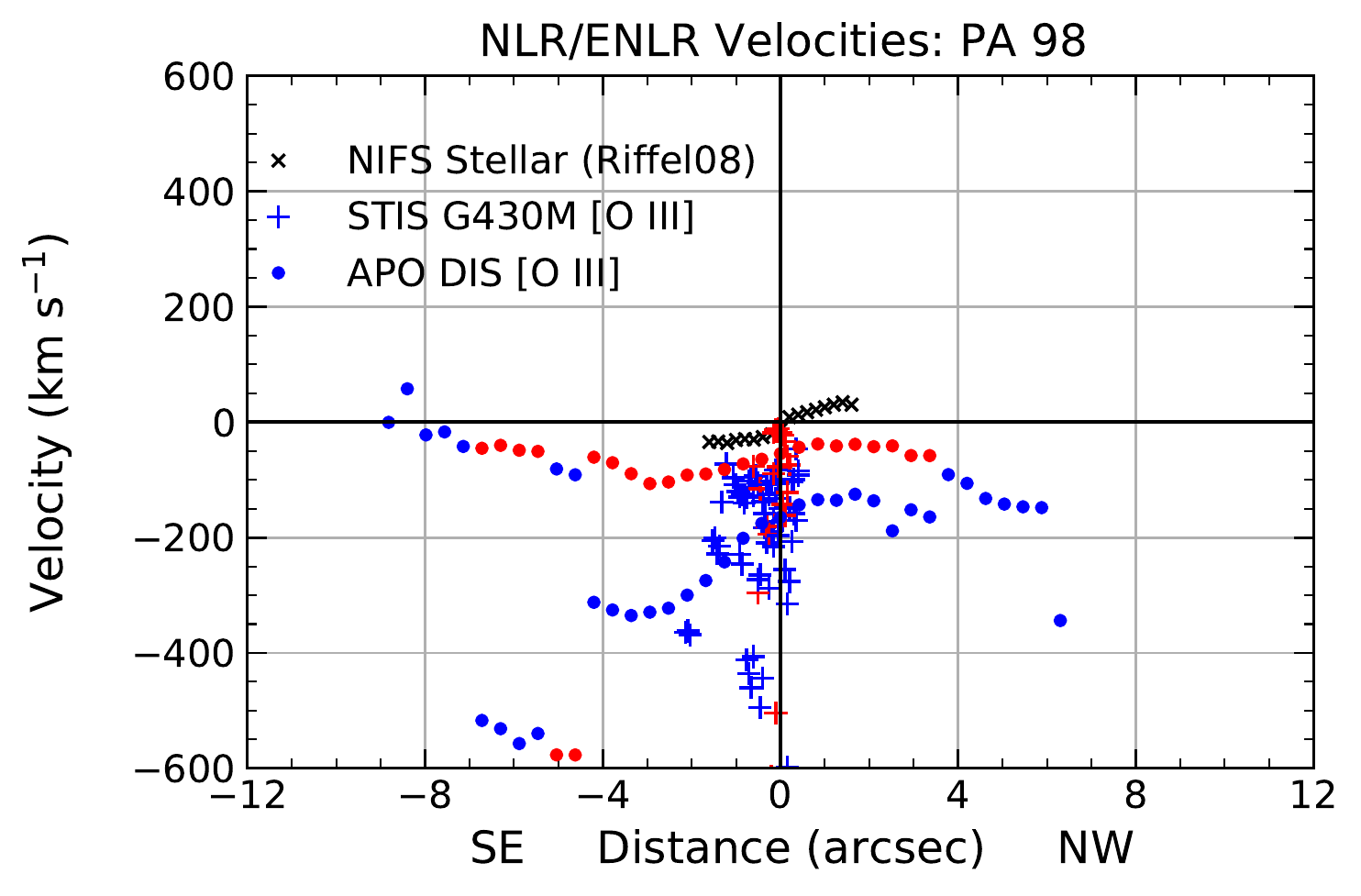}}\hspace{3ex}
\subfigure{
\includegraphics[width=0.48\textwidth]{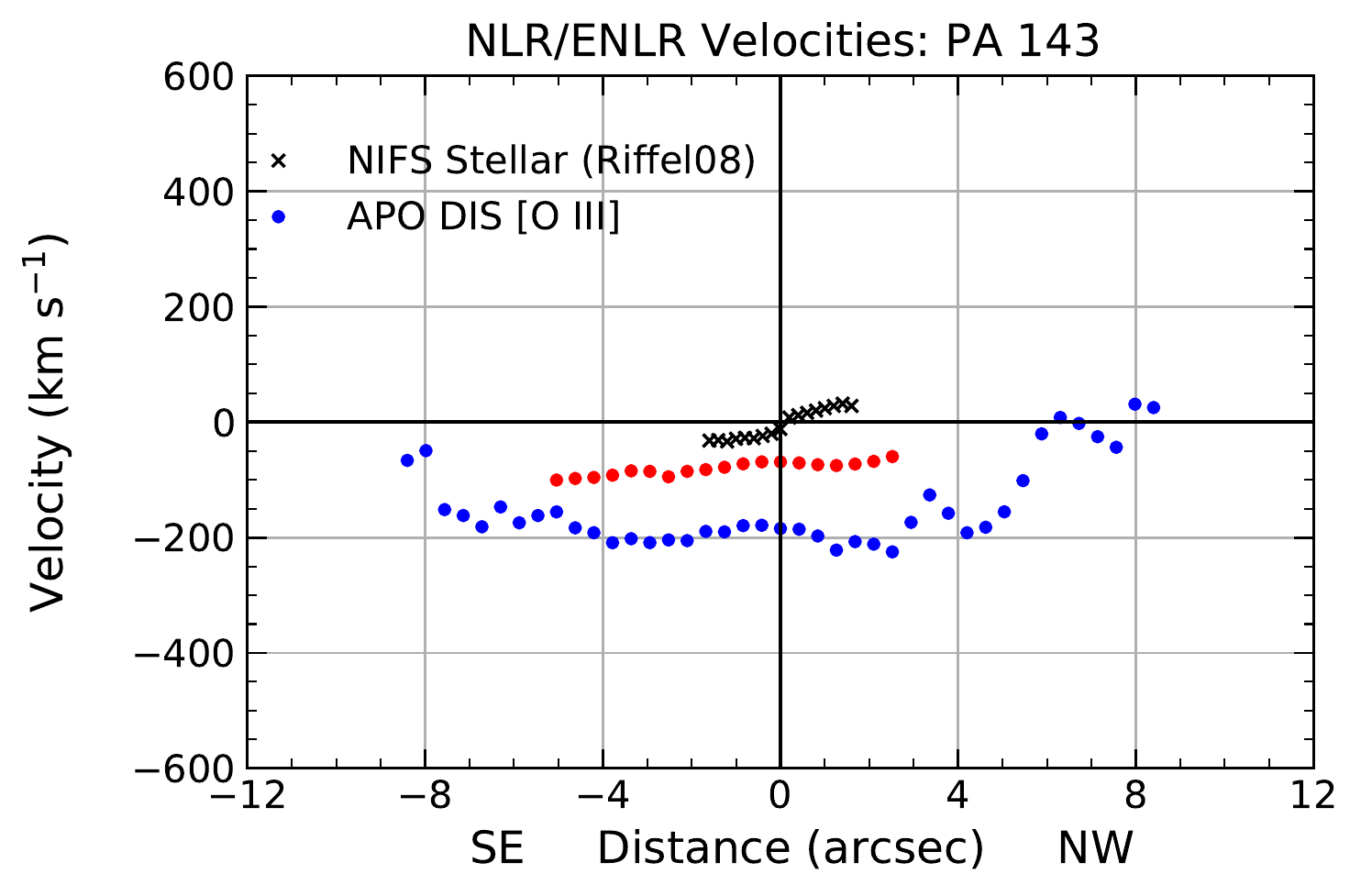}}
\caption{A comparison of stellar kinematics (black cross) from \citep{Riffel2008} with \othree ionized gas kinematics from STIS G430M (high to low FWHM in blue-red cross points) and APO DIS (high to low FWHM in blue-red dots) observations. NIFS stellar rotation curve is projected for the line of sight velocities onto APO PA $=$ 98\arcdeg\ and 143\arcdeg. The observed STIS velocities for two parallel slits (PA 89.9\arcdeg) are plotted along PA 98\arcdeg as the wide APO slits (2\arcsec) covers the entire region inside both STIS slits. No STIS observation is available to compare the kinematics with the other APO slits at PA 143\arcdeg, 188\arcdeg\ and 233\arcdeg.}.
\label{fig:apo_nifs}
\end{figure*}

To separate a possible rotational component from the outflow component of the ionized gas, we need to compare the observed kinematics with a rotational curve obtained at the same radii. \cite{Riffel2008} present the stellar kinematics (GEMINI/NIFS K-band observations) within 3\arcsec\ of the nucleus with turnover radius at $\sim$1\arcsec. The amplitude of the stellar rotation curve varies from -40 \kms to +40 \kms from SE to NW with a major axis given at 120\arcdeg, close to the photometric axis of the large scale disk at PA $=$ 130\arcdeg\ \citep{Kaneko1997}. Figure~\ref{fig:apo_nifs} shows a comparison between stellar and \othree ionized gas kinematics within the NLR. We projected the stellar velocities (PA $=$ 120\arcdeg) to APO PAs of 98\arcdeg\ and  143\arcdeg\ to identify any gas clouds that are part of the host galaxy rotation and not radially driven outflows. From Figure~\ref{fig:apo_nifs}, we don't see any APO points that are close to the stellar velocity curve. There are a few STIS points close to the nucleus that could be part of the rotation curve, although they do not show a distinct rotation pattern so they are ambiguous.

\begin{figure*}[hbt!]
\centering
\subfigure{
\includegraphics[width=0.48\textwidth]{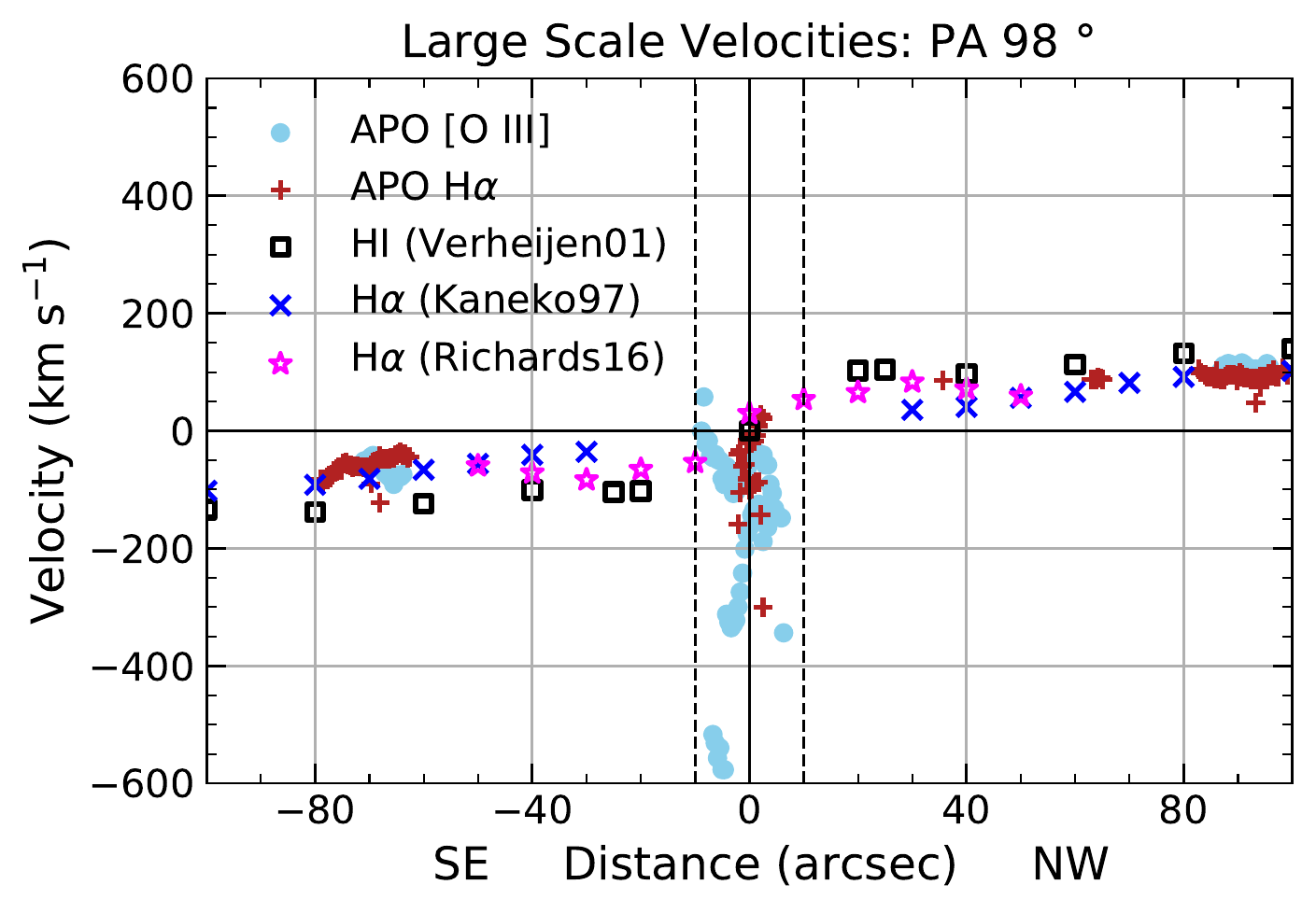}}\hspace{3ex}
\subfigure{
\includegraphics[width=0.48\textwidth]{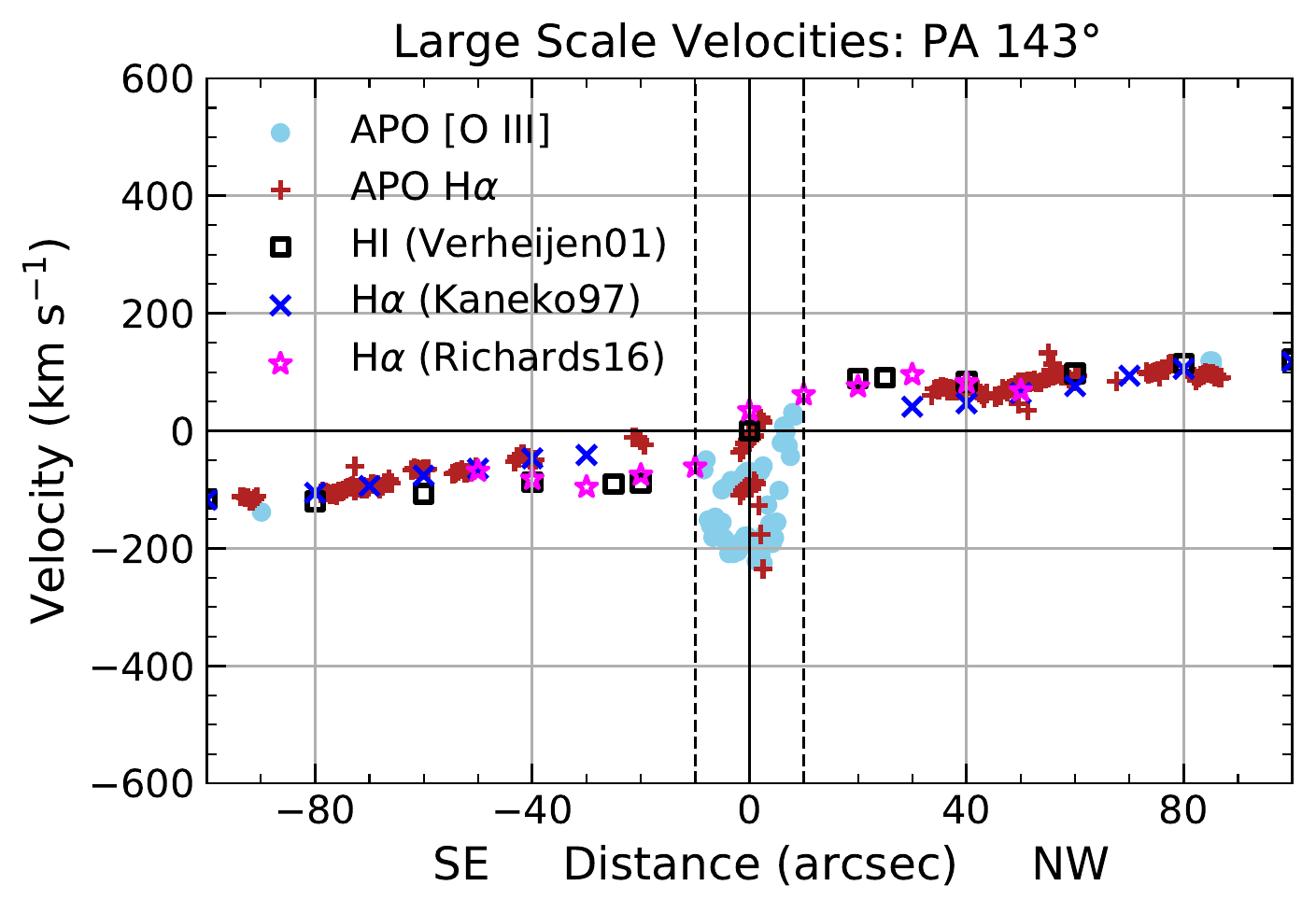}}
\caption{A comparison of the neutral (HI 21-cm) and ionized gas (\othree and H$\alpha$) kinematics for PAs $=$ 98\arcdeg\ and 143\arcdeg. The APO \othree and \halpha observations (this work) are compared with projected HI \citep{Verheijen2001} and \halpha \citep{Kaneko1997, Richards2016} rotation curves within 100\arcsec\ ($\sim$8 kpc) of the nucleus of NGC~4051. The dashed vertical lines give the extent of the NLR [O~III] and H$\alpha$ emission detected by APO long slits.}
\label{fig:rotation}
\end{figure*}

We derived the rotation curve for the host galaxy using the kinematics of extended ionized gas at distances 10\arcsec\ -- 100\arcsec. We calculated the mean velocities by fitting Gaussian profiles (see section \S \ref{subsec:gaussfit}) to relatively narrow (FWHM $\leq$ 150 km s$^{-1}$) \othree and \halpha lines separately from spectra extracted at those distances for all four long slits. In all cases, the Gaussian fits required only one kinematic component.

We  then compared our observed line of sight velocities with neutral hydrogen (HI) 21-cm observations \citep{Verheijen2001} and ionized gas (H$\alpha$) kinematics \cite{Kaneko1997, Richards2016}. Figure~\ref{fig:rotation} shows the APO \othree and \halpha velocities along with HI and ionized gas velocity fields that were obtained from the corresponding papers. We projected these extracted values to the APO slit positions. The adopted major axis was given as 310\arcdeg, 130\arcdeg, and 311\arcdeg\ and inclinations adopted were 49\arcdeg, 37\arcdeg, 44.6\arcdeg\ by \cite{Verheijen2001,Kaneko1997} and \cite{Richards2016}, respectively. Our kinematics are in excellent agreement with these previous works, particularly for PA 143\arcdeg, which is close to the major axis of the galaxy. By comparing the large scale rotation curves to the velocities within $\pm$10\arcsec\ of the nucleus (dashed vertical boundaries in Figure~\ref{fig:rotation}), the points that lie above $\sim$100 \kms are clearly part of the outflowing gas. We find no evidence for AGN driven outflow at distances greater than $\sim$10\arcsec\ ($\sim$800 pc) from the nucleus.

\begin{figure*}[hbt!]
\centering
\subfigure[]{
\includegraphics[width=0.45\textwidth]{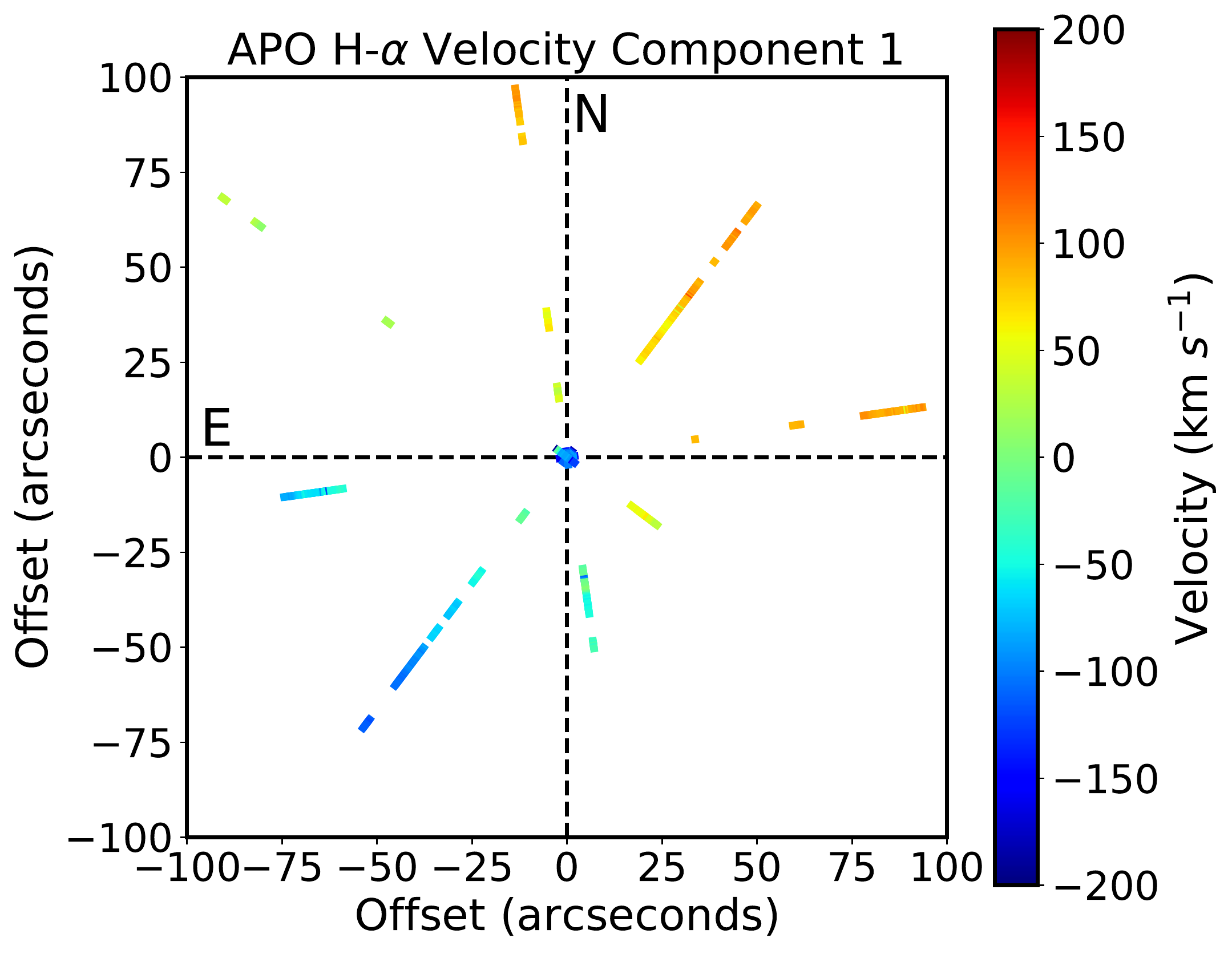}}\hspace{5ex}
\subfigure[]{
\includegraphics[width=0.45\textwidth]{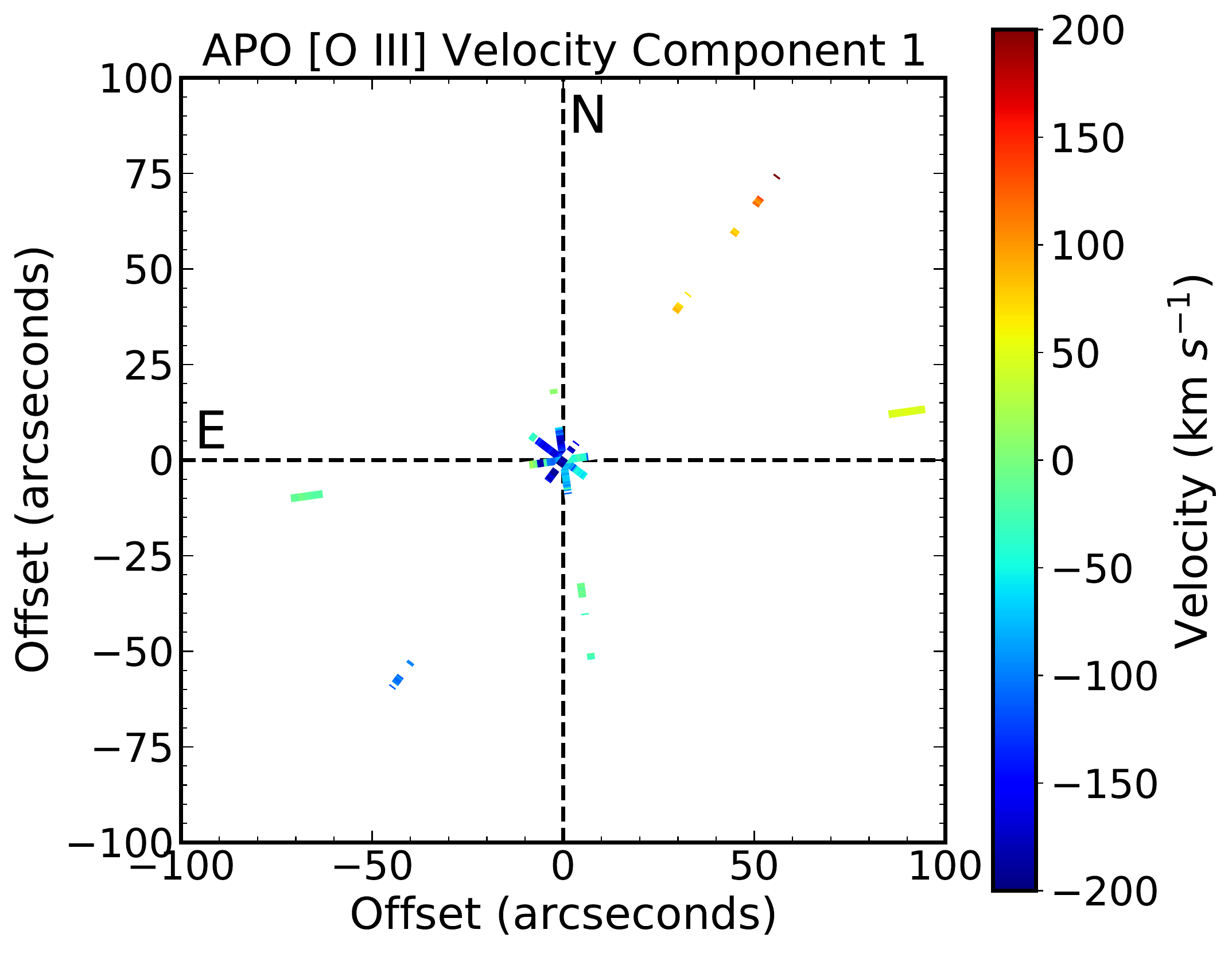}}
\subfigure[]{
\includegraphics[width=0.45\textwidth]{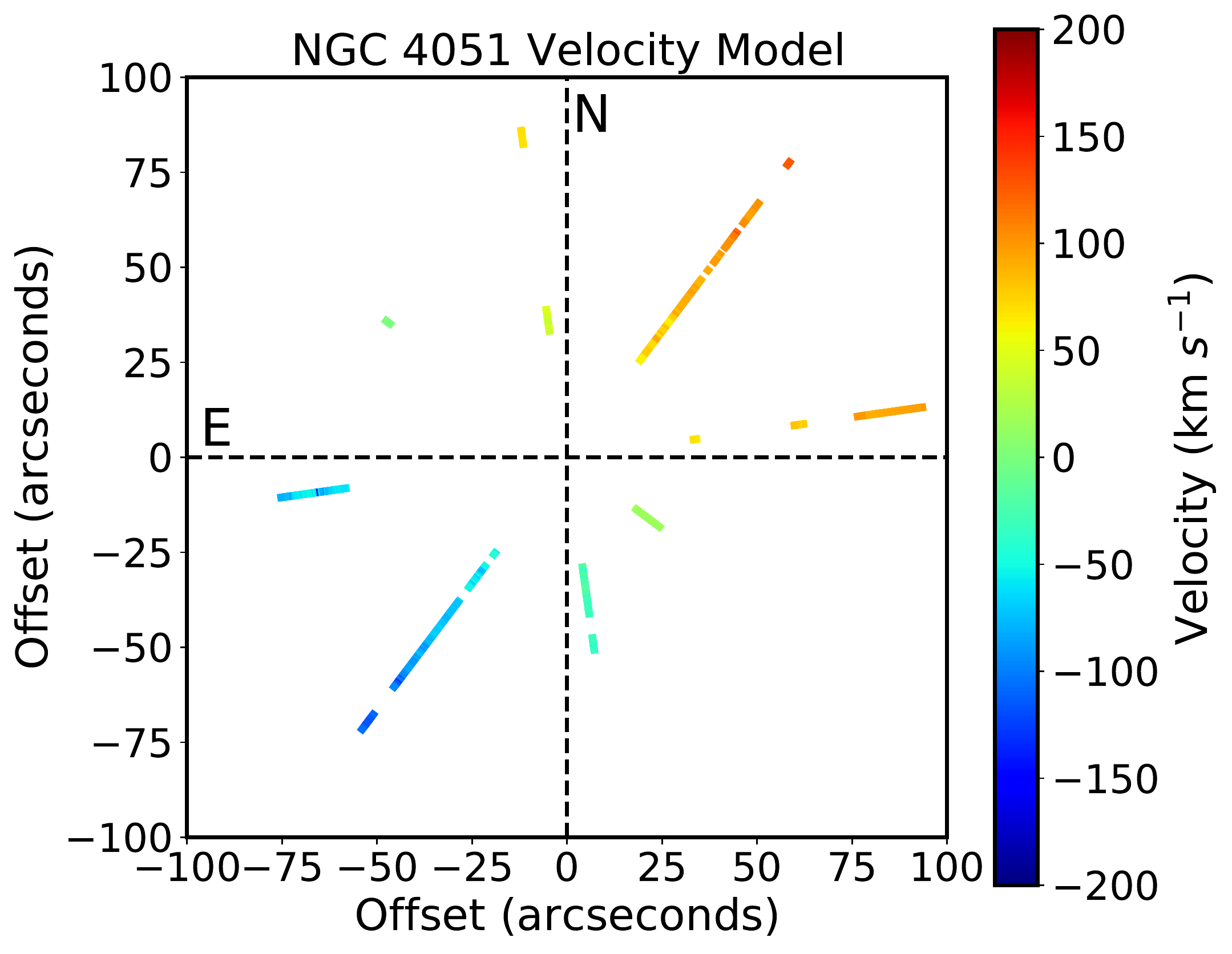}}\hspace{5ex}
\subfigure[]{
\includegraphics[width=0.45\textwidth]{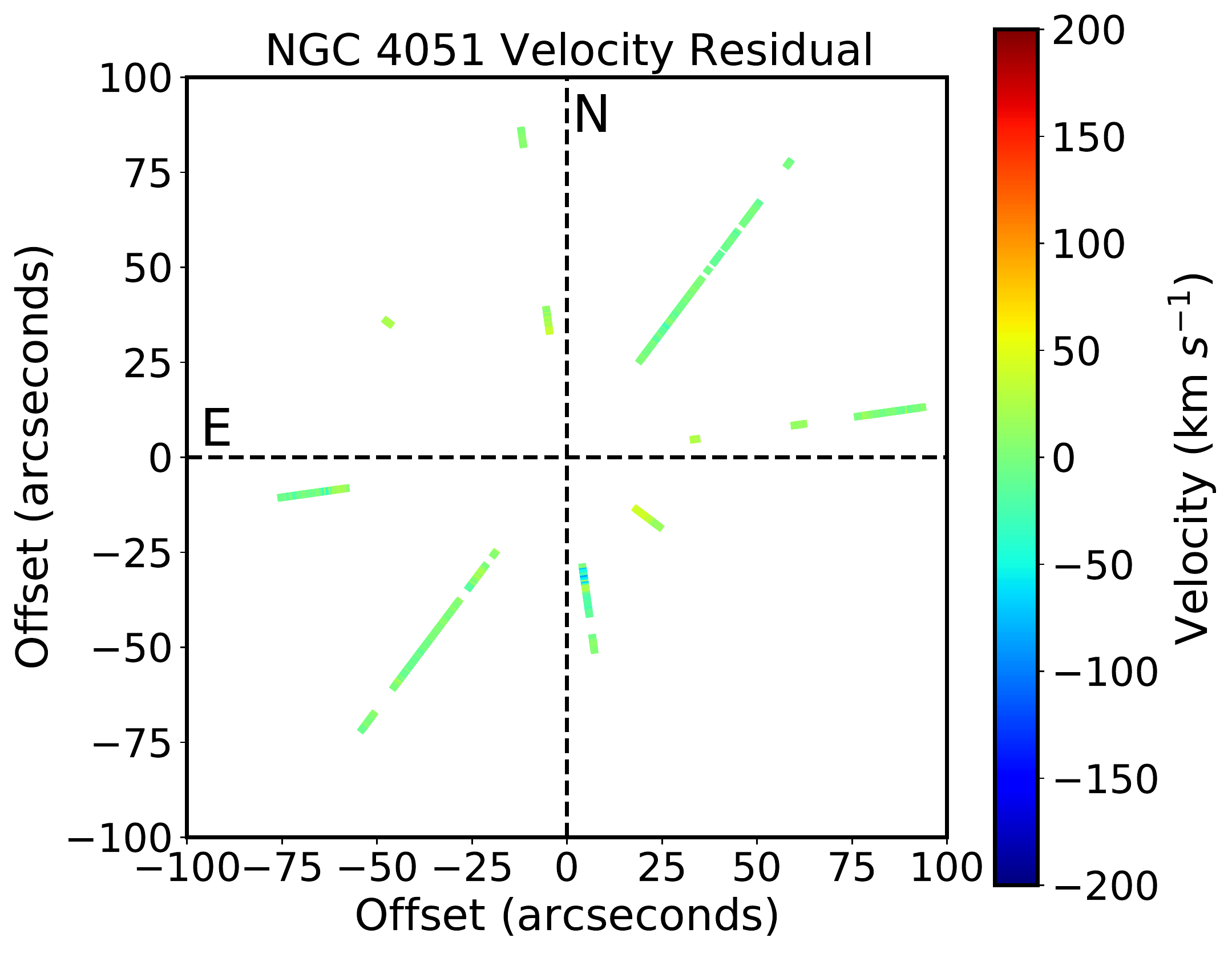}}
\caption{Figure (a) and (b) show a comparison of large scale \halpha and \othree kinematics in NGC~4051. A rotational velocity signature can be noticed in both emission lines although \halpha is more dominant and extended than [O~III]. The extended [O~III] kinematics were too weak to generate sufficient number of data points to model the velocity fields. Therefore, we used \halpha kinematics to fit the circular (rotational) velocities of the host galaxy using the Diskfit software package. Figure (c) and (d) show the modeled and residual velocity maps derived by Diskfit.}
\label{fig:velmaps_diskfit}
\end{figure*}

We also took an opportunity to model the host galaxy major axis and inclination using our observed (extended) gas kinematics. The modelling was performed with the publicly available Diskfit \citep{Spekkens..Sellwood2007, Sellwood..Spekkens2015, Kuzio2012soft,Peters2017} software package. Diskfit is generally used to fit simple symmetric models to disk galaxies using either photometric data or kinematic maps. We derived a rotation curve to NGC~4051 using APO DIS long slit kinematics. The extended (up to $\pm$100\arcsec\ from the nucleus) velocity maps for \halpha and \othree ionized gas are shown in the top panel of Figure~\ref{fig:velmaps_diskfit}. The \halpha velocity fields are much stronger and extend to larger distances than [O~III], which is ascribed to robust star formation \citep{Bartunov1994,van_Dyk1996,Anderson2012} that produces strong \halpha and \hbeta emission lines, but weaker \othree at those distances.

We fit a simple rotational model to the host galaxy using extended \halpha velocities and assumed a circular motion of the disk. As discussed above, we were unable to claim any points in the inner $\pm$10$''$ as part of the disk galaxy rotation with certainty, so it was not possible to model a rotation curve for the circumnuclear regions. To fit a rotation model to the extended galaxy, we removed the velocities from the inner $\pm$10\arcsec. The modeled velocity field and residuals are shown in Figure~\ref{fig:velmaps_diskfit} as the bottom two panels. The Diskfit model provides a host galaxy PA of 135$\pm$2 and inclination of 43$\pm$1 in degrees, within 5\arcdeg\ and 2\arcdeg\ of the values determined by \citet{Kaneko1997}, respectively (see section \S \ref{subsec:outflow}). The systemic velocity was corrected from 0 \kms to 13 km s$^{-1}$, indicating the accuracy of our velocities when compared to the adopted systemic redshift from 21-cm observations.

\subsection{Ionization Diagnostics of Observed Emission lines}\label{subsec:BPT}

\begin{figure*}[ht!]
\centering
\subfigure{
\includegraphics[width=0.33\textwidth]{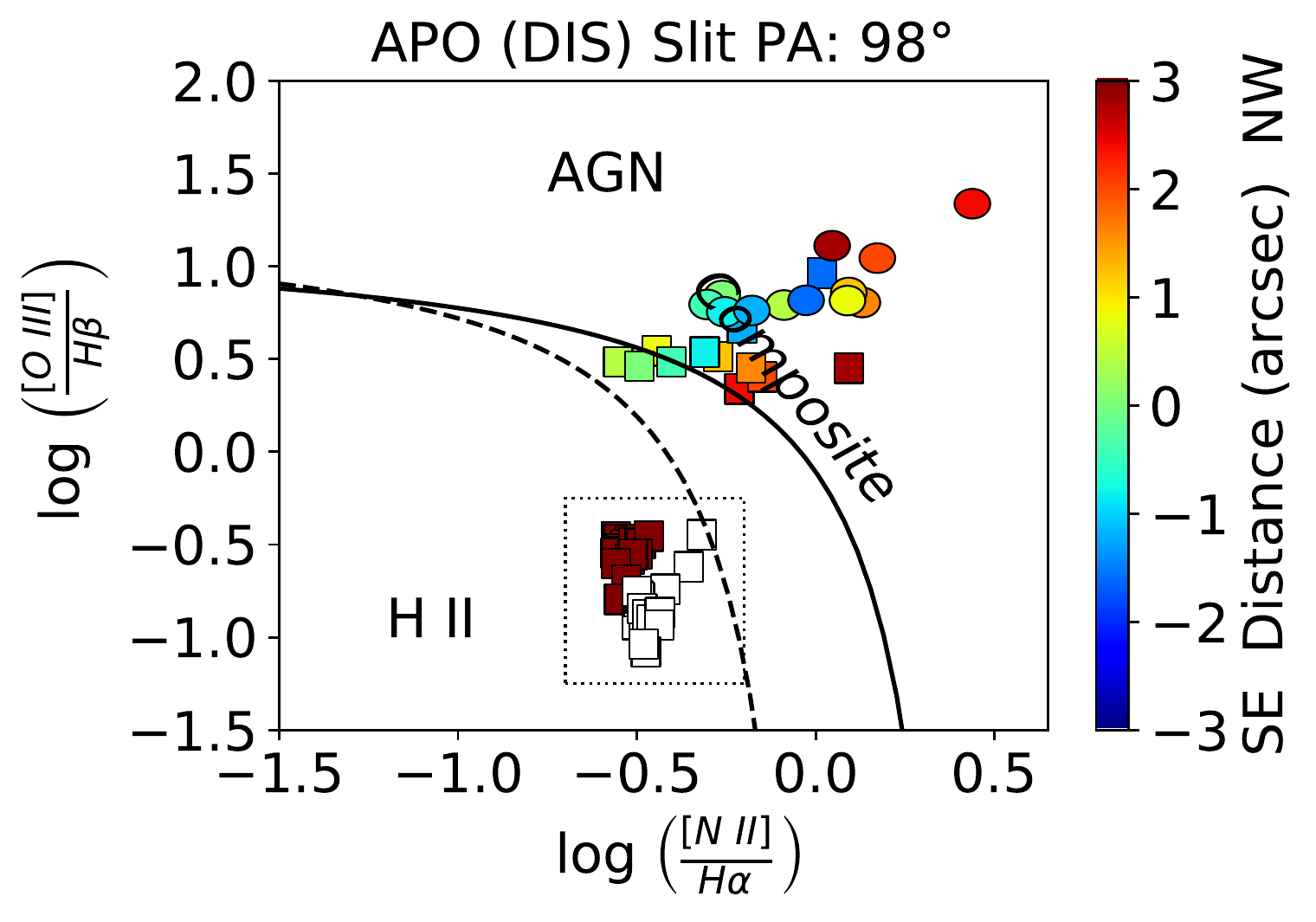}}
\subfigure{
\includegraphics[width=0.33\textwidth]{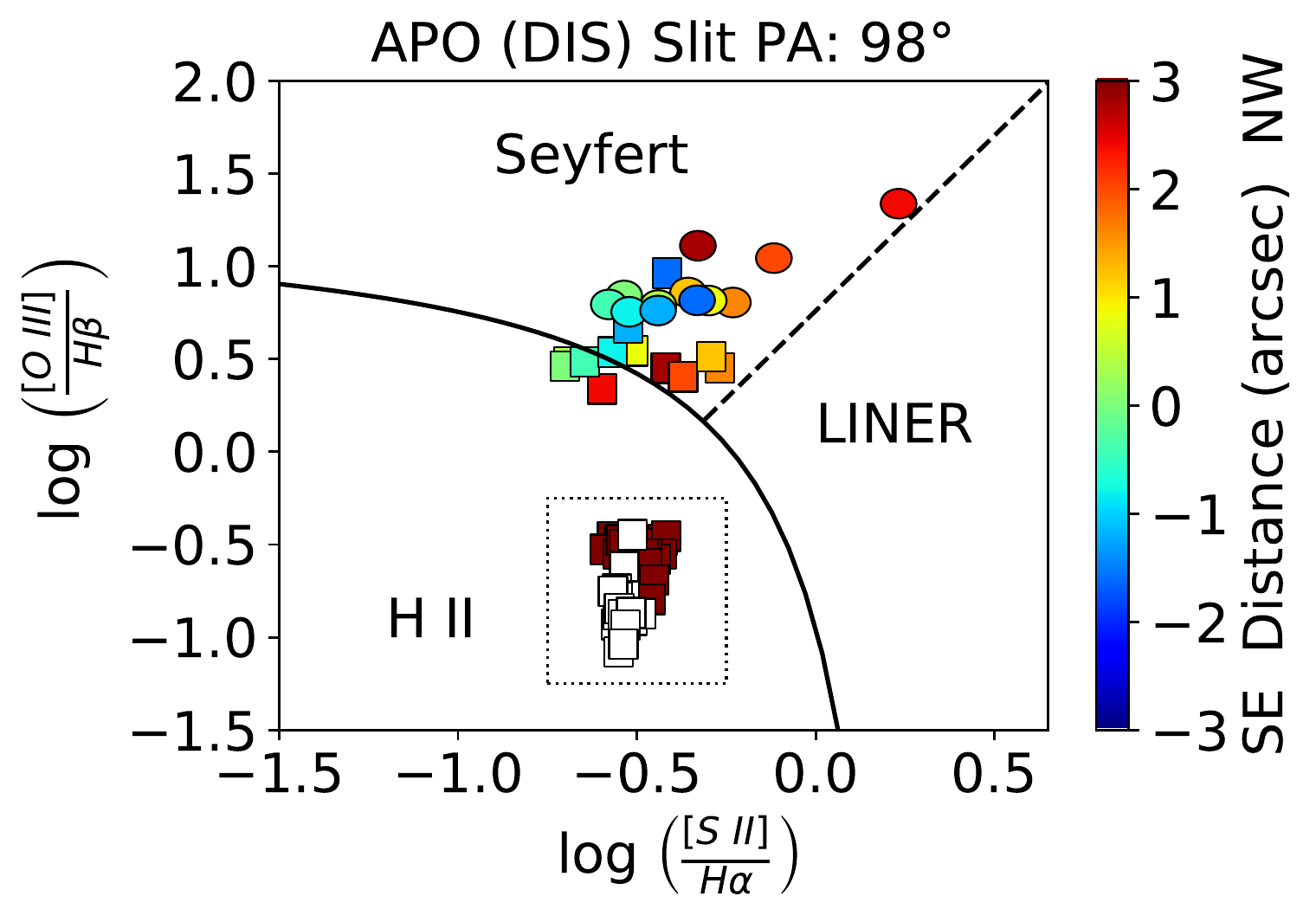}}
\subfigure{
\includegraphics[width=0.31\textwidth]{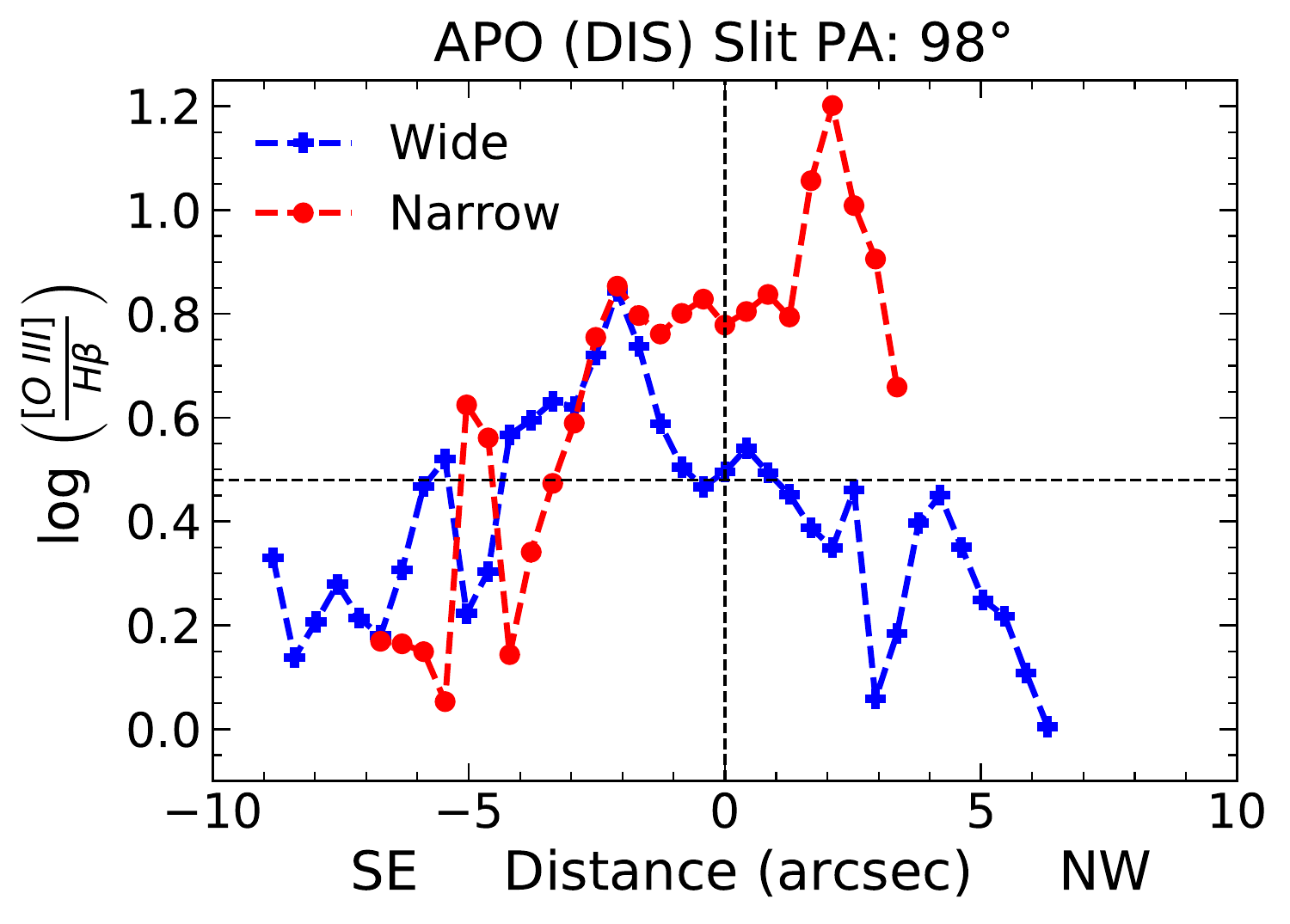}}

\subfigure{
\includegraphics[width=0.33\textwidth]{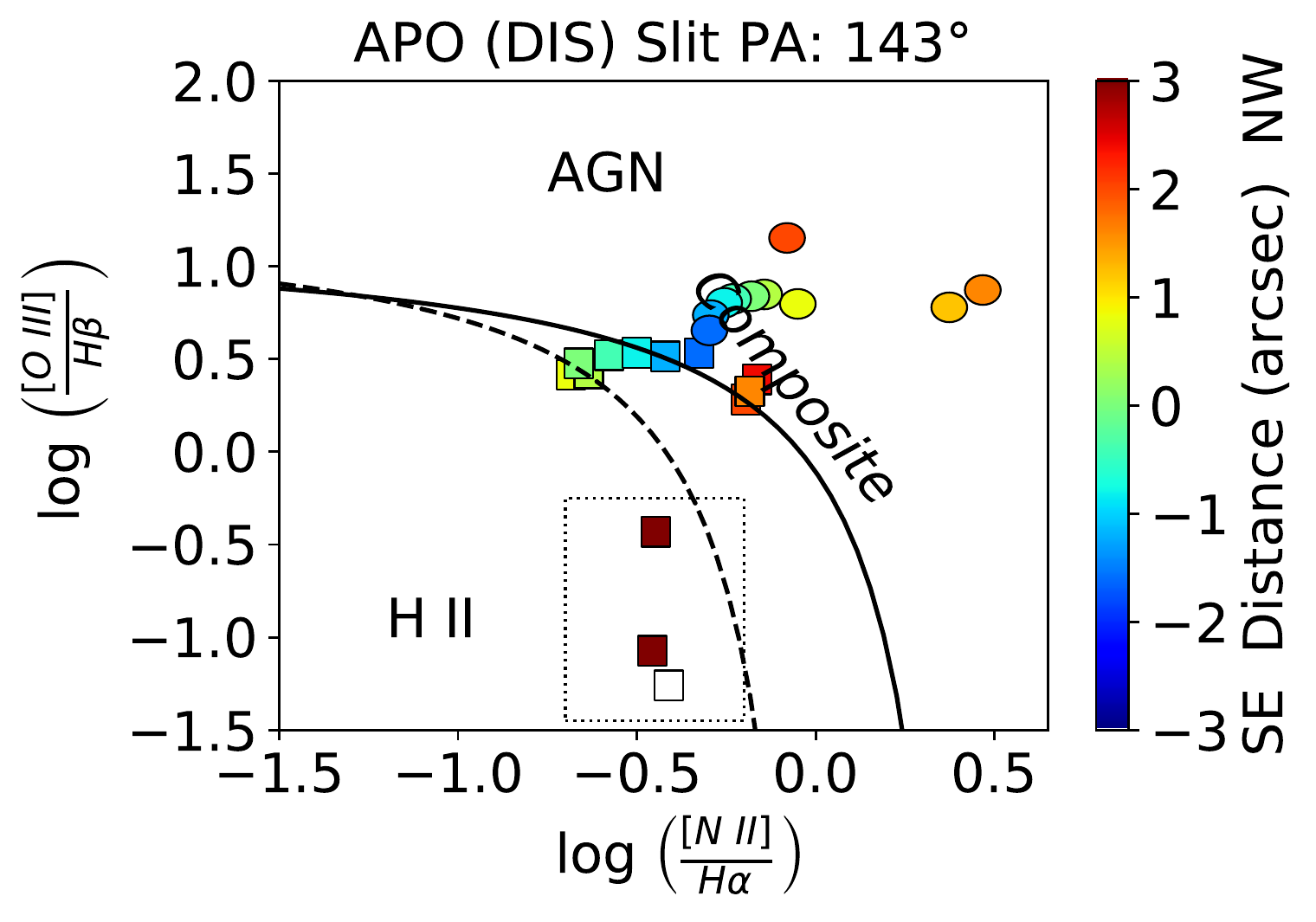}}
\subfigure{
\includegraphics[width=0.33\textwidth]{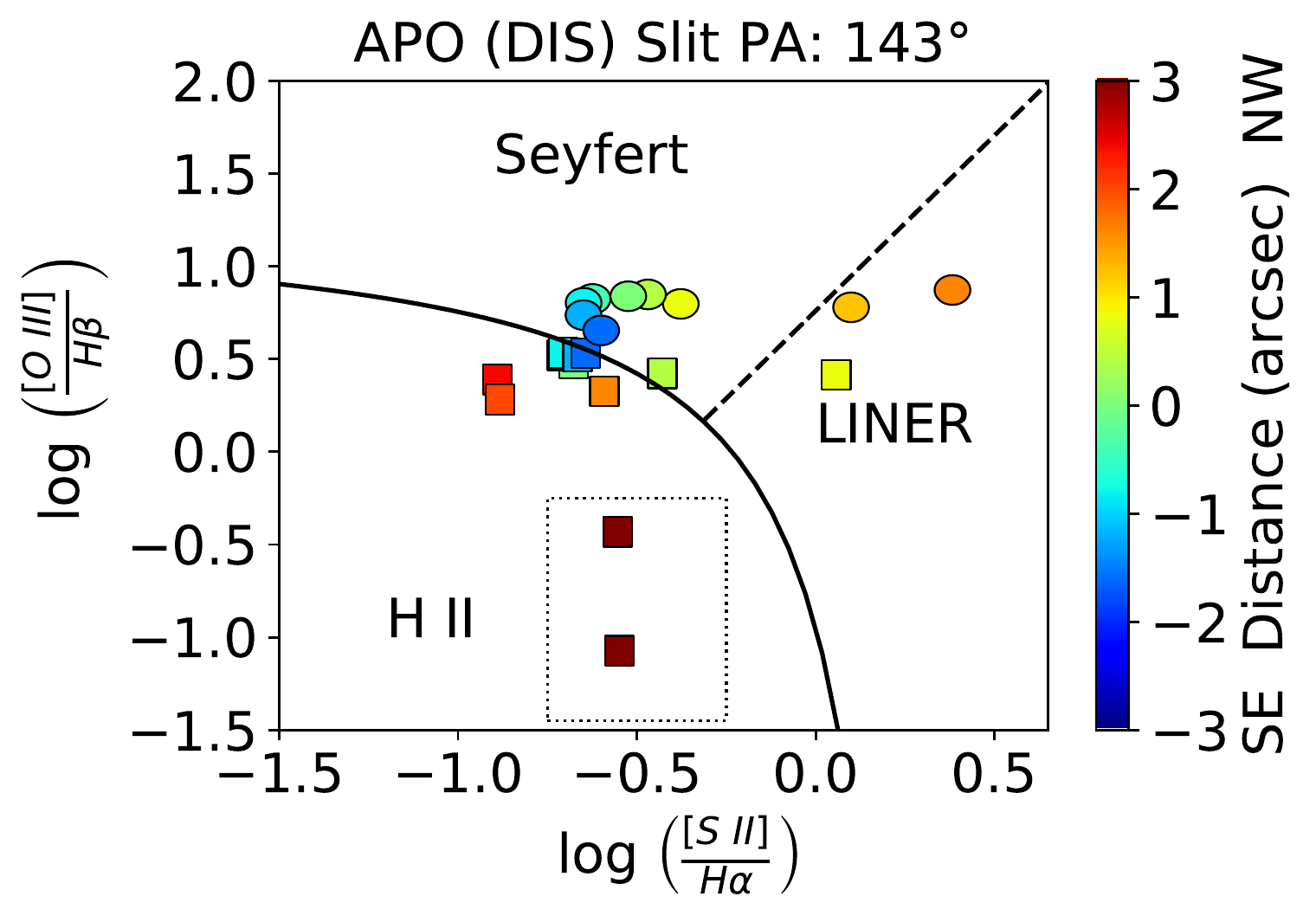}}
\subfigure{
\includegraphics[width=0.31\textwidth]{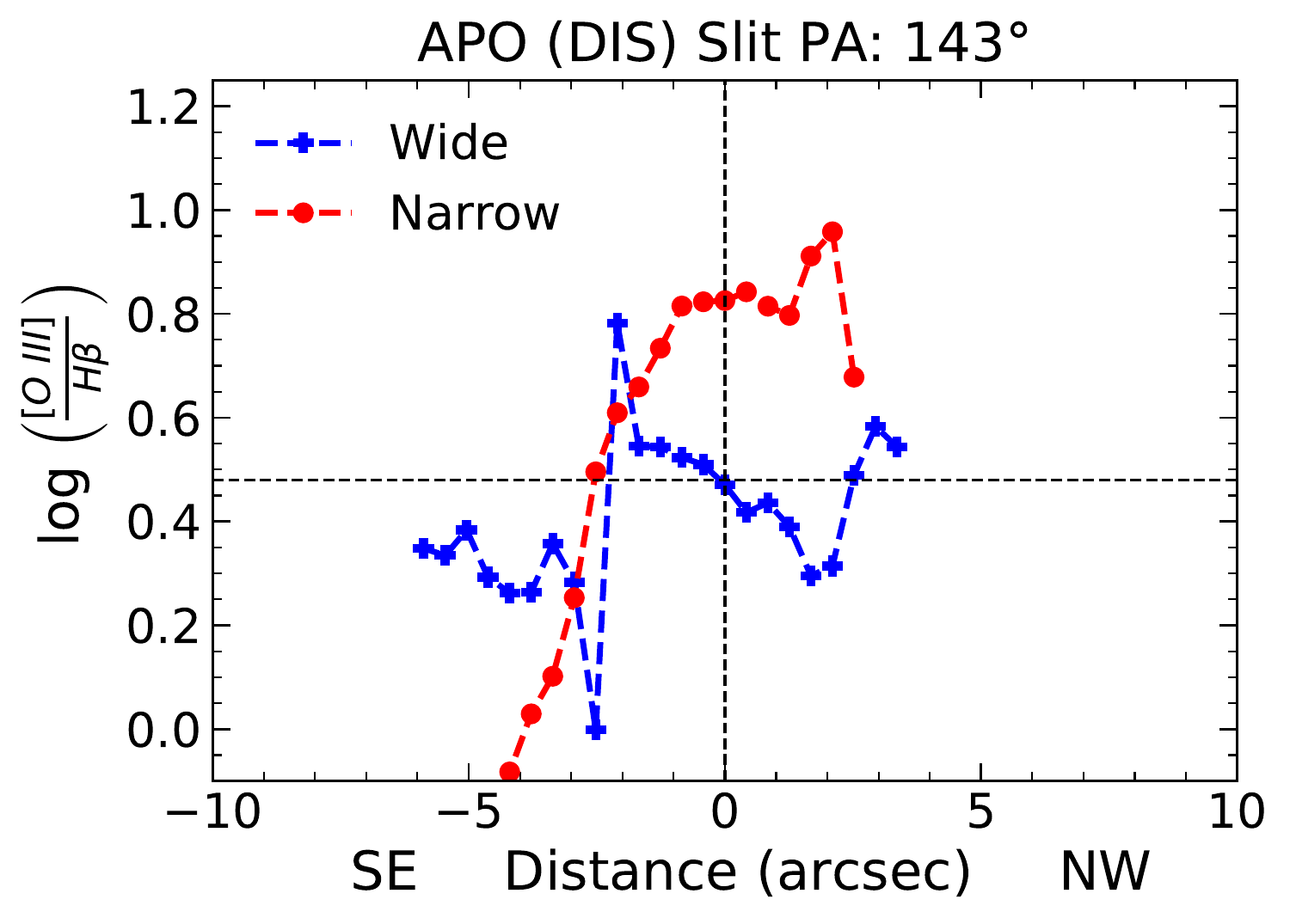}}

\subfigure{
\includegraphics[width=0.33\textwidth]{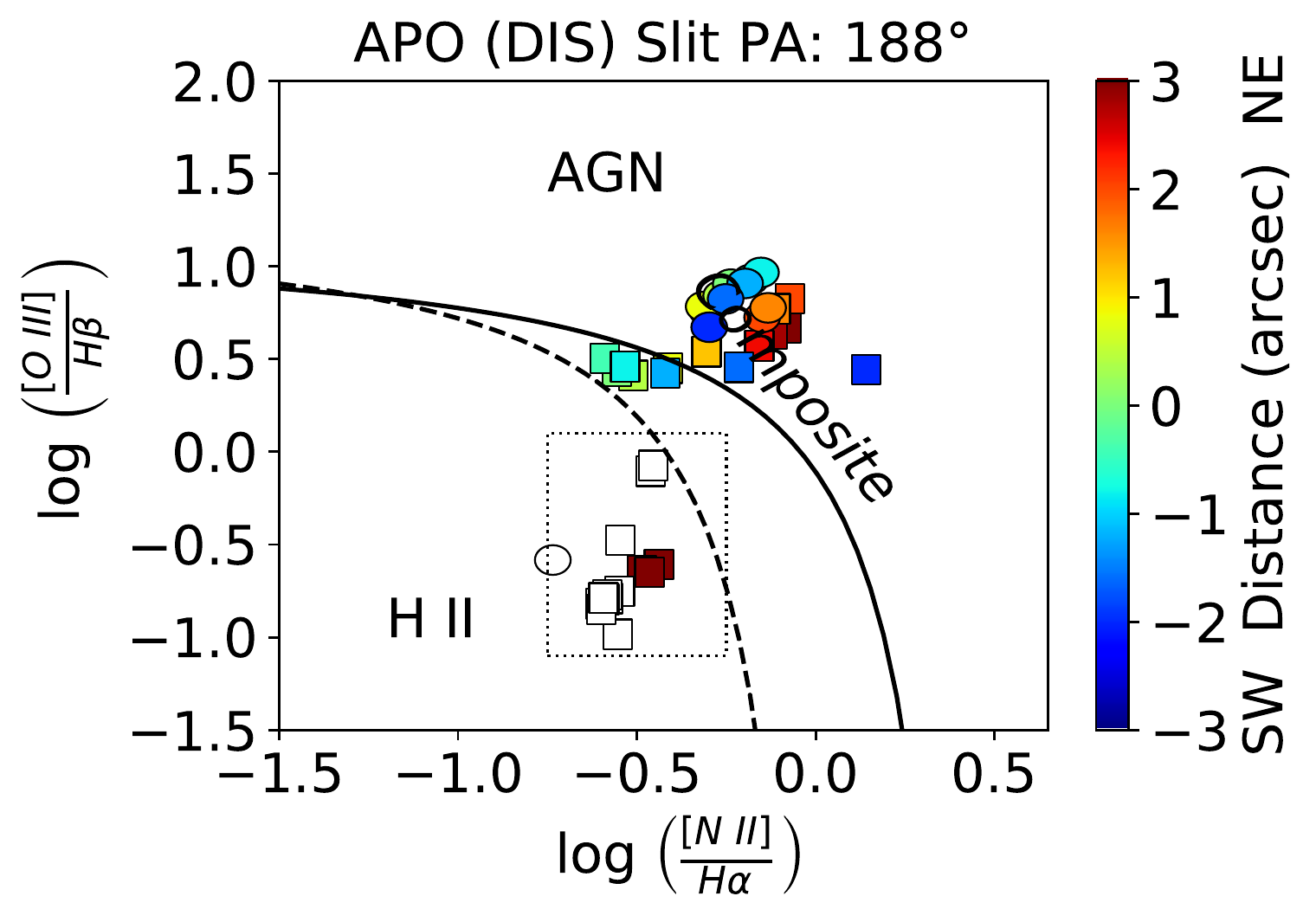}}
\subfigure{
\includegraphics[width=0.33\textwidth]{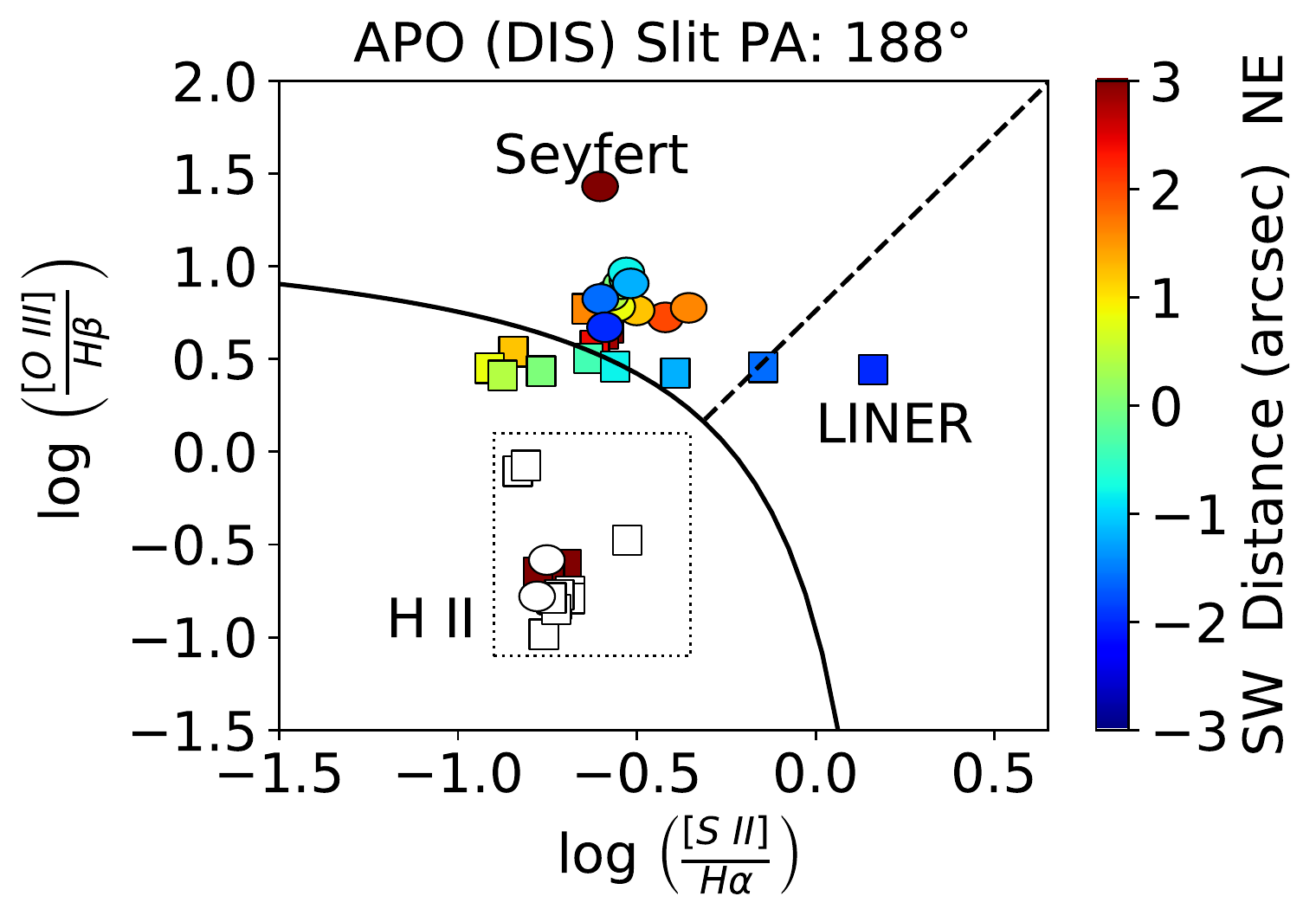}}
\subfigure{
\includegraphics[width=0.31\textwidth]{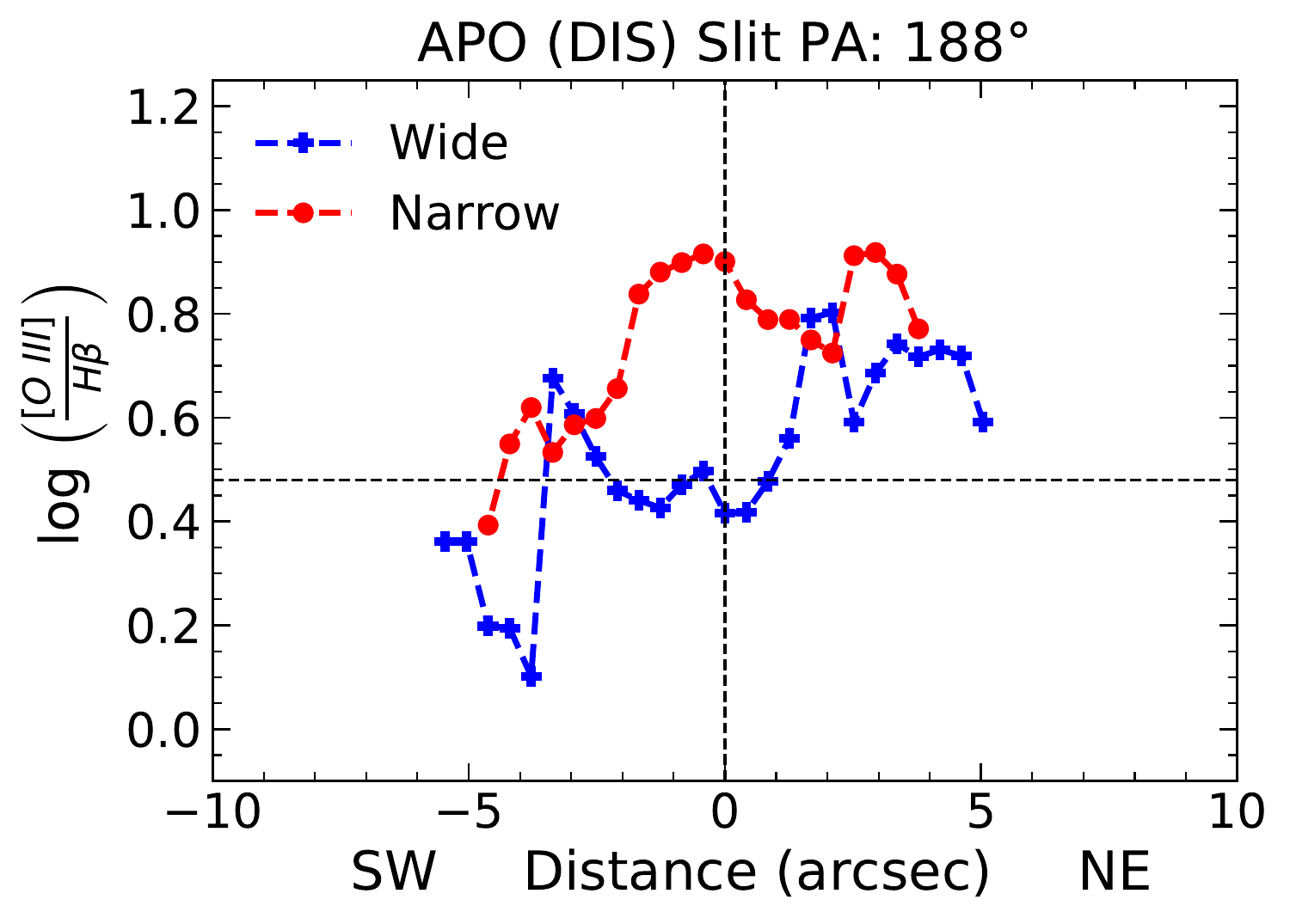}}

\subfigure{
\includegraphics[width=0.33\textwidth]{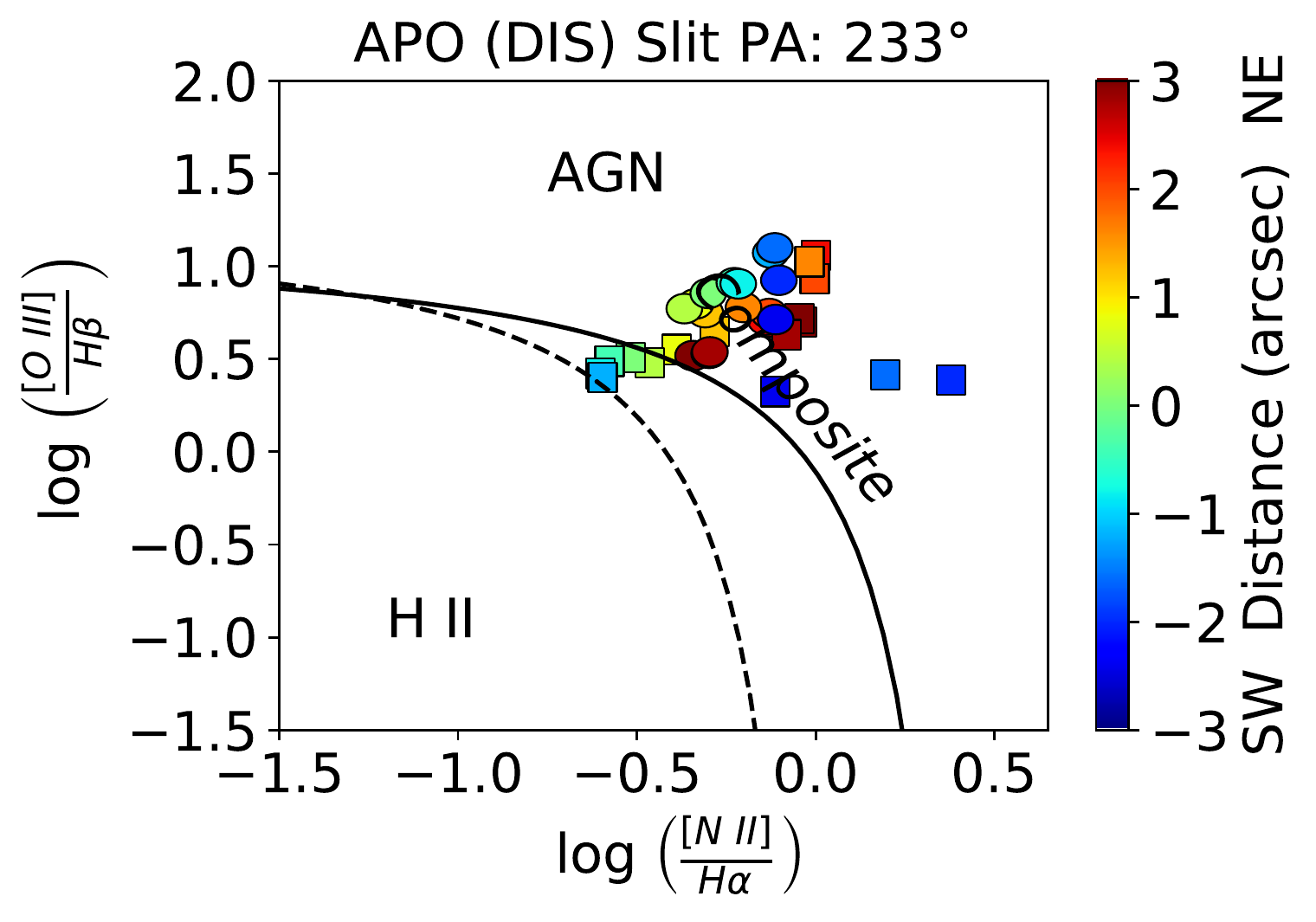}}
\subfigure{
\includegraphics[width=0.33\textwidth]{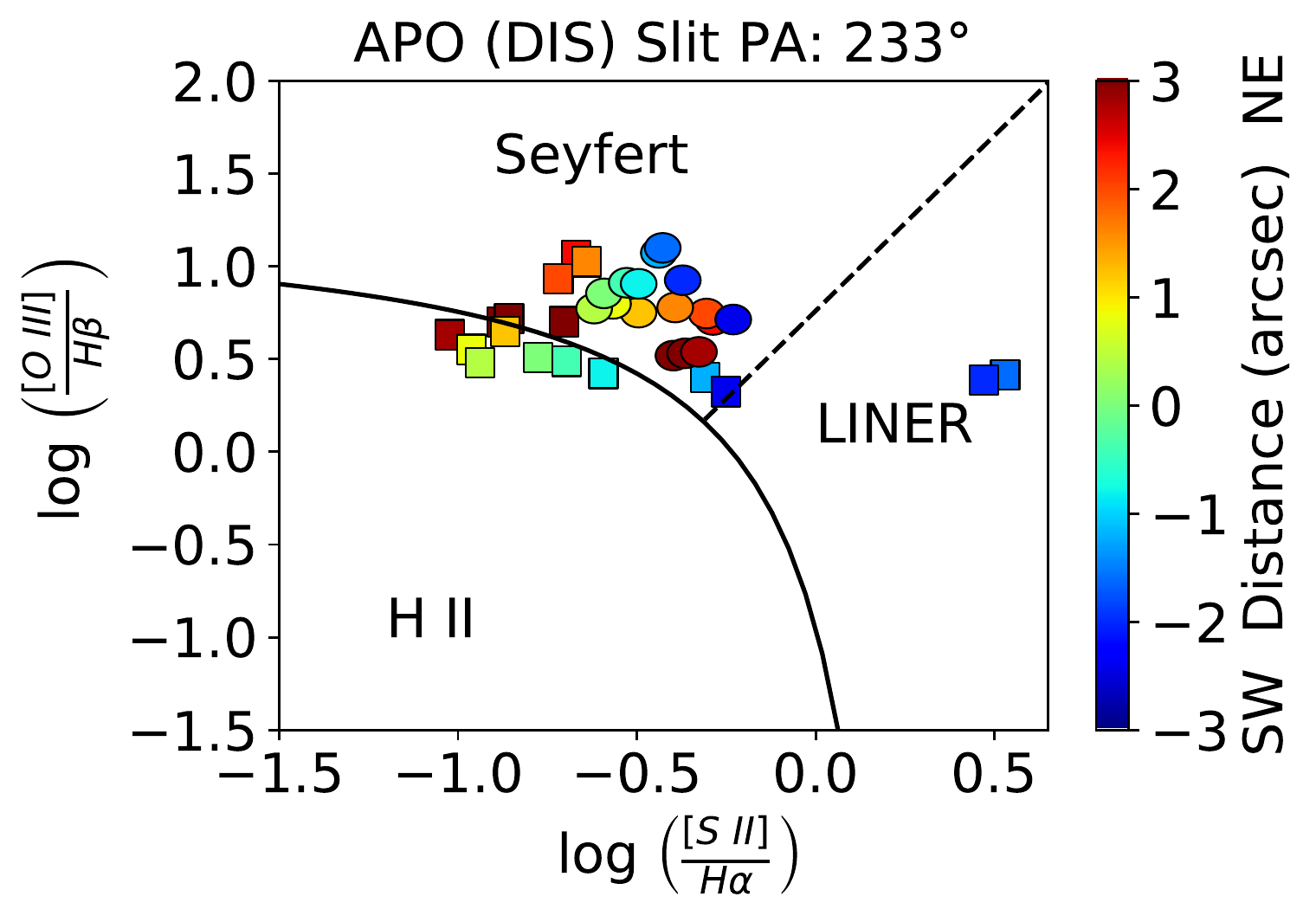}}
\subfigure{
\includegraphics[width=0.31\textwidth]{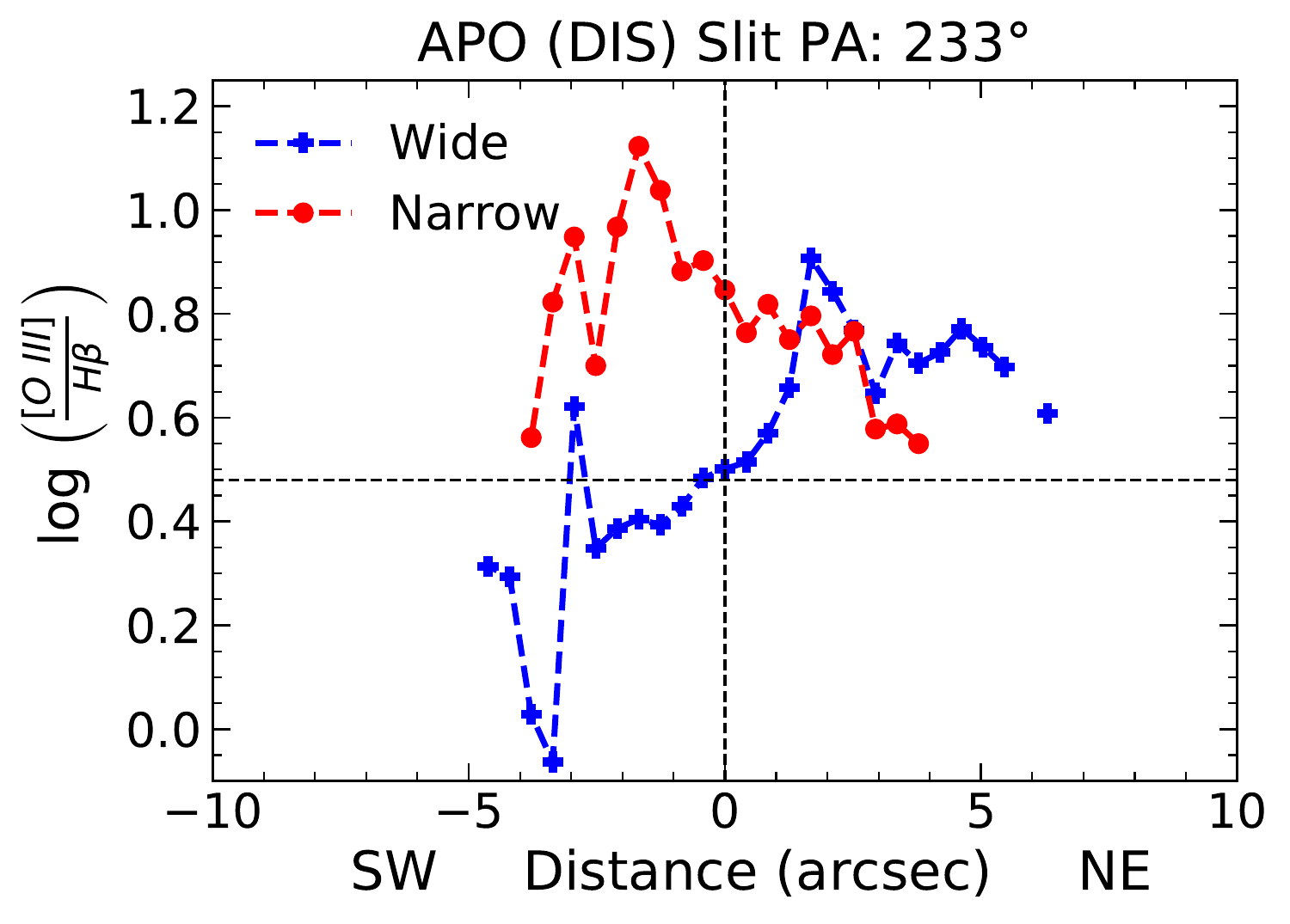}}

\caption{BPT ionization diagnostics using the APO DIS observations (PAs 98\arcdeg, 143\arcdeg, 188\arcdeg, and 233\arcdeg) for the two kinematic components. The first and second columns show the BPT diagrams using [O~III] $\lambda$5007, \hbeta$\lambda$4861, \halpha$\lambda$6563, [N~II] $\lambda$ 6584, and [S II] $\lambda\lambda$6716,6731 emission lines. The squares and circles represent the wide (higher FWHM) and narrow (lower FWHM) velocity components. The color of each point represent the distance from the nucleus as given in the colorbars on the right. H~II regions correspond to star forming gas. The dark red and white squares inside the dashed rectangular boxes show the star forming regions at 20\arcsec\ to 100\arcsec\ to the north (red) and south (white) of the center, which lie in the spiral arms that can be matched with APO DIS slit positions on the color composite ARCTIC image in Figure~\ref{fig:arctic}. The third column shows the radial variation in [O~III]/H$\beta$ ratio for the wide (blue) and narrow (red) components. The horizontal dashed line corresponds to [O~III]/H$\beta$ = 3, which is an approximate discriminator between star-forming (below the line) and AGN (above the line) regions.}
\label{fig:BPTs}
\end{figure*}

To accurately identify the extents of AGN driven outflows and fit the best model to the outflow bicone, it is important to determine whether the observed gas is ionized by the AGN, star formation, or both, over a range of radial distances from the nucleus. To determine the source of ionization, we generated spatially resolved Baldwin-Phillips-Terlevich (BPT) diagrams \citep{Baldwin1981, Veilleux1987} for the APO long slit observations. Two separate diagnostics were made using the demarcation lines defined by \cite{Kewley2001,Kewley2006} and \cite{Kauffmann2003}, as shown in Figure~\ref{fig:BPTs} for [O~III]/\hbeta versus [N~II]/\halpha (first column) and [S~II]/\halpha (middle column) line ratios. The line ratios were computed for each point in the spatially resolved kinematic components as discussed in section \S \ref{subsec:gas_kinematics} (see Figure \ref{fig:stis_apo_vel}).

The BPT plots in Figure~\ref{fig:BPTs} show that for all four APO slits, the majority of the narrow components lie completely in the AGN ionized section, while a horizontal spread is noted for some of the wide kinematic components specifically towards the SW and NW.
This spread could be due to the uncertainties in Gaussian fitting of the blended \halpha + [N II] lines while isolating the broad and intermediate \halpha components near the nucleus.

Possible circumnuclear star formation in NGC~4051 has been claimed previously, using observations of dense molecular gas \citep{Sani2012,Kohno2007} and weak polycyclic aromatic hydrocarbon (PAH) emissions \citep{Rodriguez2003,Sani2010} as close as 1\arcsec\ -- 2\arcsec\ from the nucleus.
This may explain the points near the AGN/H~II composite line in the BPT diagrams.
The dark red and white squares plotted inside the dashed box are points that are associated with the emission lines measured between 20\arcsec to 100\arcsec\ from the center and include parts of the spiral arms, and hence show the large scale star formation in NGC~4051.   

For our BPTs, we excluded the points at 3\arcsec\ -- 10\arcsec\ on both sides of the nucleus, due to the scattering observed at those parts in red channel spectra (see section \S \ref{subsec:apo-obs}), which includes the [N~II], H$\alpha$, and [S~II] lines. For those distances, we plotted the [O~III]/H$\beta$ line ratio as a function of distance (see last column of Figure~\ref{fig:BPTs}) using APO blue channel spectra, which is unaffected by instrument scattering. 
Although, the complimentary [N~II], H$\alpha$\, and [S~II] lines are important to provide the best judgement for ionizing sources, in the absence of these lines, we rely upon the [O~III]/H$\beta$ emission line ratio alone. In Figure~\ref{fig:BPTs}, the last column shows the spatial distribution of [O~III]/H$\beta$ for four slits.

The points with [O~III]/H$\beta$ $\geq$ 3 (shown as a horizontal dashed line at log$(\frac{\othree}{H\beta})=0.48$) show likely AGN ionization \citep{Shuder1981} and points with lower values of [O~III]/H$\beta$ are most likely starburst or composite regions. We notice a gradual increase in this ratio towards the north particularly in slit PAs 188\arcdeg\ and 233\arcdeg\, which contain the majority of the AGN ionization cone as seen in the \othree image in Figure~\ref{fig:F502N}. These regions also coincide with the high velocity clouds as seen in the APO kinematics in Figure~\ref{fig:stis_apo_vel}. Points in the SE and SW below the dashed line, with [O~III]/H$\beta$ $=$ 1 to 3, are consistent with a composite AGN/SF ionization as seen in the left and middle panels of Figure~\ref{fig:BPTs}. 

Thus, Figure~\ref{fig:BPTs} shows that the AGN either dominates or is a strong contributor to the ionization of gas to a projected distance of $\sim$10$''$ ($\sim$800 pc) from the nucleus, which also corresponds to the extent of the outflows. If we use these two criteria (AGN ionization and evidence for outflows) to define the NLR, then we have not clearly detected an ENLR (AGN ionization plus rotation) in NGC~4051.

\subsection{Outflow Model}\label{subsec:outflow}

\begin{figure*}[htp!]
\centering
\subfigure[PA 188\arcdeg]{
\includegraphics[width=0.48\textwidth ]{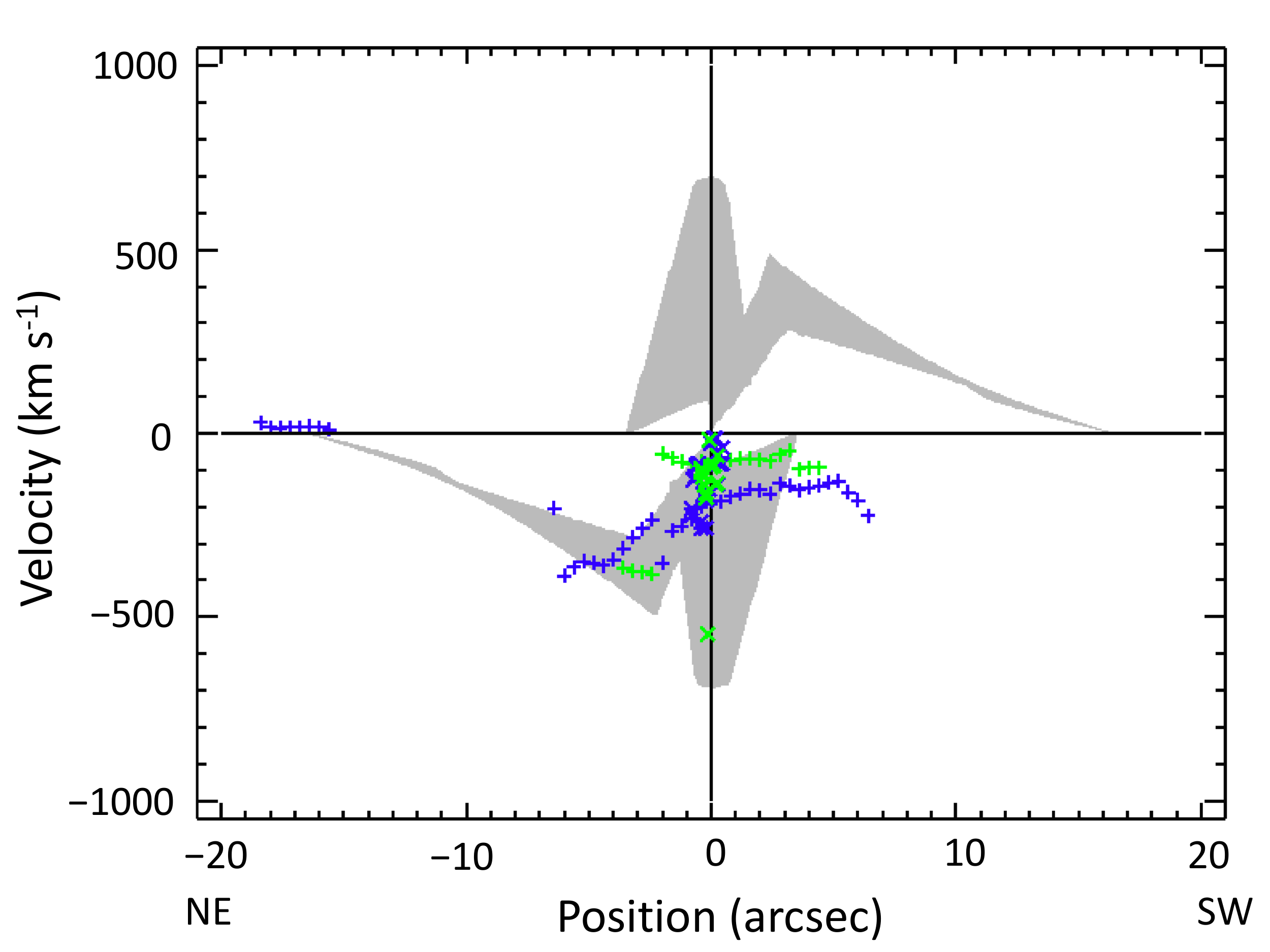}}
\subfigure[PA 233\arcdeg]{
\includegraphics[width=0.48\textwidth ]{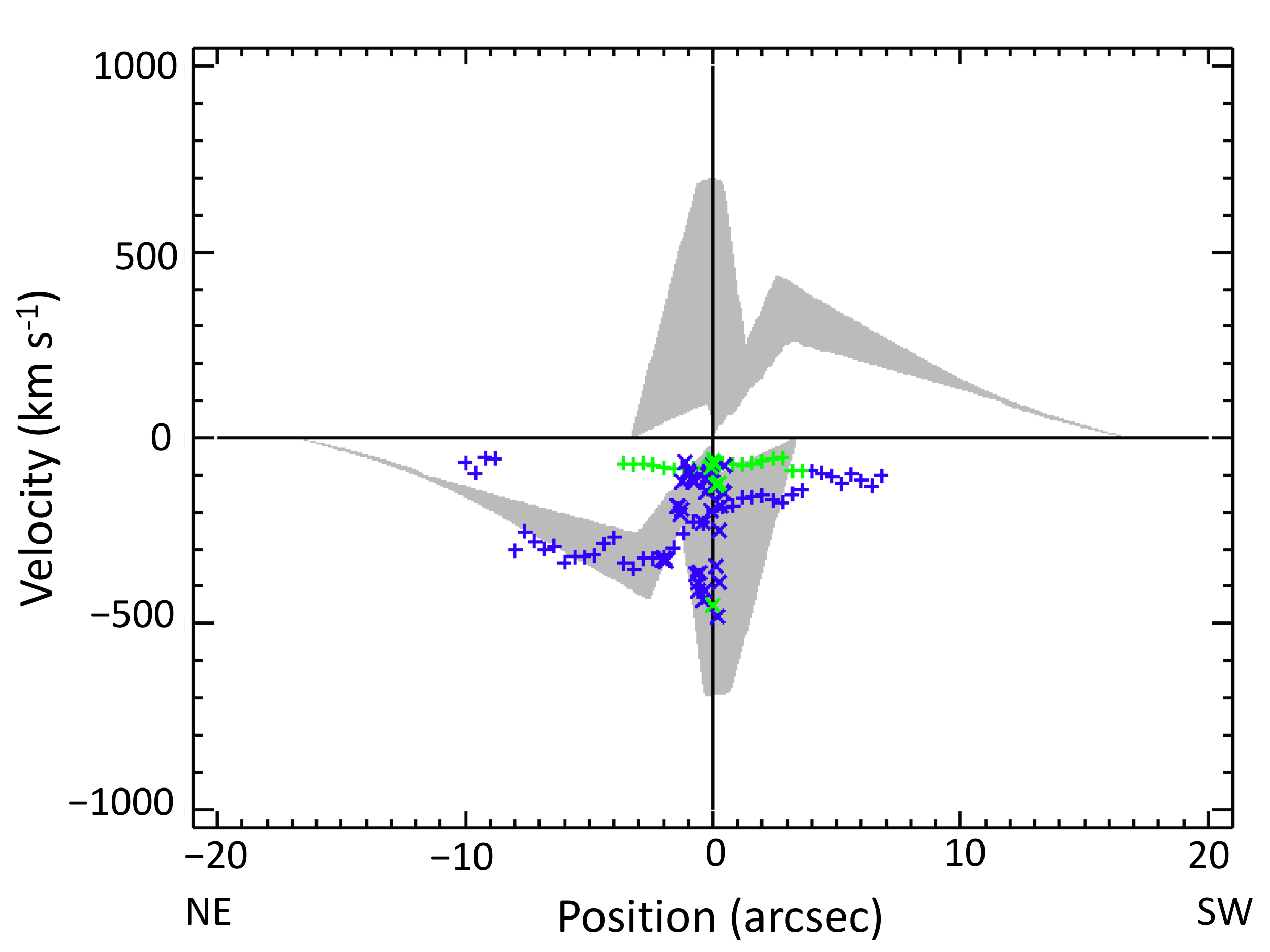}}

\begin{minipage}[t]{0.50\linewidth}\vspace{0pt}
\centering
\subfigure[Bicone model]{\includegraphics[width=\linewidth]{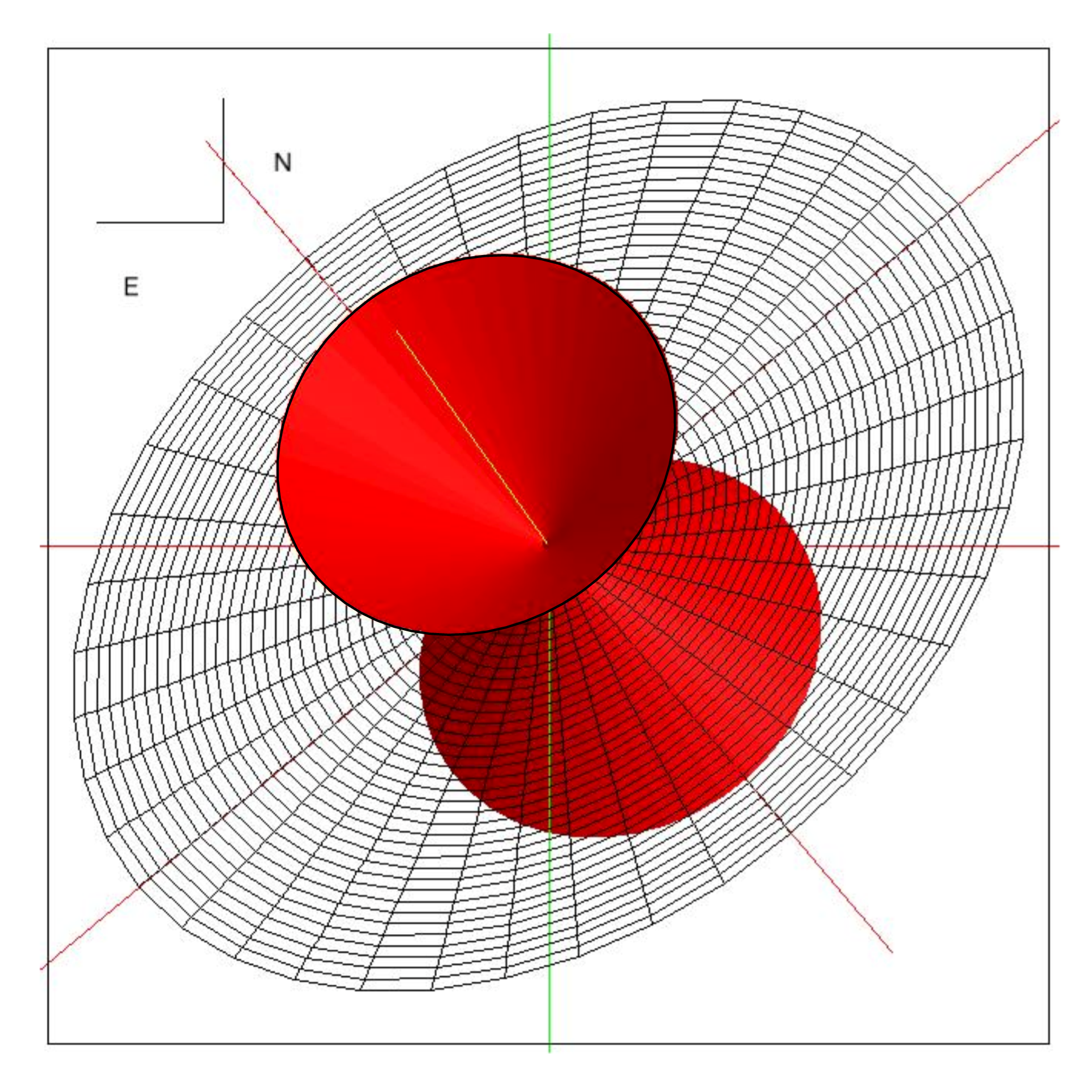}}
\end{minipage}
\hfill
\begin{minipage}[t]{0.47\textwidth}
\tablenum{3}
\captionof{table}{NGC~4051 Outflow Model Parameters \label{tab:bicone}}
\hspace{-5ex}
\resizebox{\textwidth}{!}{%
\renewcommand{\arraystretch}{1.60}
\begin{tabular}{|l|l|l|}
\hline\hline
        Parameters & Fischer+2013 & This Work \\\hline
        PA (galaxy)   & 50\arcdeg\ &   130\arcdeg\ (N-S in SE)  \\
        Inclination (galaxy)    &   5\arcdeg\    &   45\arcdeg\ (SW closer) \\
        PA (bicone) &   80\arcdeg\  &   35\arcdeg \\
        Inclination (bicone)    &   78\arcdeg\    &   60\arcdeg\ (NE closer)  \\
        HOA (max)   &   25\arcdeg\  &   40\arcdeg\  \\
        HOA (min)   &   10\arcdeg\  &   25\arcdeg\  \\
        Turnover  &  30 pc   & 380 pc  \\
        Max Height  &   100 pc  &   1090 pc    \\
        Max Velocity    &   550 km/s    &   700 km/s    \\
        \hline
\end{tabular}%
}
\centering
\setlength{\leftskip}{2ex} \tablecomments{Rows are (1) major axis of galaxy from \cite{Kaneko1997}, (2) inclination of galaxy from \cite{Kaneko1997}, (3) major axis of bicone axis, (4) inclination of bicone axis relative to plane of sky, (5) maximum half opening angle relative to bicone axis, (6) minimum half opening angle, (7) turnover radius, (8) maxiumum height of bicone along its axis, and (9) maximum outflow velocity, located at turnover radius. }
\end{minipage}
\caption{Figure (a) and (b) show the comparison of the kinematic model of biconical outflow in the NLR of NGC~4051 (gray shaded areas) with APO (``+'') and \hst (``$\times$'') radial velocities. Velocities are color coded to represent high (blue) and low (green) FWHM components. (a) Model extraction and velocities at APO DIS slit PA 188\arcdeg\ plus those from STIS slit ``A'' projected onto this PA. (b) Model extraction and velocities at APO DIS slit PA 233\arcdeg\ plus those from STIS slit ``B'' projected onto this PA. Figure (c) represents the geometry of the biconical outflow model for NGC~4051 as seen from Earth, showing the outer surface of the bicone and the disk of the host galaxy, based on the parameters in Table~\ref{tab:bicone}. Given the inner opening angle our view is through the wall of the near, blueshifted cone for this narrow-line Seyfert 1 galaxy and our view of the far, redshifted cone is blocked, presumably by dust in the host galactic plane.}
\label{fig:bicone}
\end{figure*}    

In order to de-project the observed distances and radial velocities of the emission-line knots in the NLR of NGC~4051 and obtain their true distances and velocities, we need a kinematic model of the outflow. We presented a biconical outflow model of NGC~4051 based on the \hst images and spectra alone in \cite{Fischer2013}. The APO spectra, extending the [O~III] kinematics to larger distances, and an enhanced version of the \hst [O~III] image, shown in Figure~\ref{fig:F502N}, allow us to revise the biconical outflow model. The enhanced image indicates a larger opening angle than originally thought and the APO spectra, which cover more of the projected bicone as shown in Figure~\ref{fig:F502N}, indicate that the outflows extend to larger distances than seen in the STIS spectra.

We therefore used the kinematic modeling code originally presented by \cite{Das2005} to match the new observational constraints. Consistent with the models of other AGN outflows in \cite{Fischer2013}, the best model appears to be a symmetric bicone that is evacuated along its core, characterized by a minimum and maximum half-opening angle (HOA), and a simple velocity profile that increases linearly from zero at its center to a maximum velocity at a particular turnover radius, followed by a roughly linear decline to near zero (or systemic) at a maximum height along the bicone axis. The best fit comes from varying these parameters and the inclination of the bicone axis from the plane of the sky, which affects the amplitudes of blueshifts and redshifts on either projected side of the nucleus. Radial velocities extracted from the 3D kinematic model through a pseudo slit at the observed PA result in ``envelopes'' (due to the thickness of the bicone) that are compared to the observed radial velocities at that PA.

The parameters of the biconical outflow models were adjusted to provide a reasonable fit to the \hst [O~III] image (constraining the PA and maximum opening angle of the bicone) and the APO observed radial velocities obtained at the two PAs that lie along the edges of the bicone at PA $=$ 188\degr\ and 233\degr, as well as the projected STIS velocities. These comparisons are shown in Figure~\ref{fig:bicone}. For the model, the STIS data along the near edge of the blueshifted cone helped to constrain the maximum velocity, and the APO data helped to constrain the opening angles, inclination, and extent of the outflow.

The final parameters of the NLR kinematic model and the adopted parameters of the galactic disk are given in Table~\ref{tab:bicone} along with the original values from \citep{Fischer2013}. Although the new PA, inclination, and HOAs differ somewhat from the original values, our view is still along the edge of the bicone. The adopted host galaxy parameters differ as well, but the overall picture of the host galaxy blocking the far, redshifted cone is the same. We have increased the maximum velocity somewhat, but the most significant change is the realization that the outflows extend to much larger distances. 

\subsection{Radiative Driving of the AGN}
\label{subsec:radiative_driving}

From the BPT diagnosis in section \S \ref{subsec:BPT} we can infer that the high-velocity clouds seen in the \textit{HST} STIS and APO DIS observations are primarily AGN ionized and belong to the outflow bicone (Figure~\ref{fig:bicone}). Considering a radiative driving mechanism \citep{Castor1975,Abbott1982,Proga2000} is likely responsible for pushing the gas away from the nucleus, we can compare our observed outflow kinematics to an analytical model proposed by \cite{Das2007}. This model employs forces due to AGN induced radiative acceleration and gravitational deceleration from the enclosed mass in the host galaxy, to obtain the velocity profiles of the outflowing gas as a function of distance.
These methods were employed by \cite{Fischer2017} and \cite{Fischer2019} for the Seyfert 2 galaxies Mrk~573 and 2MASX J0423 to determine the effects of radiative driving in the NLR/ENLR of these galaxies. \cite{Garcia2021} used the same approach to reproduced the ionized gas outflow as seen in MUSE- IFU observations of Seyfert galaxy NGC 5643.

The gravitational deceleration part of this gravity-radiation relationship can be measured from the enclosed mass at a given distance $r$:

\begin{equation}
a(r) = -\frac{\mathrm{G} M(r)}{r^2}
\label{eq:grav_dec}
\end{equation}
\noindent
where we derived the enclosed mass $M(r)$ as a function of distance from the nucleus using surface brightness profiles of the host galaxy and $\mathrm{G}$ is the universal gravitational constant (6.67 $\times$ 10$^{-8}$~cm$^{3}$ g$^{-1}$ s$^{-2}$).

We obtained the parameters to fit a 2-dimensional model to the surface brightness of NGC~4051 from \cite{Bentz2009}. The image decomposition was done by \cite{Bentz2009} using GALFIT \citep{Peng2002} to fit variations of Sérsic \citep{Sersic1968} profiles for different brightness components of the galaxy. The radial distribution of surface brightness of a component using a Sérsic profile can be given as \citep{Peng2010}:

\begin{equation}
\Sigma(r) = \Sigma_{e} \exp\left[-\kappa\left(\left(\frac{r}{r_{e}}\right)^{1/n} -1\right)\right]
\label{eq:sersic}
\end{equation}
\noindent
where $r_{e}$ is the effective radius for the given component, $\Sigma_{e}$ is the surface brightness of that component at $r_{e}$, $n$ is the Sérsic index that defines a power law and the value of constant $\kappa$ is calculated such that half of the total flux lies within $r_{e}$. We retrieved the four Sérsic components and the associated parameters from \cite{Bentz2009}. The first and second components are identified as two inner bulges or nuclear stellar clusters with $n$ = 1.07 and 0.31 and $r_{e}$ = 0.03 kpc and 0.07 kpc respectively. The third and fourth components are given as a bulge ($n$ = 1.80, $r_{e}$ = 0.86 kpc) and disk ($n$ = 1.00, re = 4.24 kpc).

We calculated the mass distribution of each Sérsic component with the obtained parameters by using Equation 4-6 and A2 in \cite{Terzi2005} assuming a spherical geometry. See \cite{Fischer2017,Fischer2019} for more details. We calculated a mass to light ratio ($\mathrm{M}/L$) of 2.86 for the bulge components and 0.63 for the disk using the $V - H$ color provided in \cite{Bentz2018} and the relationship given in Table~1 of \cite{Bell2001}, for the same $V$ band image as in \cite{Bentz2009} that is adopted for this work.

\begin{figure*}[hbt!]
\centering
\subfigure[]{
\includegraphics[width=0.48\textwidth]{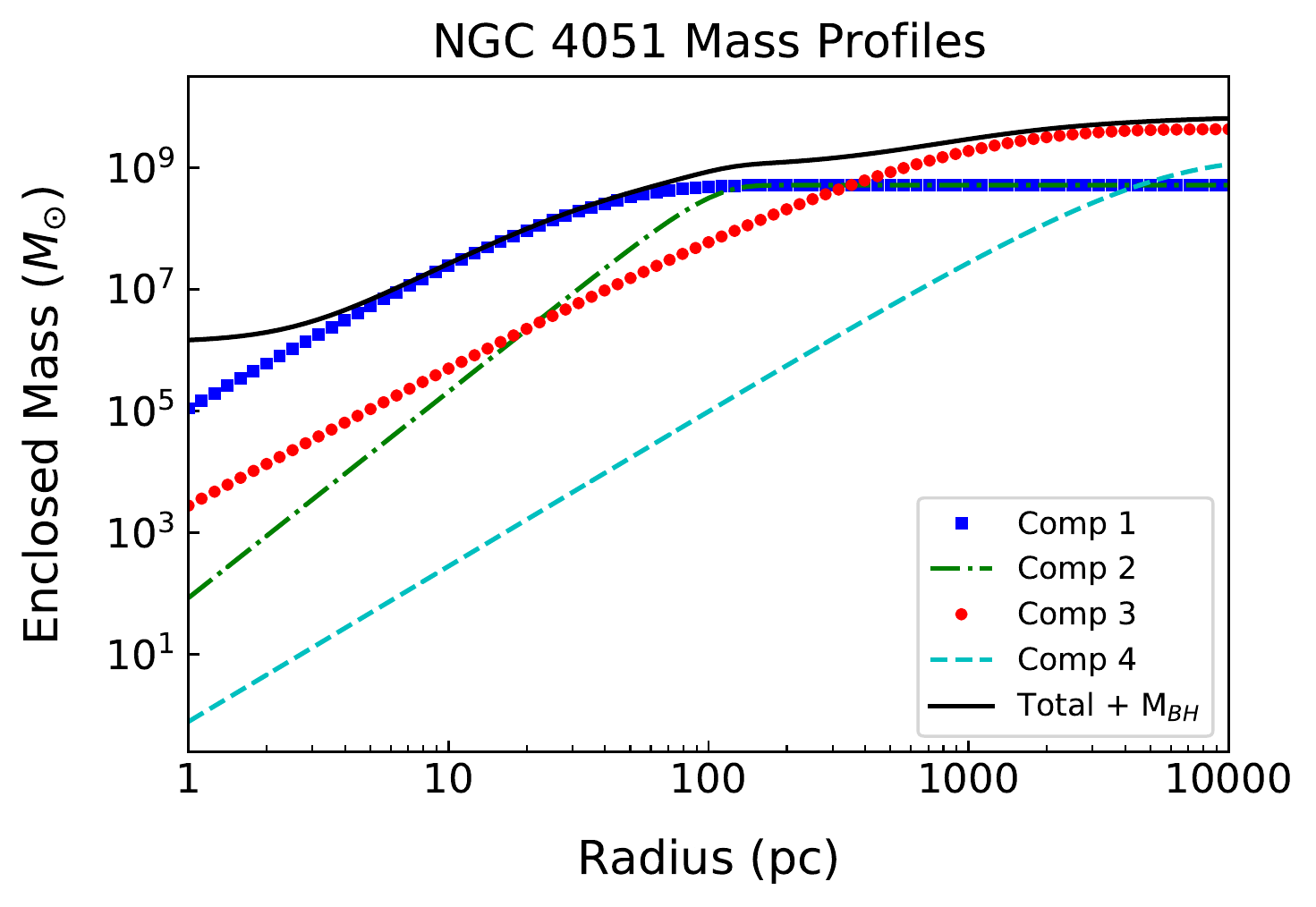}}\hspace{2ex}
\subfigure[]{
\includegraphics[width=0.48\textwidth]{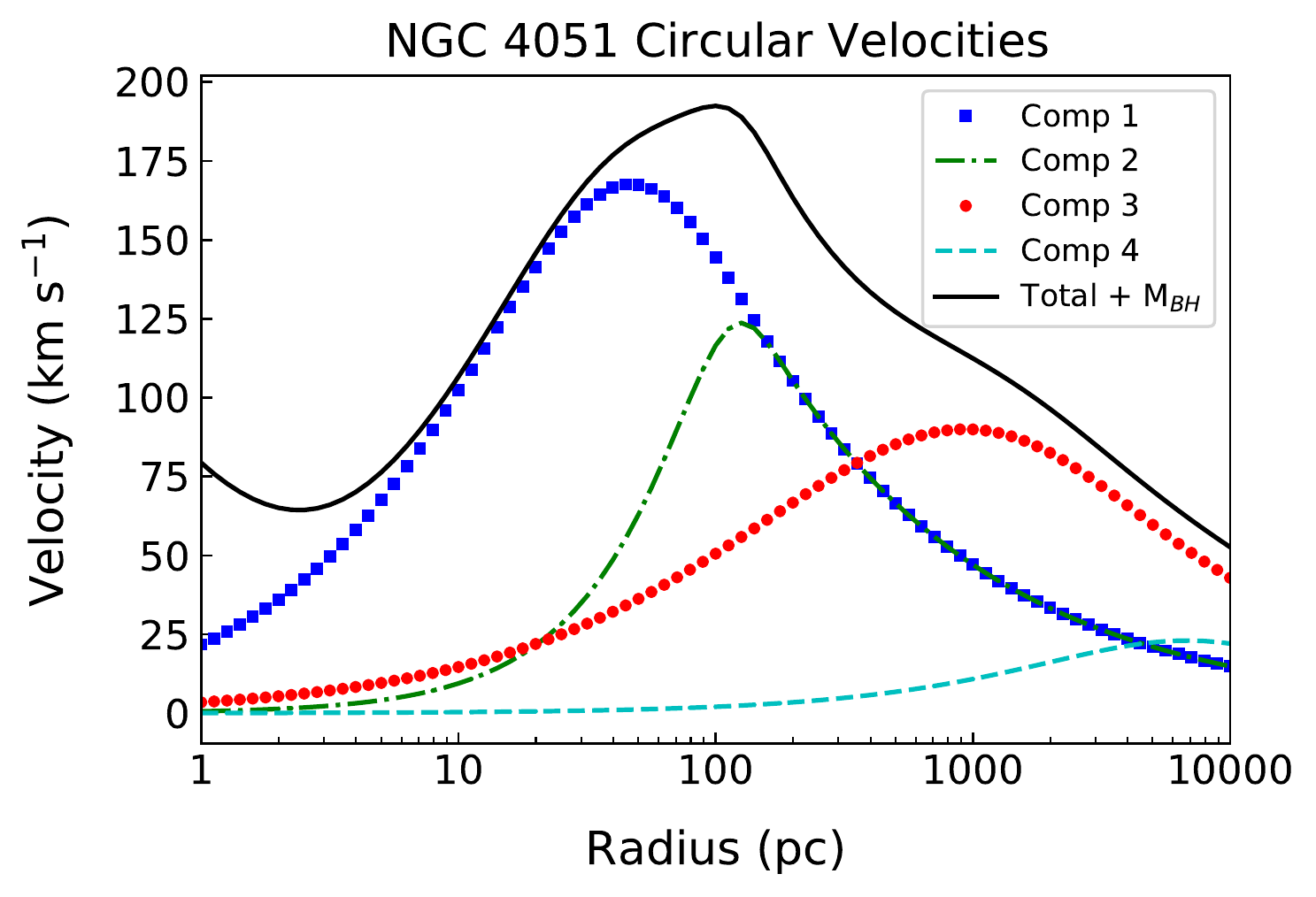}}
\subfigure[]{
\includegraphics[width=0.48\textwidth]{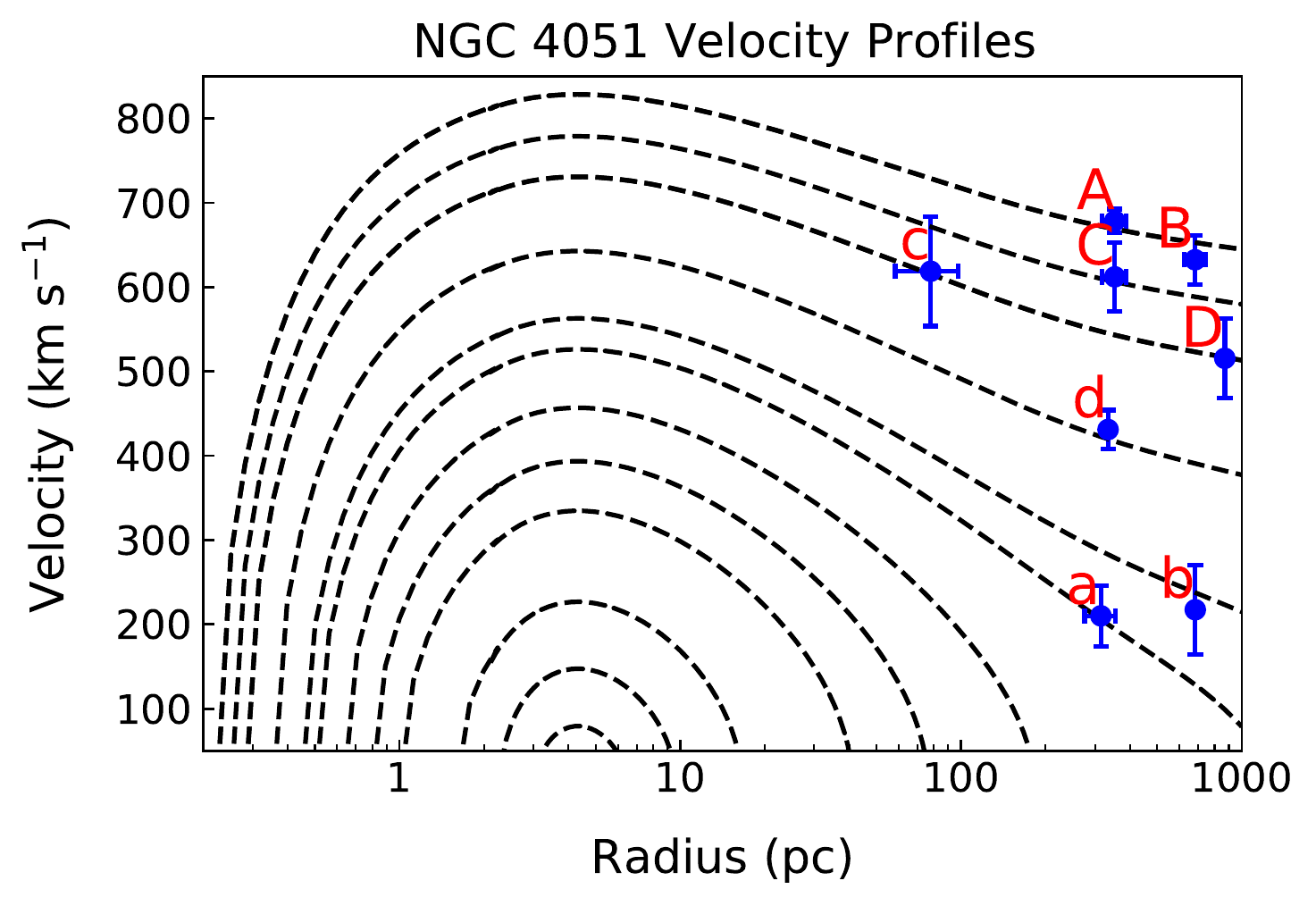}}\hspace{2ex}
\subfigure[]{
\includegraphics[width=0.48\textwidth]{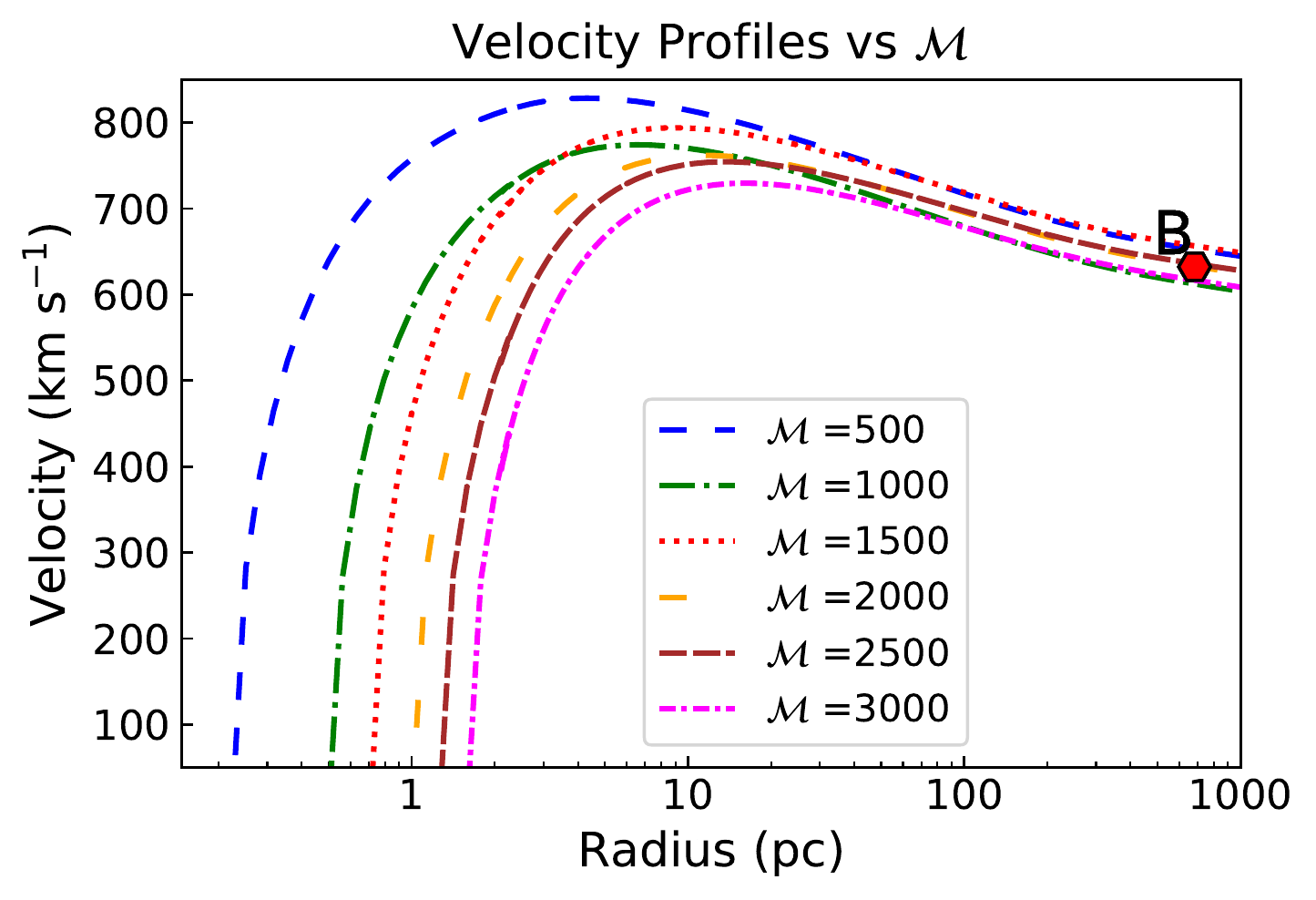}}
\caption{(a) Mass Profiles of NGC~4051, calculated using Sérsic components defined in \cite{Bentz2009}. (b) The rotation velocities are for each mass component separately and for the total baryonic mass inside 10 kpc. (c) Velocity profiles generated with the dynamical model given in Equation \ref{eq:vel_pro}. The capital letters are associated with the APO outflow knots and lowercase letters are assigned to the knots seen in \textit{HST} STIS. 
The velocities and positions are determined using the average of selected kinematic points in Figure~\ref{fig:stis_apo_vel} that are assigned to each knot. The error bars are calculated from the minimum and maximum values of distance and velocity associated with the points in a given knot. The true velocities and distance are calculated using de-projection parameter from biconical model geometry given in section \S \ref{subsec:outflow}. (d) An example of change in launch distance of the observed outflow cloud (B) for different values of force multiplier $\mathcal{M}$.}
\label{fig:rad_drive}
\end{figure*}

Figure~\ref{fig:rad_drive}a shows the radial distributions of the enclosed mass of the four components that were identified by \cite{Bentz2009} and the total mass $M(r)$, which also includes the contribution from the SMBH as a point mass ($M_{\mathrm{BH}}$ = 10$^{6.13} M_{\odot}$) at $r$ $<$ 0.01 pc. From the mass profiles, it can be noted that the SMBH dominates the inner 1 pc and then the first and second components (nuclear stellar clusters) build up the total enclosed mass to $\sim$ 100 pc before the mass from the outer bulge and the disk accumulates. The total enclosed mass at 1 kpc is close to the total Baryonic mass ($=$ 10$^{9.7} M_{\odot}$) at $\sim$ 10 kpc which accounts the total gas (based on HI cm observation) and stellar mass as reported by \cite{Robinson2021}. We also determined the rotation curve for individual components using the corresponding mass profiles as well as for the total mass using the equation:

\begin{equation}
V(r) = \sqrt{\mathrm{G} \frac{M(r)}{r}}.
\label{eq:rot}
\end{equation}
\noindent

Figure~\ref{fig:rad_drive}b shows the rotational circular velocities of each component separately as well as for the sum of components (including BH mass).
The mass from a dark matter component is not included in these calculations, and is not expected to be a major contributor at distances $<$ 1 kpc. The total circular velocity at 1 kpc is $\sim$115 km s$^{-1}$, close to that rotational velocity \halpha ionized gas $\sim$ 130 \kms (\citealp{Richards2016}, also see Figure \ref{fig:rotation}), providing confidence in our enclosed mass model and mass to light ratios.

The radiative acceleration \citep{Arav1994,Chelouche2001,Crenshaw2003} due to a central AGN with a bolometric luminosity $L$ on a mass at a distance $r$ is:

\begin{equation}
a(r) = \frac{L \sigma_{T} \mathcal{M}}{4 \pi r^{2} \mathrm{c} \mu m_{\mathrm{p}}}
\label{eq:rad_acc}
\end{equation}
\noindent
where $\sigma_{T}$ is the Thompson scattering cross section for the electron, $\mathrm{c}$ is the speed of light, $\mu$ is the mean atomic weight of a proton ($\sim$ 1.4 for typical solar abundances) and $m_{\mathrm{p}}$ is the mass of the proton. $\mathcal{M}$ is the force multiplier, which incorporates the bound-bound, bound-free and continuum opacity in addition to Thomson scattering, and depends on the ionization parameter ($U$) for a given spectral energy distribution. Hence the force multiplier is the ratio of total acceleration $a(r)$ to the acceleration due to Thomson scattering \citep{Arav1994,Chelouche2001}.

Considering spherical symmetry and assuming that no other forces (hydrodynamic, thermal, drag, and magnetic) are in play, the total acceleration on a point mass at a distance $r$ can be calculated as the sum of radiative acceleration (Equation \ref{eq:rad_acc}) and gravitational deceleration (Equation \ref{eq:grav_dec}). By substituting for the constants and solving for differential velocity, a model for the radial velocity of an outflowing cloud in the NLR (\citealp{Das2007}, Equations 19-22) is given by the following expression:
\begin{equation}
v(r) = \sqrt{ \int_{r_{1}}^{r} \Big[ 4885 \frac{L_{44} \mathcal{M}}{r^{2}} - 8.6 \times 10^{-3}\frac{M(r)}{r^{2}}} \Big] dr
\label{eq:vel_pro}
\end{equation}
where the $v(r)$ is the outflow velocity (in km s$^{-1}$) of a cloud at a distance $r$ (in pc), $L_{44}$ is the bolometric luminosity in $10^{44}$ erg s$^{-1}$, and $r_{1}$ is the launch distance of the cloud from the central SMBH. $M(r)$ is in the units of solar masses ($M_{\odot}$). The AGN of NGC~4051 is optically variable, which leads to a range of luminosities that have been provided by various observations in the past. Therefore, for our analysis, we determined the bolometric luminosity $L_{bol}$ of the AGN using the \othree luminosity $L_{\mathrm{[O~III]}}$ of the NLR and the relationship provided in \cite{Heckman2004} as $L_{bol} = 3500 \times L_{\mathrm{[O~III]}}$. We calculated $L_{\mathrm{[O~III]}}$ by measuring the total flux (not corrected for extinction) in the continuum-subtracted F502N \othree image within the NLR. Using a distance of 16.6 Mpc \citep{Yuan2020}, we calculate a value of $L_{bol}$ = 42.9$\pm$0.3 erg s$^{-1}$, which is in excellent agreement with the value of 42.95 erg s$^{-1}$ given in \cite{Bentz2013}.

We used Equation~\ref{eq:vel_pro} to numerically solve for the launch distances of the some of the observed outflowing emission-line knots as seen in our \textit{HST} STIS and APO DIS kinematics (all of the other values in this equation are constrained). Figure~\ref{fig:rad_drive}c shows the velocity profiles of selected knots as modeled using Equation~\ref{eq:vel_pro}. We chose knots from the observed kinematics as shown in Figure~\ref{fig:stis_apo_vel} for STIS parallel slits (A and B) and APO slits at PA 188\arcdeg\ and PA 233\arcdeg, which are within the blueshifted outflow cone.
The positions and velocities of outflowing knots were chosen such that a distinct group of kinematic points in Figure~8 with similar FWHM and/or flux distribution are defined as a knot. The selected knots are labeled in Figure~\ref{fig:stis_apo_vel}. The observations, the assigned name of the knot, their observed distances, and their velocities are given in Table~3.


We determined the true distances and velocities of these clouds using the parameters for the outflow bicone's geometry; these are given in columns (5) and (6) of the table. Specifically, we determined the distance and velocity de-projection factors by assuming that the STIS data points are from the near side of the outflow cone and the APO data are from the far side, as shown in Figure~\ref{fig:bicone}. We then calculated the approximate distances $r_{1}$ from which the clouds were launched/originated, using the dynamical model from Equation~\ref{eq:vel_pro}.

We adopted a conservative force multiplier $\mathcal{M}$ $=$ 500 for the model velocity profiles shown in \ref{fig:rad_drive}c, based on the discussion in Trindade Falcao (in preparation), which considers the effects of decreasing $\mathcal{M}$ and ionizing flux as you move deeper in the cloud, as well as the increased radiation pressure from dust in a typical [O~III]-emitting NLR cloud.
For comparison, these authors use a baseline value of $\mathcal{M}$ $=$ 1040, and \citet{Fischer2017} use a value of $\mathcal{M}$ $=$ 3300 from the ionized face of a NLR cloud. Figure \ref{fig:rad_drive}d shows the change in the velocity curve and launch radius $r_{1}$ with different values of $\mathcal{M}$. For a given position and velocity of observed cloud B$^\prime$, the launch distance increases with increasing $\mathcal{M}$ almost linearly.

Figures~\ref{fig:rad_drive}c and \ref{fig:rad_drive}d show that the relatively low-luminosity AGN in NGC~4051 is capable of driving the ionized gas in the NLR up to a distance of $\sim$1 kpc with de-projected velocities up to 700 km s$^{-1}$. Interestingly, the emission-line clouds originate from near the nucleus at a distance of $<$ 1 pc. It can be noted from ~\ref{fig:rad_drive}c that the outflowing gas quickly achieves high velocities within $\sim$10 pc of the center due to the low enclosed mass before being slowed down by the gravitational potential of the host galaxy. A higher value of $\mathcal{M}$ may increase the launch distance up to $\sim$3 pc.
However, based on the condition that radiative acceleration and gravitational deceleration are responsible for the observed velocities, Figure~\ref{fig:rad_drive}c shows that no outflowing clouds can be produced beyond a distance of $\sim$3 pc in this low-luminosity AGN. Hence, unlike Mrk 573 (\cite{Fischer2019}, we see no evidence of in-situ acceleration at distances of tens to hundreds of parsecs from the host galaxy disk.

\setlength{\tabcolsep}{0.1in}
\renewcommand{\arraystretch}{1.0}
\tabletypesize{\small}
\begin{deluxetable*}{cccccccccc}[htb!]
\tablenum{4}
\tablecolumns{8}
\tablecaption{Parameters for the Radiative Driving Models\label{tab:rad_drive}}
\tablehead{
\colhead{Observation} & \colhead{Knot} & \colhead{Component} & \colhead{Observed} & \colhead{Observed} & \colhead{True} & \colhead{True} & \colhead{Model} & \colhead{Launch} & \colhead{Travel} \vspace{-2ex}\\
\colhead{} & \colhead{Name} & \colhead{} & \colhead{Distance} & \colhead{Velocity} & \colhead{Distance} & \colhead{Velocity} & \colhead{Velocity} & \colhead{Distance} & \colhead{Time} \vspace{-2ex}\\
\colhead{} & \colhead{} & \colhead{} & \colhead{(\arcsec)} & \colhead{(km s$^{-1}$)}  & \colhead{(pc)}  & \colhead{(km s$^{-1}$)} & \colhead{(km s$^{-1}$)} & \colhead{(pc)} & \colhead{(10$^6$ years)}
}
\startdata
HST (Slit A) &    a & narrow & -0.41 &  -206 & 315 &    210 &     205 &     0.50 &  1.09 \\
HST (Slit A) &    b &   wide & -0.88 &  -213 & 683 &    217 &     237 &     0.45 &  2.27 \\
HST (Slit B) &    c &   wide &  0.10 &  -606 &  78 &    618 &     614 &     0.28 &  0.12 \\
HST (Slit B) &    d &   wide & -0.43 &  -422 & 334 &    431 &     419 &     0.355 &  0.69 \\
APO (PA 188) &    A & narrow &  3.15 &  -376 & 353 &    678 &     668 &     0.23 &  0.49 \\
APO (PA 188) &    B &   wide &  6.09 &  -351 & 682 &    632 &     652 &     0.23 &  0.98 \\
APO (PA 233) &    C &   wide &  3.15 &  -340 & 353 &    612 &     606 &     0.25 &  0.54 \\
APO (PA 233) &    D &   wide &  7.77 &  -286 & 870 &    515 &     516 &     0.28 &  1.55 \\
\enddata
\tablecomments{The columns list (1) the observation (\hst/APO), (2) given name for the observed knots with associated kinematics as seen in Figure \ref{fig:stis_apo_vel}, (3) the observed distances (in arcsec, (-) and (+) signs indicate a respective SW $\&$ NE direction for APO or E $\&$ W for STIS slits in the sky), (4) observed velocities, (5) the true/de-projected distances (in pc), (6) true/de-projected velocities using the parameters obtained for outflow bicone geometry (from kinematics model in section \S \ref{subsec:outflow}), (7) the velocities using the radiative driving model, (8) the calculated launch distance of these knots and (9) the total time traveled for each from launch distance to the true distance for the model velocity.}
\end{deluxetable*}

\section{SUMMARY AND Discussion} \label{sec:discuss}

\subsection{NLR Outflow and Geometry}\label{subsec:AGN_outflow}

We identify a distinct conical structure towards the NE in a high-resolution \hst \othree image of NGC~4051 that resembles the 9\arcsec\ long wedge-like structure seen in \othree ground-based observations \citep{Christopoulou1997}. The cone with an axis at PA $=$ 35\arcdeg\ defines the NLR in NGC~4051 and is in the same direction as the extended radio add{structure} at PA $\approx$ 41\arcdeg\ \citep{Ho2001}, rather than the previously proposed E-W orientation from \textit{HST} STIS \citep{Fischer2013} and GEMINI observations \citep{Barbosa2009}.

From spatially-resolved kinematics of the circumnuclear gas in NGC~4051 using \textit{HST} STIS and APO DIS long slit spectra, we identify up to two velocity components in the NLR, both of which are blueshifted with respect to the host galaxy. We find clouds with radial (line of sight) velocities up to $-$600 \kms in both STIS and APO observations towards the NE with respect to the central SMBH. Components with high radial velocities tend to have large widths, with FWHM up to $\sim$800 km s$^{-1}$. While the ionized gas kinematics for STIS show high flux, high velocity clouds close to the nucleus (inner 3\arcsec), the four 2\arcsec\ wide orthogonal long slits map the \othree ionized gas kinematics to extended distances in low flux \othree $\lambda\lambda$4959,5007 and $\lambda$6563 H$\alpha$ emission lines (See Figures \ref{fig:stis_apo_vel} and \ref{fig:rotation}) . The observed kinematics are dominated by outflow and do not show a detectable rotational component in the inner $\pm$10\arcsec\ ($\pm$800 pc). However the large scale \halpha kinematics closely follows the HI 21-cm rotation curve at projected distances $>$ 10\arcsec\ (Figure~\ref{fig:rotation}, \citealp{Kaneko1997}), indicating that the dominant kinematic component switches from outflow to rotation at the de-projected (true) distance of $\sim$1 kpc.

Kinematic modeling of the combined \hst and APO spectra shows that observed radial velocities in the inner $\pm$10\arcsec\ can be matched by conical outflow with an axis inclined by 60\arcdeg\ with respect to the plane of the sky and with minimum and maximum opening angles of 25\arcdeg\ and 40\arcdeg, refining the earlier values by \citep{Fischer2013}. Comparison with the host galaxy geometry confirms the \cite{Fischer2013} result that the presumed redshifted cone in the SW is likely hidden by dust in the galactic plane, thus explaining our detection of only blueshifted velocities with respect to the systemic velocity of the host.

BPT diagrams and the radial distribution of  [O~III]/H$\beta$ ratios show that AGN ionization dominates up to $\sim$10\arcsec\ from the nucleus towards the NE and NW. The moderately low [O~III]/H$\beta$ towards the SE and SW indicates star formation is a major contributor to the ionization at distances up to 5\arcsec, likely because these regions are outside of the nominal bicone or in areas where the redshifted cone is mostly extincted by the galactic disk, thereby reducing the AGN contribution. BPT diagrams show that the ionized gas beyond 10\arcsec\ is completely dominated by star formation, up to a projected distance of $\sim$100\arcsec\ ($\sim$8 kpc).

The NLR in NGC~4051 is well defined by AGN ionized outflows with high velocities and dispersion (intrinsic FWHM $>$ 250 km s$^{-1}$) up to 10$''$ from the SMBH. The gas at projected distances $>$ 10\arcsec\ is stellar ionized, has low velocity dispersion (FWHM $<$ 150 km s$^{-1}$), and is rotating in the plane of the host galactic disk. We find no evidence for a significant ENLR in NGC~4051, which we define as AGN ionized gas that is rotating in the plane, and is often characterized by large FWHM (identified as ``disturbed gas'' by \citealp{Fisher2018}).
An explanation is provided by our kinematic model of the outflow bicone. The axis of the bicone is pointed mostly out of the plane of the galaxy and therefore there is almost no interaction of the ionizing bicone with the rotating gas at larger distances in the host galaxy.

Given the inclination (60\arcdeg), inner (25\arcdeg), and outer (40\arcdeg) opening angles of our biconical outflow model, our view to the NLS1 nucleus of NGC~4051 is through the wall of the blueshifted cone, which suggests that many, if not all of the unusually large number of ionized blueshifted absorption systems detected in \hst UV spectra of NGC~4051  \citep{Collinge2001} are due to outflowing NLR clouds in our line of sight to the nucleus as suggested by \citet{Kraemer2012}. Supporting evidence comes from the range of outflowing velocities for the UV components, 0 to $-$650 km s$^{-1}$, which is essentially the same as that of the NLR clouds.

\subsection{Intermediate Line Region} \label{subsec:ILR}

\begin{figure*}[ht!]
\centering
\subfigure{
\includegraphics[width=0.90\textwidth]{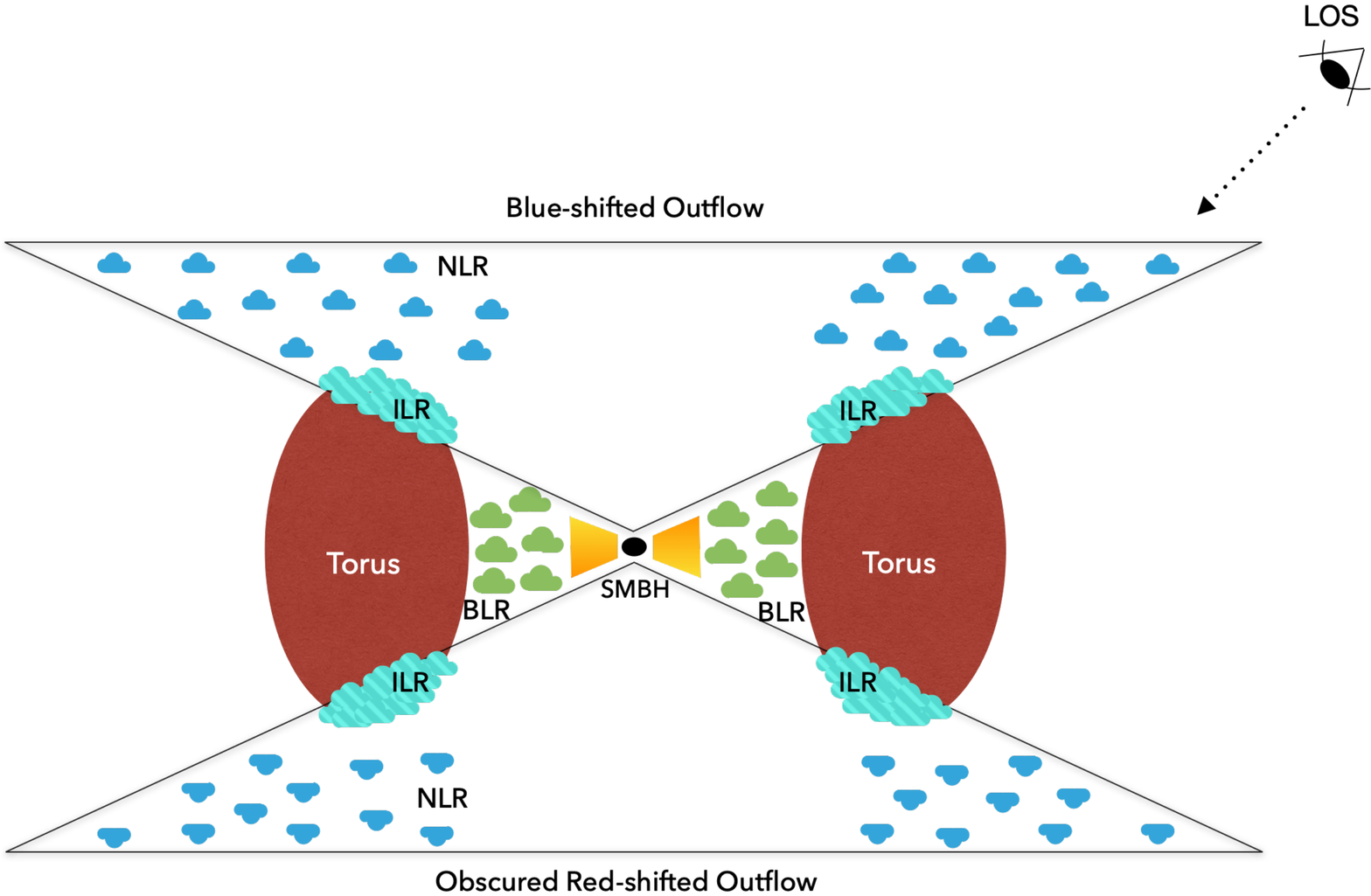}}
\caption{A cross-section interpretation of the nuclear regions in NGC 4051 which include the SMBH (black dot) with the accretion disk (yellow), BLR (green clouds), torus (brown), ILR (cyan clouds) and NLR outflows (blue clouds). The various regions in this cartoon are not on a linear scale. A \hbeta reverberation mapped BLR radius has been given by \cite{Denney2009} as R$_{BLR}$ = 1.87$^{+0.54}_{-0.50}$ {light days} and the inner edge of the torus is approximately located at 0.01-0.03 pc \citep{Suganuma2006,Krongold2007,Kishimoto2009} from SMBH. As discussed in section \S \ref{subsec:ILR}, we measured the ILR distance at $r_{\scaleto{ILR}{3pt}}$ $\approx$ 0.5 pc, which is either located close to the outer edge of the torus or is part of it's fueling material. The NLR expands from close to the ILR up to 1 kpc, pointing out of the host galaxy disk.}
\label{fig:cartoon}
\end{figure*}

We detect a spatially unresolved (FWHM $<$ 0.1\arcsec, kinematic component close to the nucleus with an intermediate width (FWHM $=$ 1010 km s$^{-1}$) between those of the broad and narrow emission lines in both \hst and APO spectra, similar to that found by \citet{Kraemer2012} and \citet{Yang2013}.
We find a weak ILR component in spectral fits of \othree $\lambda\lambda$4959,5007 as well as \hbeta lines in our observations, with a significantly small ratio for the integrated fluxes of the ILR lines of [O~III]/H$\beta$
$\approx$ 0.55. Considering a typical [O~III]/H$\beta$ ratio of $\sim$10 for low density gas in the NLR, and a value of $\sim$5 at the critical density ($n_{e}$ $= 6.8 \times 10^{5}$ cm$^{-3}$) of \othree $\lambda$5007 emission \citep{Osterbrock_Ferland2006}, the observed ratio indicates an ILR gas density of $n_{e} \approx 10^{7.5}$ cm$^{-3}$. Now, the size of the ILR can be calculated using the ionization parameter (U), and the number of ionizing photons per sec emitted by the AGN ($Q(\mathrm{H})$)\footnote{$Q(\mathrm{H})~=~\int_{\nu_0}^{\infty}
(L_{\nu}/h\nu) d\nu$, where $L_{\nu}$ is the luminosity of the AGN as a function of frequency (the SED), $h$ is Planck's constant, and $\nu_0 = 13.6 eV/h$ is the ionization potential of hydrogen \citep[\S 14.3]{Osterbrock_Ferland2006}.} \citep{Osterbrock_Ferland2006}.

\begin{equation}
U = \frac{Q(\mathrm{H})}{4 \pi r^2 n_\mathrm{H} \mathrm{c}},
\label{ionparam}
\end{equation}
where $r$ is the distance from the central AGN (cm), $n_{\mathrm{H}}$ is the hydrogen number density (cm$^{-3}$), and $\mathrm{c}$ is the speed of light. We adopted a value of log($Q(\mathrm{H})$) = 52.80 photons s$^{-1}$ which was previously computed for NGC~4051 by \cite{Kraemer2012}. Based on the absence of low ionization lines such as [S II] from the ILR, we assume an ionization parameter of log(U) $=$ $-$2. Using these values and estimated hydrogen density in Equation \ref{ionparam}, we calculate the distance of the  ILR from AGN to be $r_{\scaleto{ILR}{3pt}}$ $\approx$ 0.24 pc, considerably less than our upper limit based on the ILR being unresolved by \hst STIS.

The inner edge of the torus is often defined by the dust sublimation radius ($r_{sub}$) which has previously been calculated as $\sim$0.01 pc based on the UV luminosity of NGC~4051 \citep{Krongold2007}.
Reverberation mapping of the V and K band variability provides a similar value of 0.011$\pm$0.004 pc \citep{Suganuma2006}.
An ``effective'' inner radius of 0.032$\pm$0.003 pc has been determined by fitting ring model to K-band interferometry \citep {Kishimoto2009}.
However, it is not clear how far the torus extends and how it may connect to the fueling flow from the galaxy.
Thus, it is likely that the ILR originates from the outer portion of the dusty torus \citep{Garcia2019} or from the molecular flow that is fueling the torus.
Although the ILR is unresolved, it has a net blueshifted velocity of $\sim$ $-$110 km/sec, indicating that it may have an outflow component as well.

\citep{Riffel2008} find a blueshifted (up to $-$100 km s$^{-1}$), slightly curved structure in their Gemini NIFS observations of warm H$_2$ emission extending up to 2\arcsec\ along the eastern edge of the ionized bicone. They present arguments for and against inflow and outflow of the H$_2$ gas, and slightly prefer the inflow interpretation. We suggest that outflow is more likely, because the H$_2$ emission is nearly coincident with the ionized gas, suggesting a common origin as found in our NIFS studies of Mrk~573 \citep{Fischer2017} and Mrk~3 \citep{Gnilka2020}.

\subsection{Outflow Origins} \label{subsec:outflow_origin}

\begin{figure*}[htb!]
\centering
\subfigure{
\includegraphics[width=0.95\textwidth]{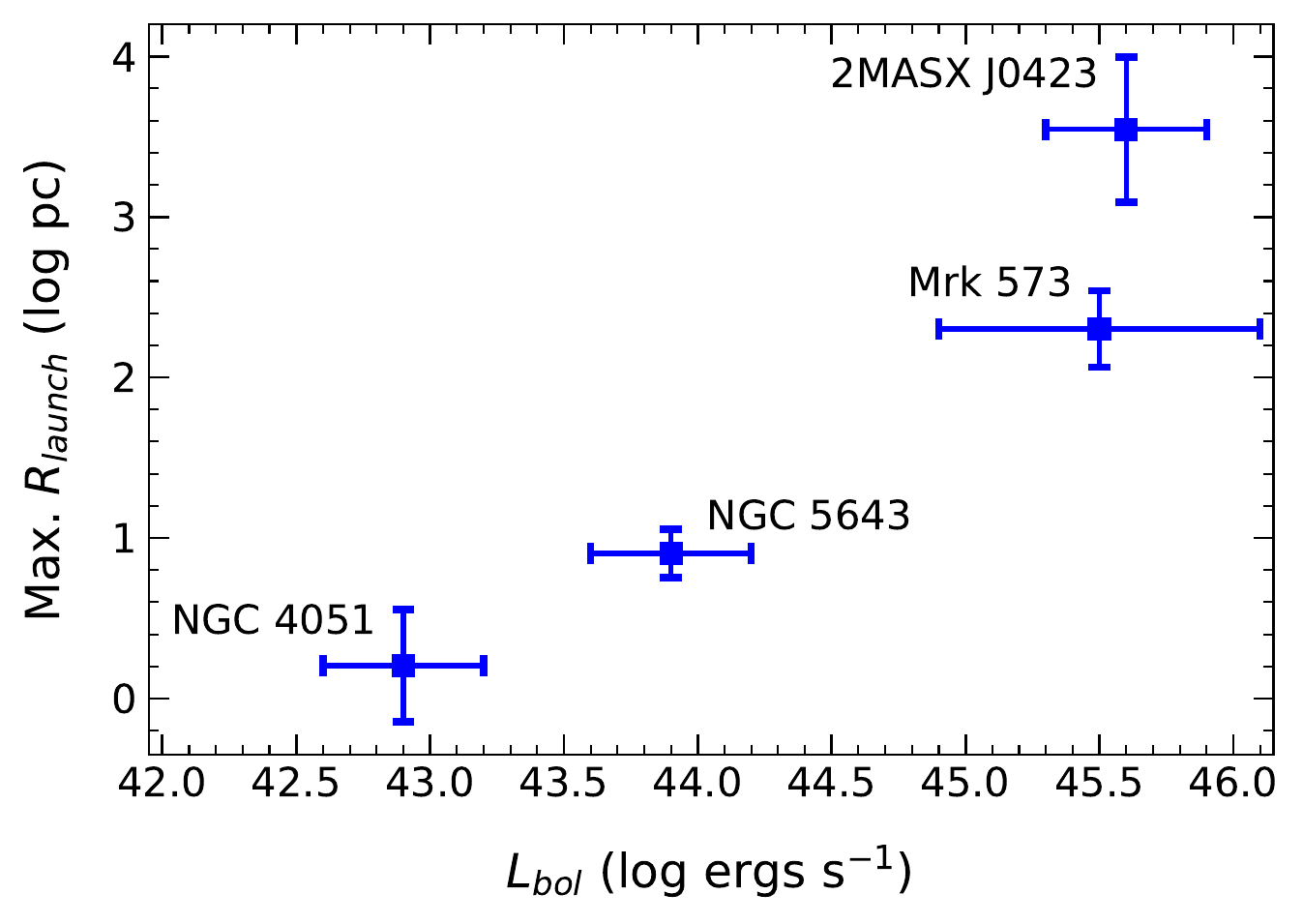}}
\caption{A comparison of the maximum launch radius, $R_{launch}$ from which the outflow clouds can be radiatively driven in the four AGN of different bolometric luminosities ($L_{bol}$). The launch distances for NGC~4051 are calculated from this work and rest were extracted from \cite{Garcia2021} (NGC 5643), \cite{Fischer2017} (Mrk 573) and \cite{Fischer2019} (2MASX J0423).  The uncertainties in the launch distances were determined based on $\mathcal{M}$ variation. For NGC 4051, Mrk 573 and 2MASX J0423, lower limit corresponds to $\mathcal{M}$ = 500 and an upper limit corresponds to $\mathcal{M}$ = 3300. For NGC 5643, $\mathcal{M}$ was given in the range of 200 to 400 (See Figure 12, \citealp{Garcia2021}). The uncertainties in bolometric luminosities was adopted as $\pm$0.3 dex for NGC~4051 and 2MASX J0423 from uncertainties in the [O~III] to bolometric luminosity relationship and correction factor. The bolometric luminosity for NGC~5643 is  the same as chosen by \cite{Garcia2021} analysis, which lead to uncertainties = $\pm$0.3 dex from scattering in L$_{14-195~\mathrm{keV}}$ \citep{Ricci2017}. The uncertainties for $L_{bol}$ for Mrk 573 is given as $\pm$0.6 dex in \cite{Kraemer2009}.}
\label{fig:compare}
\end{figure*}

As seen in Figure~\ref{fig:rad_drive}c, the clouds launched within $\sim$ 0.5 pc of the nucleus (for $\mathcal{M}$ = 500) can travel up to $\sim$1 kpc in NGC~4051 However, as we increase the launch distance, the clouds start to turn over and reach the systemic velocity of the galaxy within 100 pc. This is different than the case of Mrk~573, which has a much higher luminosity (log($L_{bol}$) = $45.5\pm0.6$ erg s$^{-1}$), where the gas clouds can be launched from within 600 pc (see Figure~15 in \citealp{Fischer2017}). Interestingly, though, nearly all of the bright clouds have traveled for distances $<$ 100 in Mrk~573, compared to hundreds of pc in NGC~4051. The maximum launch radius increases in Seyfert 2 galaxy 2MASX J0423 ($L_{bol}$ = $10^{45.55}$ erg s$^{-1}$) by almost 10 times, due to the lower gravitational potential of the host disk \citep{Fischer2019}. Recently, \citep{Garcia2021} showed that the ionized gas outflows as seen in MUSE/IFU observations of NGC 5643 ($L_{bol}$ = $8.14 \times 10^{43}$ erg s$^{-1}$) were launched within 4 pc of the nucleus.

This trend is shown in Figure \ref{fig:compare}, which illustrates that for a comparable mass distribution there is a luminosity dependence of the maximum distance to which the AGN is capable of driving the gas away in the form of ionized outflows, as expected from our radiative driving model.
Even though intermediate and low luminosity AGN can apparently drive the gas up to hundreds of parsecs, these clouds are originating from within a few pc of the nucleus and presumably not disrupting the gas reservoir outside those radii. On the other hand a brighter AGN is more successful in radiatively driving the gas from within hundreds of pc and thus disrupting the molecular reservoirs and impacting nuclear star formation on a large scale. Nonetheless, the total outflow travel distances tend to be much smaller then the full extents of AGN ionization and typical bulge sizes in these galaxies. That said, AGN driven outflows may potentially produce negative feedback by evacuating some fraction of the gas in bulges and affect nuclear star formation in galaxies with brighter AGN. However, this hypothesis requires further analysis of AGN on the higher end of luminosity range, as well as determinations of spatially-resolved mass outflow and kinetic energy rates over a broad range in AGN luminosity \citep{Revalski2021}.

Figure \ref{fig:cartoon} shows a geometric representation of the BLR, torus, ILR, and NLR locations in the nuclear regions of NGC 4051. It is possible that the ILR is not physically separated from the NLR but defines the very inner boundaries of the NLR outflows. 
Comparing the launch distances as given in Table \ref{tab:rad_drive} with the ILR radius ($\sim$0.24 pc) as discussed in section \S \ref{subsec:ILR}, it is likely that the outflowing clouds originated close to the ILR.
Thus in NGC 4051, both ILR and NLR were potentially formed as a result of AGN radiation interaction with the molecular reservoirs in the vicinity of the torus. In this scenario, the ILR represents the location of recently launched clouds.

\section{CONCLUSIONS} \label{sec:conclusion}
We present a detailed analysis of ionized gas outflow properties in the low luminosity NLS1 galaxy NGC~4051 and compare the effectiveness of radiative driving of the gas in the gravitational potential of the host galaxy with previous studies of higher luminosity AGN. Our main conclusions are:

\begin{enumerate}
\item The spatially resolved kinematics of ionized gas in the NLR of NGC~4051 show two blueshifted velocity components but no redshifted emissions. We find AGN ionization and outflow extends up to a projected distance of $\sim$10\arcsec\ ($\sim$800 pc) towards the NE. The outflow structure follows the \othree cone seen in a high resolution \textit{HST} WFC3 image and the alignment of the extended radio structure.

\item We developed a biconcial outflow model to determine the true velocities and distances of the observed outflowing clouds. The maximum extent of the observed outflows is $\sim$1 kpc, which is much larger than the $\sim$100 pc from \citep{Fischer2013} due to the addition of the APO data. The peak velocities are $\sim$680 \kms at a distance of $\sim$350 pc from the nucleus.

\item A radiation pressure driven wind plus gravitational deceleration model for the AGN ionized gas shows that the observed outflowing clouds in NGC~4051 originate within $\sim$0.5 pc (for a force multipler $\mathcal{M}$ $=$ 500, up to 3 pc for $\mathcal{M}$ $=$ 3000) from the SMBH and can travel up to $\sim$1 kpc. However the radiative driving fails beyond $\sim$3 pc radius from the center and no outflowing clouds can be produced at larger distances. Hence, unlike higher luminosity AGN, there is no evidence of in situ ionization and acceleration of gas from large distances in the host disk.

\item A comparison of maximum launch distance for this low luminosity AGN with those that were previously measured for higher luminosity AGN shows a strong positive correlation with luminosity. This means, for a comparable mass distribution, a higher luminosity AGN is more efficient in ionizing the gas and producing the outflows at larger distances from the nucleus. This implies that the AGN feedback is more efficient for brighter AGN. However a larger sample over a wider luminosity range is required to quantify any correlation.

\item We find a moderate-density intermediate line region (ILR) component in the \hbeta and \othree emission lines, with density $n_{\mathrm{H}}$ $\approx$ 10$^{7.5}$ cm$^{-3}$ and radius of $\sim$ 0.24 pc. The ILR component has FHWM  $=$ 1010 \kms and is blueshifted by $-$110 \kms with respect to the systemic velocity of the host galaxy. This component is unresolved in both \textit{HST} STIS and APO DIS spectra, with a brightness profile that follows the instrument/seeing PSF. The similar locations of the ILR and the outflow launch distances of the NLR outflows suggest they have a common origin, which is the gas evacuation caused by AGN radiation acting on the torus or its fueling flow.

\item Our line of sight passes through the wall of our biconical outflow model, which suggests that most, if not all, of the unusually large number of ionized UV absorbers seen in \hst spectra of NGC~4051 are due to NLR clouds seen in absorption. The nearly identical range in outflow velocity (0 to $-$650 km s$^{-1}$) of the NLR clouds and UV absorbers supports this interpretation.

\item Beyond $\sim$ 1 kpc, the gas is stellar ionized, rotating in the galactic plane, and has low FWHM. There is no significant evidence for AGN ionized gas rotating in the plane, and therefore no evidence for a significant ENLR in NGC~4051. This can be explained by our kinematic model of biconical outflow, as the near, blueshifted cone points up out of the galactic plane and therefore does not create a region of extended, AGN ionized gas in the plane.
\end{enumerate}

\acknowledgements

The authors would like to thank the anonymous referee for their constructive feedback on this paper. B.M. is also grateful of Dr. Misty Bentz and Justin H. Robinson for their helpful advice and discussions. 
This research has made use of NASA's Astrophysics Data System. This research has made use of the NASA/IPAC Extragalactic Database (NED), which is operated by the Jet Propulsion Laboratory, California Institute of Technology, under contract with the National Aeronautics and Space Administration. IRAF is distributed by the National Optical Astronomy Observatories, which are operated by the Association of Universities for Research in Astronomy, Inc., under cooperative agreement with the National Science Foundation

\facilities{APO(DIS, ARCTIC), \hst (STIS, WFC3)}

\software{IRAF \citep{Tody1986, Tody1993},
MultiNest \citep{Feroz2019}, SAOImage DS9 \citep{ds92000},
Mathematica \citep{Mathematica}, Python (\citealp{VanRossum2009}, \url{https://www.python.org}), Astropy \citep{astropy:2013}, Interactive Data Language (IDL, \url{https://www.harrisgeospatial.com/Software-Technology/IDL}).}

\bibliography{refs.bib}{}
\bibliographystyle{aasjournal}

\end{document}